\documentclass[english,11pt,oneside]{report}
\usepackage[a4paper,hmargin=3.5cm,vmargin=5.3cm,head=1.5cm,foot=3.2cm]{geometry}
\usepackage{url}          

\providecommand{\currentpdfbookmark}[2]{}

\usepackage{ifxetex}
\usepackage{amssymb,amsthm}
\usepackage{fontspec,unicode-math}
\defaultfontfeatures{Scale=MatchLowercase,Ligatures=TeX}
\usepackage{dsfont}

\ifxetex
\fi

\DeclareMathOperator*{\argmin}{arg\,min}
\DeclareMathOperator*{\argmax}{arg\,max}

\newcommand{\ket}[1]{\left|#1\right\rangle}
\newcommand{\bra}[1]{\left\langle#1\right|}
\newcommand{\braket}[1]{\left\langle#1\right\rangle}
\newcommand{\llbracket}{[\![}
\newcommand{\rrbracket}{]\!]}

\newtheorem{theorem}{Theorem}[chapter]
\newtheorem{corollary}[theorem]{Corollary}

\theoremstyle{definition}

\usepackage[english]{babel}    
\usepackage[bibencoding=utf8,backend=biber,sorting=none,bibstyle=phys]{biblatex}  
\usepackage{csquotes}

\usepackage{setspace}   
\renewcommand\footnotesize{\fontsize{9pt}{11pt}\selectfont} 
\usepackage{afterpage}    
\usepackage{titlefoot}    
\usepackage[explicit]{titlesec}  
\usepackage[titles]{tocloft}  
\usepackage[page]{appendix}  
\usepackage[normalem]{ulem}  

\usepackage{graphicx}  
\usepackage{subcaption} 
\usepackage{enumitem}   
\usepackage[figuresright]{rotating}  
\usepackage{placeins}    
\usepackage{adjustbox}    
\usepackage{algorithm}      
\usepackage{algpseudocode}  

\usepackage{array}       
\usepackage{booktabs}     
\usepackage{multicol} 		
\usepackage{multirow} 		

\usepackage{xcolor}			  
\usepackage[super]{nth}		
\usepackage{siunitx}		  

\usepackage[noabbrev,nameinlink]{cleveref}		

\DeclareSourcemap{
  \maps[datatype = bibtex]{
    \map{
       \step[fieldsource = abstract,
          match = \regexp{([^\\])\%},
          replace = \regexp{$1\\\%}]
    }
  }
}

\AtEveryBibitem{\clearfield{abstract}}
\AtEveryBibitem{\clearlist{abstract}}

\AtEveryBibitem{\clearfield{issn}}
\AtEveryBibitem{\clearlist{issn}}

\AtEveryBibitem{\clearfield{language}}
\AtEveryBibitem{\clearlist{language}}

\AtEveryBibitem{\clearfield{doi}}
\AtEveryBibitem{\clearlist{doi}}

\AtEveryBibitem{\clearfield{url}}
\AtEveryBibitem{\clearlist{url}}

\AtEveryBibitem{%
  \ifentrytype{online}
    {}
    {\clearfield{urlyear}\clearfield{urlmonth}\clearfield{urlday}}}

\newcommand\blfootnote[1]{%
  \begingroup
  \renewcommand\thefootnote{}\footnote{#1}%
  \addtocounter{footnote}{-1}%
  \endgroup
}

\newcommand{\chaptersource}[1]{%
  \begingroup
  \setlength{\fboxsep}{10pt}%
  \noindent\fbox{%
    \parbox{\dimexpr\linewidth-2\fboxsep-2\fboxrule\relax}{#1}%
  }%
  \endgroup
}

\begin{document}
\setcounter{tocdepth}{2} 


\newgeometry{hmargin=3.5cm,vmargin=5.3cm,head=1.5cm,foot=3.2cm}
\newcommand{\coverTitle}{When Quantum Meets AI: Quantum Methods for Machine Learning and Machine Learning Methods for Quantum Systems}
\newcommand{\coverName}{Tak Hur}


\begin{titlepage}
\clearpage\thispagestyle{empty}
\begin{center}
\vspace*{1.5cm}

{\fontsize{18pt}{22pt} \selectfont \coverTitle \vspace{3.5cm} \par}

{\fontsize{18pt}{22pt} \selectfont \coverName \vspace{3.5cm} \par}
{\fontsize{18pt}{22pt} \selectfont
Department of Statistics and Data Science \par
Graduate School  \par
Yonsei University \par}
\vfill

\end{center}
\end{titlepage}

\currentpdfbookmark{Cover Page}{coverPage}  


\newgeometry{hmargin=3.5cm,vmargin=5.3cm,head=1.5cm,foot=3.2cm}

\newcommand{\thesisTitle}{When Quantum Meets AI: Quantum Methods for Machine Learning and Machine Learning Methods for Quantum Systems}
\newcommand{\mythesisTitle}{When Quantum Meets AI: Quantum Methods for Machine Learning and Machine Learning Methods for Quantum Systems}
\newcommand{\yourName}{Tak Hur}
\newcommand{\yourMonth}{August}
\newcommand{\yourYear}{2026}


\begin{titlepage}
\clearpage\thispagestyle{empty}
\begin{center}
\vspace*{0.8cm}

{\fontsize{16pt}{20pt} \selectfont \mythesisTitle \vspace{1.3cm} \par}
{\fontsize{16pt}{20pt} \selectfont Advisor : Prof. Daniel K. Park \vspace{1.3cm} \par}
{\fontsize{16pt}{16pt}
\selectfont
    A Dissertation Submitted \par
    to the Department of Statistics and Data Science \par
    and the Committee of the Graduate School \par
    of Yonsei University in Partial Fulfillment of the \par
    Requirements for the Degree of \par
    Doctor of Philosophy \\
    \vspace{1.5cm}
    \par}

    {\fontsize{16pt}{18pt}
    \selectfont
    \yourName \par
    \vspace{0.8cm}
    \yourMonth{} \yourYear{} \par
}
\vfill
\par

\end{center}
\end{titlepage}

\currentpdfbookmark{Title Page}{titlePage}  



\clearpage
\newgeometry{hmargin=3.5cm,vmargin=5.3cm,head=1.5cm,foot=3.2cm}
\begin{centering}
{\fontsize{13pt}{16pt}\selectfont\textbf{ACKNOWLEDGEMENTS}}\\
\vspace{\baselineskip}
\end{centering}

I sincerely thank all the friends, colleagues, collaborators, and mentors whom I met during my PhD journey.
Looking back, what remains most valuable to me is not the results or achievements, but the memories of struggle and joy I shared with the people around me.
I would especially like to thank my advisor, Daniel K. Park, who not only guided me through the research, but also helped me grow to be a better person. 
I feel fortunate to have first met him when I was still an undergraduate.
I have learned so much from him in the years since, and I can only hope to one day pass on to others what he has given me.

I also thank my family, who have supported me throughout my life. 
Although I don't express it often, knowing that you are proud of what I do means more to me than I can say.
I would not be who I am today without your endless love and support. 

Finally, I thank my wife and best friend, Eun Hee. 
You have been by my side throughout my academic journey, from London to Seoul to Paris. 
The ups and downs we have shared along the way are among my most precious memories.
I am excited for our next chapter together, wherever it may take us.
If this thesis had a coauthor, it would be you.
\pagenumbering{gobble}
\thispagestyle{empty}
\clearpage


\newgeometry{hmargin=3.5cm,vmargin=5.3cm,head=1.5cm,foot=3.2cm,footskip=1cm}
\pagenumbering{roman}
\setcounter{page}{1} 
\renewcommand{\cftchapdotsep}{\cftdotsep}  
\renewcommand{\cftchapfont}{\bfseries}  
\renewcommand{\cftchappagefont}{}  
\renewcommand{\cftchappresnum}{Chapter }
\renewcommand{\cftchapaftersnum}{:}
\renewcommand{\cftchapnumwidth}{7em}
\renewcommand{\cftchapafterpnum}{\vskip\baselineskip} 
\renewcommand{\cftsecafterpnum}{\vskip\baselineskip}  
\renewcommand{\cftsubsecafterpnum}{\vskip\baselineskip} 
\renewcommand{\cftsubsubsecafterpnum}{\vskip\baselineskip} 

\renewcommand{\cftpartfont}{\bfseries\Large}
\renewcommand{\cftpartpagefont}{\bfseries}
\renewcommand{\cftpartafterpnum}{\vskip\baselineskip}

\titleformat{\chapter}[display]
{\normalfont\bfseries\fontsize{16pt}{20pt}\selectfont\filcenter}{\chaptertitlename\ \thechapter}{0pt}{\MakeUppercase{#1}}

\renewcommand\contentsname{Table of Contents}

\begin{singlespace}
\tableofcontents
\end{singlespace}

\currentpdfbookmark{Table of Contents}{TOC}

\clearpage


\addcontentsline{toc}{chapter}{List of Figures}
\begin{singlespace}
\setlength\cftbeforefigskip{\baselineskip}  
\listoffigures
\end{singlespace}

\clearpage

\addcontentsline{toc}{chapter}{List of Tables}
\begin{singlespace}
	\setlength\cftbeforetabskip{\baselineskip}  
	\listoftables
\end{singlespace}

\clearpage

\addcontentsline{toc}{chapter}{Abstract}
\clearpage
\begin{centering}
{\fontsize{13pt}{16pt}\selectfont\textbf{ABSTRACT}}\\
\vspace{\baselineskip}
\end{centering}

\begin{centering}
{\fontsize{13pt}{16pt}\selectfont\textbf{\mythesisTitle}}\\
\vspace{\baselineskip}
\end{centering}

\begin{flushright}
    Tak Hur  \\
    Department of Statistics and Data Science \\
    The Graduate School, Yonsei University
\end{flushright}

{\fontsize{10pt}{12pt}\selectfont
This thesis studies the intersection of quantum computing and artificial intelligence in two directions.
The first direction, \emph{Quantum for AI}, asks how quantum models can be used for machine learning.
The second direction, \emph{AI for Quantum}, asks how machine learning can help solve problems that arise in quantum error correction and quantum many-body physics.

The Quantum for AI part begins with background on quantum computing, supervised learning, quantum neural networks, and quantum kernels.
It then presents Neural Quantum Embedding, a method for learning the data embedding used before quantum classification.
The trace distance between embedded class ensembles sets a floor on the empirical risk achievable by the downstream classifier. And learning the embedding substantially raises this distinguishability, markedly improving classification accuracy on noisy quantum hardware.
A DQC1-compatible training objective based on the Hilbert--Schmidt inner product extends the method to ensemble quantum systems and is demonstrated on an NMR quantum processor.
The next contribution turns to generalization in quantum machine learning.
It establishes a margin-based generalization bound for quantum neural networks, shows empirically that margin distributions predict generalization more reliably than parameter-count metrics, and links achievable margins to the trace distance between class ensembles, connecting generalization to quantum state discrimination.

The AI for Quantum part applies neural methods to two quantum problems.
For quantum error correction, the thesis develops a Mamba-based neural decoder for the surface code whose inference cost scales quadratically rather than quartically with code distance. The decoder matches the accuracy of a Transformer baseline in memory experiments on simulated and real hardware data. In real-time decoding under an explicit decoder-induced-noise model, the Mamba decoder outperforms the Transformer baseline, achieving lower logical error rates and a higher effective error threshold.
For neural quantum states, the thesis analyzes stochastic reconfiguration, a standard optimization method in variational Monte Carlo.
It shows that the diagonal shift acts as a statistical spectral filter that trades bias against sampling variance under finite Monte Carlo sampling. A multi-shift variant combines independent regularized SR solves to reduce checkpoint-local validation residuals and update variance relative to standard SR at the training shift, at additional computational cost.

Together, these contributions position the intersection of quantum computing and machine learning not as a one-way application of techniques, but as a bidirectional exchange in which each field supplies principled tools for the other's hardest problems.
}

\blfootnote{Keywords: Quantum AI, Quantum Machine Learning, Quantum Error Correction, Neural Quantum States, Stochastic Reconfiguration, Generalization, Quantum Embeddings}

%
%


\clearpage
\pagenumbering{arabic}
\setcounter{page}{1} 

\newcommand{\chapternamefont}{\fontsize{16pt}{20pt}\selectfont}
\newcommand{\chaptertitlefont}{\fontsize{16pt}{20pt}\selectfont}
\titleformat{\chapter}[display]
{\normalfont\bfseries\chapternamefont\raggedright}{\MakeUppercase{\chaptertitlename\ \thechapter}}{0pt}{\MakeUppercase{#1}}  
\titlespacing*{\chapter}
  {0pt}{\topskip}{30pt}	

\titleformat{\section}{\normalfont\bfseries\fontsize{13pt}{16pt}\selectfont}{\thesection}{1em}{#1}

\titleformat{\subsection}{\normalfont\bfseries\fontsize{13pt}{16pt}\selectfont}{\thesubsection}{1em}{#1}

\titleformat{\subsubsection}{\normalfont\bfseries\itshape\fontsize{13pt}{16pt}\selectfont}{\thesubsubsection}{1em}{#1}




\part{Quantum for AI}


\chapter{Introduction to Quantum Machine Learning}
\label{chap:qml-introduction}


\section{Background}
\label{sec:ch1-background}

This chapter develops the technical background needed for the \emph{Quantum for AI} direction, while also introducing notation and machine learning concepts that reappear in Part~II.
Readers already familiar with quantum computing may skip \Cref{subsec:ch1-quantum-computing}, and those with a background in machine learning may skip \Cref{subsec:ch1-machine-learning}.

\subsection{Introduction to Quantum Computing}
\label{subsec:ch1-quantum-computing}

Quantum computing exploits the principles of quantum mechanics---superposition, entanglement, and interference---to process information in ways that differ fundamentally from classical computation. This section reviews the essential building blocks of quantum computation used in the subsequent chapters. For a comprehensive treatment, we refer the reader to~\textcite{nielsen2010quantum}.

\subsubsection{Qubits and Quantum States}

The fundamental unit of quantum information is the \emph{qubit}, a two-level quantum system. Unlike a classical bit, which takes a definite value $0$ or $1$, a qubit can exist in a \emph{superposition} of both computational basis states:
\begin{equation}
    |\psi\rangle = \alpha |0\rangle + \beta |1\rangle, \quad \alpha, \beta \in \mathbb{C}, \quad |\alpha|^2 + |\beta|^2 = 1.
    \label{eq:qubit}
\end{equation}
The state $|\psi\rangle$ is a unit vector in a two-dimensional complex Hilbert space $\mathcal{H} \cong \mathbb{C}^2$, and $\alpha$, $\beta$ are called \emph{probability amplitudes}.

A system of $n$ qubits lives in the tensor product space $\mathcal{H}^{\otimes n} \cong \mathbb{C}^{2^n}$. A general $n$-qubit state can be written as
\begin{equation}
    |\psi\rangle = \sum_{x \in \{0,1\}^n} c_x |x\rangle, \quad \sum_{x} |c_x|^2 = 1,
    \label{eq:nqubit}
\end{equation}
where $|x\rangle$ denotes a computational basis state. The dimension of the state space grows exponentially with $n$, meaning that describing a general quantum state requires an exponentially large number of complex amplitudes---a feature that lies at the heart of the potential power of quantum computation.

An important consequence of this tensor product structure is \emph{entanglement}: multi-qubit states that cannot be written as a product of individual qubit states. For example, the Bell state $|\Phi^+\rangle = \frac{1}{\sqrt{2}}(|00\rangle + |11\rangle)$ exhibits correlations between qubits that have no classical explanation and serve as a key resource in many quantum algorithms and communication protocols.

\subsubsection{Quantum Gates and Circuits}

The evolution of a closed quantum system is described by \emph{unitary operators}. A unitary operator $U$ satisfies $U^\dagger U = U U^\dagger = I$, which ensures that quantum states remain normalized under evolution. In the circuit model of quantum computation, algorithms are constructed by composing elementary unitary operations called \emph{quantum gates}.

Common single-qubit gates include the Pauli operators and the Hadamard gate:
\begin{equation}
    X = \begin{pmatrix} 0 & 1 \\ 1 & 0 \end{pmatrix}, \quad
    Y = \begin{pmatrix} 0 & -i \\ i & 0 \end{pmatrix}, \quad
    Z = \begin{pmatrix} 1 & 0 \\ 0 & -1 \end{pmatrix}, \quad
    H = \frac{1}{\sqrt{2}}\begin{pmatrix} 1 & 1 \\ 1 & -1 \end{pmatrix}.
    \label{eq:pauli}
\end{equation}
Other important single-qubit gates include parameterized rotation gates $R_P(\theta) = e^{-i\theta P/2}$ for $P \in \{X, Y, Z\}$. Multi-qubit entangling gates, such as the controlled-NOT (CNOT) gate, are essential for creating entanglement:
\begin{equation}
    \text{CNOT} = |0\rangle\langle 0| \otimes I + |1\rangle\langle 1| \otimes X.
    \label{eq:cnot}
\end{equation}
A fundamental result in quantum computing is that any unitary operation on $n$ qubits can be decomposed into a sequence of single-qubit gates and CNOT gates, forming a \emph{universal gate set}~\cite{nielsen2010quantum}.

A \emph{quantum circuit} is a sequence of quantum gates applied to a register of qubits, typically initialized in the state $|0\rangle^{\otimes n}$. The circuit model provides a convenient framework for designing and analyzing quantum algorithms, and it is the computational model used throughout this thesis.

\subsubsection{Measurement}

Quantum measurement extracts classical information from a quantum system. In the standard computational basis measurement, measuring an $n$-qubit state $|\psi\rangle = \sum_x c_x |x\rangle$ yields outcome $x$ with probability
\begin{equation}
    p(x) = |c_x|^2,
    \label{eq:born-rule}
\end{equation}
as dictated by the \emph{Born rule}. After the measurement, the state collapses to the observed basis state $|x\rangle$, destroying the superposition. This probabilistic and irreversible nature of measurement means that quantum algorithms must be carefully designed so that the desired answer is obtained with high probability, often by exploiting constructive and destructive interference among the amplitudes.

More generally, measurements can be described by a set of measurement operators $\{M_m\}$ satisfying the completeness relation $\sum_m M_m^\dagger M_m = I$, where outcome $m$ occurs with probability $p(m) = \langle\psi|M_m^\dagger M_m|\psi\rangle$. In this thesis, we primarily work with computational basis measurements and expectation values of observables $O$, computed as $\langle O \rangle = \langle\psi|O|\psi\rangle$.

\subsubsection{The Promise of Quantum Computation}

The exponentially large state space of $n$ qubits does not, by itself, guarantee computational advantage---after all, measurement collapses the state and yields only a single classical outcome. The power of quantum computing lies in the ability to orchestrate interference and entanglement so that useful answers emerge with high probability.

This principle has led to quantum algorithms that provably outperform the best known classical algorithms for specific problems. Shor's algorithm~\cite{shor1994algorithms} factors integers in polynomial time, an exponential speedup over known classical methods. Grover's algorithm~\cite{grover1996fast} searches an unstructured database of $N$ items in $O(\sqrt{N})$ queries, a quadratic improvement over the classical $O(N)$ lower bound. These results demonstrate that quantum computers can, in principle, solve certain problems that are intractable for classical machines.

In the near term, fully fault-tolerant quantum computers remain out of reach. Current devices, known as \emph{noisy intermediate-scale quantum} (NISQ) devices~\cite{preskill2018quantum}, contain tens to hundreds of qubits with limited coherence times and noisy gate operations. This has motivated the development of hybrid quantum-classical approaches that use shallow parameterized circuits amenable to near-term hardware, which we discuss in \Cref{sec:ch1-qml}.

\subsection{Introduction to Machine Learning}
\label{subsec:ch1-machine-learning}

Machine learning, broadly defined, is the study of algorithms that improve their performance on a task through experience~\cite{mitchell1997machine}. While the field encompasses a wide range of paradigms---including unsupervised, reinforcement, and self-supervised learning---this thesis focuses primarily on \emph{supervised learning}, where a model learns a mapping from inputs to outputs given a labeled training dataset. This section traces the key developments in machine learning that motivate the quantum extensions discussed in the remainder of this thesis.

\subsubsection{From Classical Methods to Deep Learning}

Early machine learning research focused on models with strong theoretical foundations but limited representational capacity. Linear classifiers, decision trees, and nearest-neighbor methods formed the initial toolkit. A major advance came with \emph{support vector machines} (SVMs)~\cite{cortes1995support}, which use the \emph{kernel trick} to implicitly map data into high-dimensional feature spaces where linear separation becomes possible. SVMs offered strong generalization guarantees grounded in statistical learning theory~\cite{vapnik1995nature} and dominated many benchmarks through the 2000s. Kernels and the margin theory of SVMs appear twice in this thesis: first in \Cref{subsec:ch1-quantum-kernels}, where we introduce their quantum generalization, and later in \Cref{chap:generalization}, where we leverage margin-based arguments from SVM theory to derive generalization bounds for quantum neural networks.

The deep learning revolution was catalyzed by \emph{convolutional neural networks} (CNNs). Although the core ideas date back to the neocognitron~\cite{fukushima1980neocognitron} and LeNet~\cite{lecun1998gradient}, the breakthrough came in 2012 when AlexNet~\cite{krizhevsky2012imagenet} won the ImageNet Large Scale Visual Recognition Challenge by a decisive margin, demonstrating that deep CNNs trained on GPUs could dramatically outperform hand-engineered feature pipelines. The key architectural principles---local connectivity, weight sharing, and hierarchical feature extraction---enabled CNNs to scale to high-dimensional image data while keeping the parameter count manageable. Subsequent architectures such as VGGNet~\cite{simonyan2015very}, GoogLeNet~\cite{szegedy2015going}, and ResNet~\cite{he2016deep} continued to push performance by increasing depth and introducing structural innovations like skip connections.

Beyond computer vision, \emph{recurrent neural networks} (RNNs) and their gated variants---Long Short-Term Memory (LSTM)~\cite{hochreiter1997long} and Gated Recurrent Units (GRU)~\cite{cho2014learning}---became the standard for sequential data, powering advances in machine translation, speech recognition, and natural language understanding.

\subsubsection{The Transformer Era}

The introduction of the \emph{Transformer} architecture~\cite{vaswani2017attention} marked a major shift across machine learning. By replacing recurrence with a \emph{self-attention} mechanism that computes pairwise interactions among all elements of a sequence in parallel, Transformers resolved the sequential bottleneck of RNNs and enabled efficient training on massive datasets. This architecture gave rise to large language models such as GPT~\cite{radford2018improving} and BERT~\cite{devlin2019bert}, and has since been adopted well beyond natural language processing---including computer vision (Vision Transformer~\cite{dosovitskiy2021image}), protein structure prediction (AlphaFold~\cite{jumper2021highly}), and scientific computing. Transformer-based architectures now form an important part of the modern machine learning toolkit, especially for sequence modeling and large-scale representation learning.

The Transformer paradigm also plays a central role in Part~II of this thesis.
In \Cref{chap:decoder}, we build on the AlphaQubit architecture, a Transformer-based neural decoder for quantum error correction, and propose a Mamba-based decoder that replaces the attention mechanism with a state-space model for improved scalability.
In \Cref{chap:nqs}, transformer and foundation neural quantum states provide the parameter-rich setting in which we analyze stochastic reconfiguration as a statistical spectral filter for variational Monte Carlo.

A recurring theme in these developments is that \emph{architectural design}---the choice of hypothesis class---plays a decisive role in model performance. CNNs succeeded because their inductive biases match the structure of image data; Transformers succeeded because self-attention captures long-range dependencies that recurrent models struggle with. This observation motivates a central question of this thesis: can quantum computing offer new model classes with inductive biases that are advantageous for certain learning problems?

\subsubsection{Sources of Prediction Error}
\label{subsubsec:error-decomposition}

To reason about the performance of any learning algorithm, it is useful to decompose the prediction error into distinct components~\cite{shalev2014understanding}. Consider a supervised learning problem where we seek a function that maps inputs $x \in \mathcal{X}$ to outputs $y \in \mathcal{Y}$. Let $\mathcal{R}(f) := \mathbb{E}_{(x,y) \sim \mathcal{D}}[\ell(f(x), y)]$ denote the \emph{risk} (expected loss) of a function $f$ with respect to the true data distribution $\mathcal{D}$, and let $\widehat{\mathcal{R}}(f) := \frac{1}{n}\sum_{i=1}^{n} \ell(f(x_i), y_i)$ denote the \emph{empirical risk} computed over $n$ training examples.

In practice, we do not have access to $\mathcal{R}$. Instead, we can only minimize $\widehat{\mathcal{R}}$ over a chosen hypothesis class $\mathcal{F}$. Let $f^*$ denote a population-risk minimizer over the unrestricted target class, let $f_{\mathcal{F}}^* \in \argmin_{f \in \mathcal{F}} \mathcal{R}(f)$ denote the best-in-class predictor for the population risk, let $\hat{f}_{\mathcal{F}} \in \argmin_{f \in \mathcal{F}} \widehat{\mathcal{R}}(f)$ denote an empirical-risk minimizer, and let $\hat{f}$ denote the function returned by the actual training algorithm. The excess risk of $\hat{f}$ can then be decomposed as:
\begin{equation}
    \mathcal{R}(\hat{f}) - \mathcal{R}(f^*)
    =
    \underbrace{\mathcal{R}(f_{\mathcal{F}}^*) - \mathcal{R}(f^*)}_{\text{approximation}}
    +
    \underbrace{\mathcal{R}(\hat{f}_{\mathcal{F}}) - \mathcal{R}(f_{\mathcal{F}}^*)}_{\text{estimation / generalization}}
    +
    \underbrace{\mathcal{R}(\hat{f}) - \mathcal{R}(\hat{f}_{\mathcal{F}})}_{\text{optimization}}.
    \label{eq:error-decomposition}
\end{equation}
Each term captures a distinct source of error:
\begin{itemize}
    \item \textbf{Approximation error}: residual risk from the limited \emph{expressivity} of $\mathcal{F}$.

    \item \textbf{Estimation error}: the cost of learning from finite data (\emph{generalization}).

    \item \textbf{Optimization error}: suboptimality from non-convex training.
\end{itemize}
This decomposition provides a useful organizing framework for Part~I of the thesis. Specifically, the following chapters study how quantum embeddings set a representation-induced floor on achievable training loss and how margins affect generalization in quantum machine learning.

\section{Quantum Machine Learning}
\label{sec:ch1-qml}

\subsection{History}
\label{subsec:ch1-qml-history}

Quantum machine learning has been an important area of quantum computing research for over a decade. An influential early result in this direction was the \emph{HHL algorithm}~\cite{harrow2009quantum}, proposed by Harrow, Hassidim, and Lloyd in 2009, which solves a system of linear equations $A\mathbf{x} = \mathbf{b}$ in time logarithmic in the dimension under assumptions on sparsity, conditioning, and quantum state preparation. This result suggested that quantum computers could accelerate core linear algebra subroutines underlying many machine learning methods.

Building on HHL, a wave of quantum machine learning algorithms emerged in the early 2010s. \textcite{rebentrost2014quantum} proposed a quantum support vector machine (QSVM) that performs classification in $O(\log N)$ time by leveraging quantum linear algebra to solve the least-squares formulation of SVMs. Similarly, \textcite{lloyd2014quantum} introduced quantum principal component analysis (QPCA), which extracts eigenvalues and eigenvectors of a density matrix exponentially faster than classical PCA. These results generated significant optimism that quantum computers could accelerate machine learning workloads under strong data-access and state-preparation assumptions.

However, these algorithms share a critical assumption: they require efficient \emph{quantum access} to classical data, typically through a quantum random access memory (QRAM) that can prepare arbitrary quantum states encoding the input data in superposition. Constructing a practical QRAM remains a major open challenge---the hardware overhead required to maintain coherent access to $N$ classical data entries may negate the very speedup that the algorithms promise~\cite{jaques2023qram}. Without QRAM, the cost of loading classical data into quantum states can dominate the total runtime, reducing the practical advantage to at best polynomial.

A more fundamental challenge to these early speedup claims came from a series of \emph{dequantization} results by \textcite{tang2019quantum}, who demonstrated a classical algorithm for recommendation systems that matches the quantum algorithm's performance up to polynomial factors, assuming only classical \emph{sample and query} access to the input data. This work was followed by the dequantization of quantum PCA~\cite{tang2021quantum}, which showed that much of the claimed exponential speedup was an artifact of the strong input model (quantum state preparation) rather than a genuine computational advantage of quantum mechanics. Together, these results established that quantum speedups built on the HHL framework are far more fragile than initially believed.

\begin{figure}[t]
    \centering
    \begin{subfigure}[b]{0.54\textwidth}
        \centering
        \includegraphics[width=\textwidth]{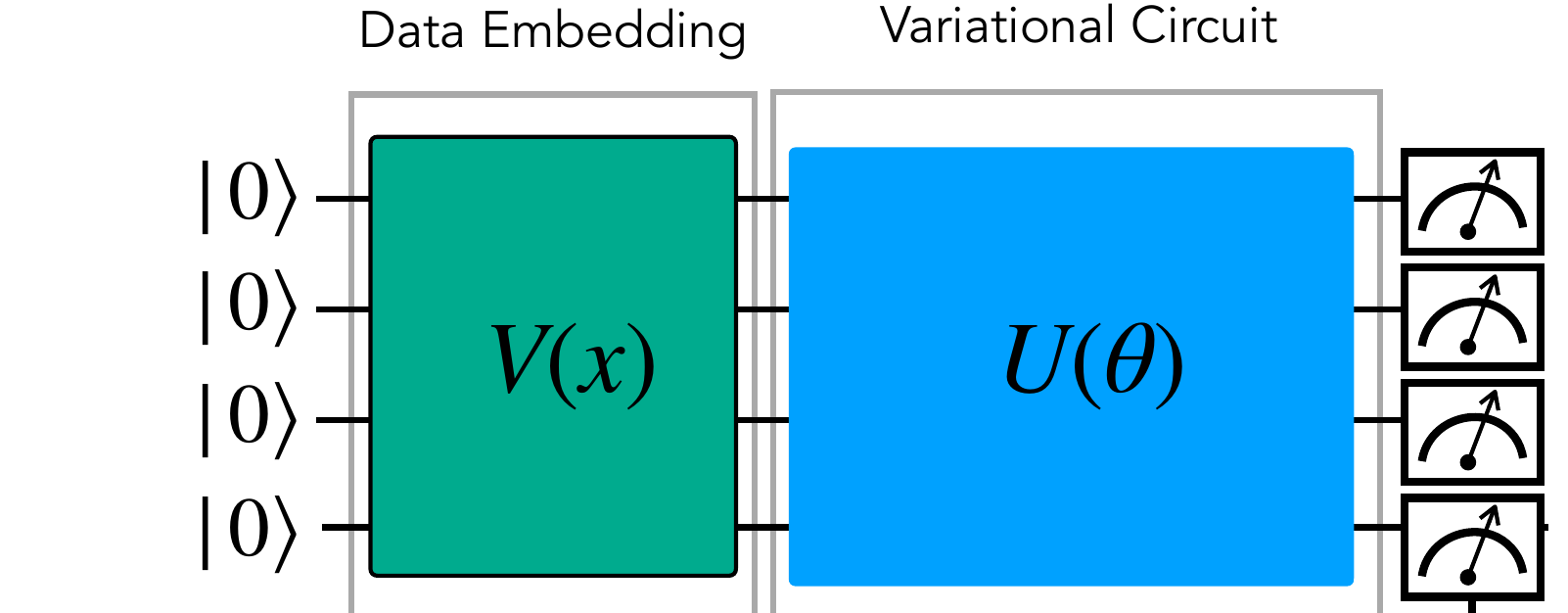}
        \caption{Quantum Neural Network (QNN)}
        \label{fig:qnn-schematic}
    \end{subfigure}
    \hfill
    \begin{subfigure}[b]{0.421\textwidth}
        \centering
        \includegraphics[width=\textwidth]{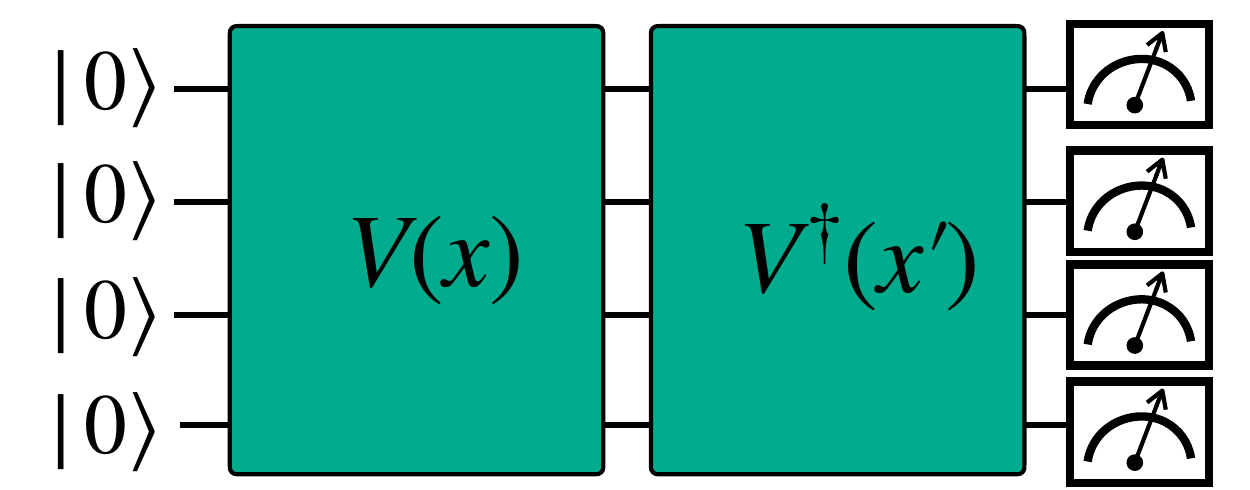}
        \caption{Quantum Kernel Method (QKM)}
        \label{fig:qkm-schematic}
    \end{subfigure}
    \caption{Schematics of the two dominant paradigms in near-term quantum machine learning. (a)~A quantum neural network uses a parameterized quantum circuit $U(\boldsymbol{\theta})$ trained via a classical optimizer. (b)~A quantum kernel method uses a fixed quantum feature map to compute kernel entries, which are then processed by a classical algorithm such as an SVM.}
    \label{fig:qnn-qkm}
\end{figure}

The QRAM bottleneck and the dequantization barrier shifted the focus of the field from fault-tolerant quantum linear-algebra subroutines toward models that can be run on noisy intermediate-scale quantum (NISQ) devices~\cite{preskill2018quantum}.
This shift did not remove the need for rigorous advantage claims. Rather, it changed the practical emphasis toward trainability, representation, finite samples, and hardware noise.
Within this NISQ-era setting, two supervised-learning paradigms are central for this thesis: quantum neural networks and quantum kernel methods.
They are illustrated in \Cref{fig:qnn-qkm} and discussed in the following subsections.

\subsection{Quantum Neural Networks}
\label{subsec:ch1-qnn}

The term \emph{parameterized quantum circuit} covers a broad family of trainable circuits used in variational quantum algorithms, including VQE for energy minimization~\cite{peruzzo2014variational} and QAOA for combinatorial optimization~\cite{farhi2014qaoa}.
In this thesis, the term \emph{quantum neural network} (QNN) refers more narrowly to supervised-learning trainable quantum models built from such parameterized circuits~\cite{cerezo2021variational}.
In a typical supervised QNN, a data encoding circuit $V(\mathbf{x})$ first embeds classical input $\mathbf{x}$ into a quantum state, followed by a trainable circuit $U(\boldsymbol{\theta})$ with parameters $\boldsymbol{\theta}$.
The prediction is obtained by measuring an observable $O$:
\begin{equation}
    f(\mathbf{x}; \boldsymbol{\theta}) = \langle 0 | V^\dagger(\mathbf{x}) \, U^\dagger(\boldsymbol{\theta}) \, O \, U(\boldsymbol{\theta}) \, V(\mathbf{x}) | 0 \rangle.
    \label{eq:qnn-prediction}
\end{equation}
When the input is quantum data (e.g., a quantum state $\rho$ from a physical system), the encoding circuit $V(\mathbf{x})$ is replaced by the quantum state itself, and the QNN acts directly on the given state. A classical optimizer iteratively updates $\boldsymbol{\theta}$ to minimize a loss function computed from the measurement outcomes. One example of the trainable circuit $U(\boldsymbol{\theta})$ is the \emph{hardware-efficient ansatz} (HEA)~\cite{kandala2017hardware}, consisting of $L$ layers of single-qubit rotation gates followed by entangling two-qubit gates:
\begin{equation}
    U(\boldsymbol{\theta}) = \prod_{\ell=1}^{L} W_{\text{ent}} \cdot \bigotimes_{i=1}^{n} R(\theta_i^{(\ell)}),
    \label{eq:hea}
\end{equation}
where $R(\theta) = R_Z(\theta_1) R_Y(\theta_2) R_Z(\theta_3)$ represents a general single-qubit rotation (up to global phase), and $W_{\text{ent}}$ denotes a fixed entangling layer such as a ladder of CNOT gates matching the hardware topology. Beyond generic hardware-efficient circuits, QNNs also include structured ansatz families that build architectural bias into the circuit.

\subsubsection{Structured QNNs: Quantum Convolutional Neural Networks}
\label{subsec:ch1-qcnn}

\begin{figure}[t]
    \centering
    \includegraphics[width=0.9\textwidth]{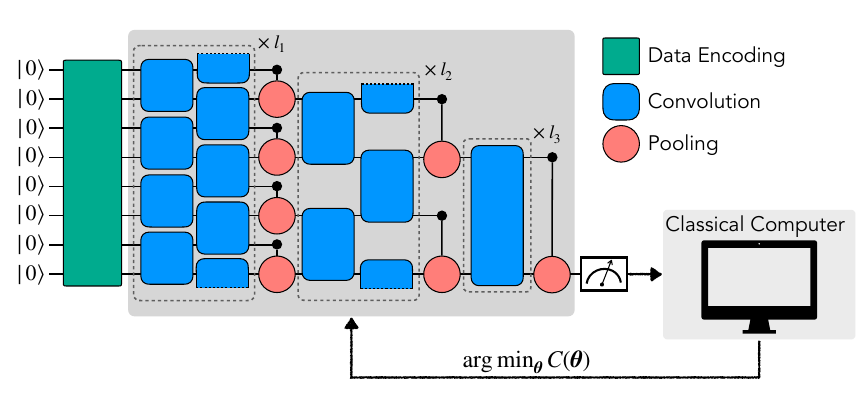}
    \caption{Schematic of a quantum convolutional neural network (QCNN). The circuit alternates between convolutional layers (two-qubit unitaries with shared parameters across neighboring pairs) and pooling layers, until a small number of qubits remain for the final measurement.}
    \label{fig:qcnn}
\end{figure}

A widely studied structured QNN architecture is the \emph{quantum convolutional neural network} (QCNN), introduced by \textcite{cong2019quantum} for classifying quantum phases of matter. QCNNs draw direct inspiration from classical CNNs by combining local receptive fields, shared parameters, and hierarchical pooling. The architecture alternates between convolutional and pooling layers, progressively reducing the system size until only a few qubits remain for measurement:
\begin{equation}
    U_{\text{QCNN}} = U_{\text{conv}}^{(L)} \circ \text{Pool}^{(L-1)} \circ \cdots \circ U_{\text{conv}}^{(2)} \circ \text{Pool}^{(1)} \circ U_{\text{conv}}^{(1)},
    \label{eq:qcnn-full}
\end{equation}
where $L$ denotes the number of convolutional layers. As illustrated in \Cref{fig:qcnn}, each convolutional layer applies a translationally invariant two-qubit parameterized unitary $U(\boldsymbol{\theta})_{i,i+1}$ across all neighboring qubit pairs, with shared parameters analogous to the weight-sharing mechanism in classical CNNs. Each pooling layer then halves the number of active qubits by measuring a subset and applying conditioned unitaries on the remaining qubits. After $\log_2(n)$ pooling operations, only $\mathcal{O}(1)$ qubits remain for the final classification measurement.

Although QCNNs were originally designed for quantum data, prior work also applied them to classical data classification by prepending a data encoding circuit~\cite{hur2022quantum}, where they serve as compact, parameter-sharing QNN classifiers.

QCNNs enjoy several favorable properties as QNN architectures. Their hierarchical structure and local operations provably avoid barren plateaus~\cite{pesah2021absence}, enabling efficient training even for large system sizes. On the other hand, several local, hierarchical QCNN architectures admit efficient classical simulation or tensor-network contraction under additional structural assumptions~\cite{bermejo2024qcnn}.

\subsubsection{Gradient Computation: The Parameter-Shift Rule}

Training QNNs requires computing gradients of the cost function with respect to the circuit parameters. Unlike classical neural networks, where backpropagation provides an efficient route to exact gradients, quantum circuits do not admit direct differentiation---measurements collapse the quantum state and prevent the propagation of gradient information. Recent work has explored quantum analogues of backpropagation~\cite{abbas2023backpropagation}, showing that achieving backpropagation-like scaling requires access to multiple copies of the quantum state via shadow tomography.

A key insight enabling gradient-based optimization of QNNs is the \emph{parameter-shift rule}~\cite{mitarai2018quantum,schuld2019evaluating}. For gates of the form $R_P(\theta) = e^{-i\theta P/2}$ where $P^2 = I$ (such as Pauli rotation gates), the partial derivative of the expectation value can be computed exactly using only two circuit evaluations:
\begin{equation}
    \frac{\partial}{\partial \theta} f(\theta) = \frac{1}{2}\left[ f\left(\theta + \frac{\pi}{2}\right) - f\left(\theta - \frac{\pi}{2}\right) \right],
    \label{eq:parameter-shift}
\end{equation}
where $f(\theta) = \text{Tr}[\rho \, U^\dagger(\theta) \, O \, U(\theta)]$ for an input state $\rho$. This formula provides the exact analytical gradient (not a finite-difference approximation) using only shifted circuit evaluations, making it naturally compatible with quantum hardware where only expectation values can be measured.

The parameter-shift rule forms the foundation for gradient-based training of QNNs, enabling the use of optimizers such as gradient descent, Adam, and their variants.

\subsubsection{The Barren Plateau Problem}

A major obstacle to training QNNs is the \emph{barren plateau} phenomenon~\cite{mcclean2018barren}, wherein the variance of the cost function gradient vanishes exponentially with the number of qubits $n$:
\begin{equation}
    \text{Var}_{\boldsymbol{\theta}}\left[\frac{\partial C}{\partial \theta_k}\right] \leq F(n) \cdot 2^{-n},
    \label{eq:barren-plateau}
\end{equation}
where $F(n)$ is a polynomial factor. When the gradient variance is exponentially small, an exponential number of measurement shots is required to distinguish the gradient from statistical noise, negating any potential quantum advantage. Barren plateaus can arise from various sources, including excessive circuit depth, global cost functions, noise, and high entanglement. Recent work has provided a unified understanding of barren plateaus through the lens of the \emph{dynamical Lie algebra} (DLA) of the circuit generators~\cite{ragone2024lie,fontana2024adjoint}. These results show that the gradient variance scales inversely with the dimension of the DLA, establishing that circuits with a polynomially scaling DLA can escape barren plateaus, while those with an exponentially scaling DLA cannot. We refer the reader to \textcite{larocca2025barren} for a comprehensive review.

\subsection{Quantum Kernels}
\label{subsec:ch1-quantum-kernels}

Kernel methods offer an alternative paradigm for quantum machine learning, one that trades the parameterized circuits of QNNs for a fixed quantum feature map and classical post-processing.
This approach provides strong theoretical guarantees, and in constructed supervised-learning settings it has yielded rigorous separations under explicit computational assumptions.
We begin with a brief review of classical kernel methods before introducing their quantum generalization.

\subsubsection{Classical Kernel Methods}

In classical machine learning, kernel methods~\cite{cortes1995support,vapnik1995nature} address the fundamental challenge of learning nonlinear decision boundaries by implicitly mapping data into a high-dimensional \emph{feature space}. Given a feature map $\phi: \mathcal{X} \to \mathcal{F}$ that embeds inputs $\mathbf{x} \in \mathcal{X}$ into a (possibly infinite-dimensional) Hilbert space $\mathcal{F}$, a kernel function computes inner products in this feature space without explicitly constructing the embedding:
\begin{equation}
    K(\mathbf{x}, \mathbf{x}') = \langle \phi(\mathbf{x}), \phi(\mathbf{x}') \rangle_{\mathcal{F}}.
    \label{eq:kernel-def}
\end{equation}
This \emph{kernel trick} enables algorithms like support vector machines to operate efficiently in feature spaces of exponential or even infinite dimension. The \emph{representer theorem} further guarantees that the optimal predictor in a kernel-regularized learning problem can be expressed as a linear combination of kernel evaluations on the training data:
\begin{equation}
    f^*(\mathbf{x}) = \sum_{i=1}^{N} \alpha_i K(\mathbf{x}, \mathbf{x}_i).
    \label{eq:representer}
\end{equation}
This elegant structure motivates the question: can quantum computers realize feature maps that are computationally intractable for classical machines, while remaining efficiently estimable?

\subsubsection{Quantum Feature Maps}

A \emph{quantum feature map} encodes classical data $\mathbf{x}$ into a quantum state by applying a unitary transformation to an initial state~\cite{havlicek2019supervised}:
\begin{equation}
    |\phi(\mathbf{x})\rangle = V(\mathbf{x}) |0\rangle^{\otimes n}.
    \label{eq:quantum-feature-map}
\end{equation}
The quantum states $|\phi(\mathbf{x})\rangle$ live in the $2^n$-dimensional Hilbert space of $n$ qubits, and the \emph{quantum kernel} is defined as the squared overlap between embedded states:
\begin{equation}
    K(\mathbf{x}, \mathbf{x}') = |\langle \phi(\mathbf{x}) | \phi(\mathbf{x}') \rangle|^2 = |\langle 0 |^{\otimes n} V^\dagger(\mathbf{x}) V(\mathbf{x}') | 0 \rangle^{\otimes n}|^2.
    \label{eq:quantum-kernel}
\end{equation}
This kernel can be estimated on a quantum computer by preparing the state $V^\dagger(\mathbf{x}) V(\mathbf{x}') |0\rangle^{\otimes n}$ and measuring the probability of obtaining the all-zeros outcome. The key insight is that if $V(\mathbf{x})$ generates states with classically intractable correlations, the resulting kernel may be hard to compute classically, yet remain efficiently estimable on quantum hardware.

\subsubsection{The ZZ Feature Map}

A concrete realization of this idea is the \emph{ZZ feature map} introduced by \textcite{havlicek2019supervised}. This feature map applies $D$ repetitions of a data-encoding unitary:
\begin{equation}
    V(\mathbf{x}) = \left[ \exp\left(i \sum_{i<j} (\pi - x_i)(\pi - x_j) Z_i Z_j \right) \exp\left(i \sum_i x_i Z_i \right) H^{\otimes n} \right]^D,
    \label{eq:zz-feature-map}
\end{equation}
where $H^{\otimes n}$ denotes Hadamard gates on all qubits, $Z_i$ is the Pauli-$Z$ operator on qubit $i$, and $\mathbf{x} = (x_1, \ldots, x_n)$ is the input vector. The two-qubit $ZZ$ interactions create entanglement that depends on the input data, embedding classical features into quantum correlations.

The structure of the ZZ feature map is designed so that the resulting kernel involves multi-qubit correlations that are believed to be classically hard to compute. While computing the kernel exactly incurs a cost that scales exponentially in the number of qubits classically, a quantum computer can estimate it in polynomial time using the overlap measurement circuit.

\subsubsection{Constructed Quantum Advantage Results}

The question of whether quantum kernels can provide a computational advantage over classical methods was rigorously addressed by \textcite{liu2021rigorous} in a constructed problem setting.
Their work constructs a learning problem whose labels are defined through the \emph{discrete logarithm problem}, an instance of the abelian hidden subgroup problem that Shor's algorithm solves in polynomial time but for which no efficient classical algorithm is known.
The key result is the existence of a data distribution $\mathcal{D}$ such that:
\begin{enumerate}
    \item A quantum kernel classifier can learn to classify samples from $\mathcal{D}$ with high accuracy.
    \item No polynomial-time classical learner can classify significantly better than random guessing, assuming the discrete logarithm problem is classically intractable; the barrier is computational rather than a lack of training data.
\end{enumerate}
This provides a rigorous separation between quantum and classical learning under the stated distributional and computational assumptions.
Importantly, the separation is shown to survive the finite-sampling (shot) noise incurred when the kernel entries are estimated from a finite number of measurements, addressing earlier concerns about the fragility of quantum speedup claims.
This establishes that a provable quantum advantage with quantum kernels is possible in principle. However, whether such an advantage extends to natural, real-world datasets remains an open question.

\subsubsection{Relationship to Quantum Neural Networks}

A fundamental connection between quantum kernels and QNNs was established by \textcite{schuld2019quantum}. They showed that QNNs with fixed data encodings are equivalent to linear models in the quantum feature space. Specifically, for a QNN with output
\begin{equation}
    f(\mathbf{x}) = \langle \phi(\mathbf{x}) | O | \phi(\mathbf{x}) \rangle = \text{Tr}[\rho(\mathbf{x}) \cdot O],
    \label{eq:qnn-as-linear}
\end{equation}
where $\rho(\mathbf{x}) = |\phi(\mathbf{x})\rangle\langle\phi(\mathbf{x})|$ is the density matrix of the embedded state and $O$ is the observable, the prediction is \emph{linear} in $\rho(\mathbf{x})$. This places QNNs within the framework of kernel methods in a reproducing kernel Hilbert space (RKHS).

The representer theorem implies that the optimal observable for minimizing the training error can be written as:
\begin{equation}
    O^* = \sum_{i=1}^{N} \alpha_i |\phi(\mathbf{x}_i)\rangle\langle\phi(\mathbf{x}_i)|,
    \label{eq:qkm-representer}
\end{equation}
where the coefficients $\alpha_i$ are determined by the training data. This yields an important insight: under the corresponding loss and regularization setting, the kernel formulation can search over a larger linear hypothesis class associated with the same encoding than a fixed parameterized observable family.

However, this does not imply that kernel methods are universally superior. The larger hypothesis class of kernel methods may lead to poorer \emph{generalization}---the model may overfit the training data. QNNs, by constraining the observable to have a specific parameterized form, effectively regularize the model. This trade-off between expressivity and generalization is a central theme in \Cref{chap:generalization}.

\subsubsection{Data Re-uploading and Model Expressivity}

The equivalence between QNNs and kernel methods breaks down when data is \emph{re-uploaded} multiple times during the circuit~\cite{perez2020data}. In a \emph{data re-uploading} circuit, input data $\mathbf{x}$ is encoded into multiple layers:
\begin{equation}
    U(\mathbf{x}, \boldsymbol{\theta}) = U_D(\boldsymbol{\theta}_D) V(\mathbf{x}) \cdots U_1(\boldsymbol{\theta}_1) V(\mathbf{x}),
    \label{eq:data-reuploading}
\end{equation}
where $V(\mathbf{x})$ encodes the data and $U_\ell(\boldsymbol{\theta}_\ell)$ are trainable layers. \textcite{perez2020data} showed that data re-uploading circuits can act as universal function approximators, expressing arbitrary continuous functions of the input.

The relationship between re-uploading circuits and kernel methods was clarified by \textcite{jerbi2023quantum}. They proved that mapping a $D$-layer data re-uploading circuit to an equivalent kernel model requires $\Omega(D)$ additional ancilla qubits. This result has important implications for the design of quantum machine learning models: while single-encoding QNNs are equivalent to kernel methods and their advantage must come from the kernel itself, re-uploading circuits can leverage their depth to access richer function classes. Understanding when and how this additional expressivity translates to practical advantages remains an active area of research.

\section{Summary}
\label{sec:ch1-summary}

This chapter has laid the foundations for the quantum machine learning investigations that follow in Part~I. We began with the essential building blocks of quantum computation---qubits, quantum gates, and measurement---and reviewed the key developments in classical machine learning that motivate quantum extensions, from kernel methods and support vector machines to deep learning and the transformer revolution.

A central organizing principle introduced in this chapter is the \emph{error decomposition} (\Cref{eq:error-decomposition}), which partitions prediction error into three components:
\begin{itemize}[leftmargin=*]
    \item \textbf{Approximation error}: determined by the expressivity of the hypothesis class.
    \item \textbf{Optimization error}: the gap between the achieved and optimal training loss.
    \item \textbf{Generalization error}: the discrepancy between training and test performance.
\end{itemize}
This framework provides a unified lens through which to analyze and improve quantum machine learning models, and it guides the structure of the subsequent chapters.

We traced the evolution of quantum machine learning from early HHL-based algorithms, whose speedup claims depend on strong input-access assumptions, to NISQ-era approaches based on parameterized circuits and quantum feature maps.
Variational quantum algorithms such as VQE and QAOA use parameterized quantum circuits for optimization problems in quantum chemistry, many-body physics, and combinatorial optimization.
The supervised-learning focus of this thesis uses related circuit technology in two main forms:
\begin{enumerate}
    \item \textbf{Quantum Neural Networks (QNNs)}: Supervised-learning models that combine data encoding, trainable parameterized circuits, and measurement-based prediction. Key challenges include the barren plateau phenomenon, in which the gradient variance vanishes exponentially in deep or highly expressive circuits.

    \item \textbf{Quantum Kernel Methods (QKMs)}: Fixed quantum feature maps combined with classical kernel algorithms. Constructed QKM problems provide rigorous supervised-learning separations under explicit assumptions, while the kernel framework also clarifies when QNNs behave as linear models in quantum feature space.
\end{enumerate}

The relationship between these approaches reveals a trade-off between expressivity, optimization, and generalization: QKMs use a fixed quantum feature map with classical kernel learning, while QNNs restrict the trainable observable or circuit family and can thereby impose useful inductive bias. Data re-uploading circuits bridge these paradigms by adding expressivity beyond the simplest fixed-kernel picture.

\paragraph{Concepts introduced in Chapter 1 and where they are used later.}
\begin{center}
\small
\begin{tabular}{>{\raggedright\arraybackslash}p{0.27\textwidth} | >{\raggedright\arraybackslash}p{0.63\textwidth}}
    \toprule
    \textbf{Concept} & \textbf{Later use in the dissertation} \\
    \midrule
    Quantum embedding and feature maps
    & Trace-distance empirical-risk bounds and Neural Quantum Embedding in \Cref{chap:nqe}. \\
    \midrule
    QCNN
    & Background architecture for downstream NQE classifiers and QPR experiments. \\
    \midrule
    Barren plateaus and trainability
    & Motivation for structured circuits and embedding-aware QNN design in \Cref{chap:nqe,chap:generalization}. \\
    \midrule
    Quantum kernel methods
    & Alternative supervised-QML paradigm and kernel interpretation of embeddings in \Cref{chap:nqe}. \\
    \midrule
    Sequence models and self-attention
    & Hardware-aware QEC decoding and the Mamba/Transformer comparison in \Cref{chap:decoder}; the attention-based ViT ansatz for neural quantum states in \Cref{chap:nqs}. \\
    \bottomrule
\end{tabular}
\end{center}

\paragraph{Roadmap for Part I.}
The concepts introduced in this chapter set the stage for Part I of this thesis, which explores how quantum computing can enhance machine learning:
\begin{itemize}
    \item \textbf{\Cref{chap:nqe}: Neural Quantum Embedding} addresses the embedding-induced training-loss limit in quantum supervised learning. Rather than using fixed feature maps, we train neural networks to learn embeddings that increase the distinguishability of quantum states from different classes.

    \item \textbf{\Cref{chap:generalization}: Generalization in Quantum Machine Learning} addresses the \emph{generalization} side by developing margin-based bounds for quantum neural networks. Margins provide a statistical-control lens for finite training data and connect the trace-distance representation view to state discrimination.
\end{itemize}

\paragraph{Roadmap for Part II.}
Part II of this thesis reverses the direction, exploring how artificial intelligence techniques can address challenges in quantum computing:
\begin{itemize}
    \item \textbf{\Cref{chap:decoder}: Neural Decoders for Quantum Error Correction} develops neural network-based decoders for quantum error correcting codes and studies their accuracy--latency trade-off in real-time decoding benchmarks.

    \item \textbf{\Cref{chap:nqs}: Neural Quantum States} explores the use of neural networks to represent quantum many-body states and analyzes the optimizer that makes these representations practical. We recast stochastic reconfiguration as tangent-space ridge regression, identify the expressivity gap as finite-sample residual noise, and develop multi-shift stochastic reconfiguration as a richer spectral filter.
\end{itemize}


\chapter{Neural Quantum Embedding}
\label{chap:nqe}


\chaptersource{The results in this chapter are based on \textit{Neural Quantum Embedding: Pushing the Limits of Quantum Supervised Learning}~\cite{hur2024nqe} and \textit{Neural Quantum Embedding via Deterministic Quantum Computation with One Qubit}~\cite{liu2025nqe_dqc1}. Code is available at \url{https://github.com/takh04/neural-quantum-embedding}.}
\bigskip

In the previous chapter, we saw that the performance of quantum machine learning classifiers depends critically on how classical data is encoded into quantum states. In this chapter, we formalize this intuition by showing that, for binary classification under the linear loss ($\ell = \tfrac{1}{2}(1 - y f(\mathbf{x}))$), the empirical risk is lower bounded by the trace distance between embedded data ensembles, a quantity determined by the choice of data embedding. This trace-distance limit is not the entire optimization problem, but it identifies a representation-induced floor on the training loss that no downstream classifier circuit can overcome. We then show that conventional deterministic embedding schemes---amplitude encoding, angle encoding, and the ZZ feature map---do not guarantee that this trace distance will be large for a given dataset. This motivates \emph{Neural Quantum Embedding} (NQE), a data-driven approach that uses classical neural networks to learn an embedding that increases the distinguishability of quantum states representing different classes. In practice, NQE is trained using fidelity and Hilbert--Schmidt losses, which serve as tractable surrogates for the trace distance.

\section{Theoretical Background}
\label{sec:ch2-theory}

\subsection{Empirical Risk and Trace Distance}
\label{subsec:ch2-optimization-trace}

We consider a quantum binary classification task.
Given a labeled dataset
\begin{equation}
    S = \{(\mathbf{x}_i^-, -1)\}_{i=1}^{N^-} \cup \{(\mathbf{x}_i^+, +1)\}_{i=1}^{N^+},
\end{equation}
with $N = N^- + N^+$ samples, a quantum embedding circuit $V$ maps each classical input $\mathbf{x} \in \mathbb{R}^m$ to a quantum state:
\begin{equation}
    |\mathbf{x}\rangle = V(\mathbf{x}) |0\rangle^{\otimes n}.
    \label{eq:ch2-embedding}
\end{equation}
A parameterized quantum circuit $U(\boldsymbol{\theta})$ is then applied, followed by measurement of an observable $O$, yielding a prediction function:
\begin{equation}
    f(\mathbf{x}; \boldsymbol{\theta}) = \langle \mathbf{x} | U^\dagger(\boldsymbol{\theta})\, O\, U(\boldsymbol{\theta}) | \mathbf{x} \rangle.
    \label{eq:ch2-prediction}
\end{equation}

This prediction process can be recast as a quantum state discrimination problem~\cite{helstrom1976quantum}. Let $O$ be a Hermitian observable with eigenvalues $\pm 1$ (equivalently, $O^2=I$), which is the case for the Pauli observables used below. We define two positive operator-valued measure (POVM) elements parameterized by $\boldsymbol{\theta}$:
\begin{equation}
    E_{\pm}(\boldsymbol{\theta}) = \frac{I \pm U^\dagger(\boldsymbol{\theta})\, O\, U(\boldsymbol{\theta})}{2},
    \label{eq:ch2-povm}
\end{equation}
which satisfy $E_+(\boldsymbol{\theta}) + E_-(\boldsymbol{\theta}) = I$ and $E_{\pm}(\boldsymbol{\theta}) \geq 0$. The probability of obtaining outcome $\pm 1$ given input $\mathbf{x}$ is:
\begin{equation}
    P(E_{\pm}(\boldsymbol{\theta}) \,|\, \mathbf{x}) = \langle \mathbf{x} | E_{\pm}(\boldsymbol{\theta}) | \mathbf{x} \rangle.
    \label{eq:ch2-measurement-prob}
\end{equation}
The natural loss function for classification is then the misclassification probability:
\begin{equation}
    \ell(f(\mathbf{x}; \boldsymbol{\theta}), y) = P(E_{\bar{y}}(\boldsymbol{\theta}) \,|\, \mathbf{x}),
    \label{eq:ch2-loss}
\end{equation}
where $\bar{y}$ denotes the complement of $y$, i.e., if $y = +1$ then $\bar{y} = -1$ and vice versa. Let $f(\mathbf{x}) = \langle \mathbf{x} | U^\dagger(\boldsymbol{\theta})\, O\, U(\boldsymbol{\theta}) | \mathbf{x} \rangle \in [-1, 1]$ be the prediction. The loss then takes the equivalent \emph{linear} form
\begin{equation}
    \ell(f(\mathbf{x}; \boldsymbol{\theta}), y) = \frac{1}{2}\left(1 - y\, f(\mathbf{x})\right).
    \label{eq:ch2-linear-loss}
\end{equation}

\paragraph{Lower bound of empirical risk.}
The empirical risk over the training set $S$ can be written as:
\begin{equation}
    L_S = \frac{1}{N}\left[\sum_{i=1}^{N^-} P(E_+(\boldsymbol{\theta}) \,|\, \mathbf{x}_i^-) + \sum_{i=1}^{N^+} P(E_-(\boldsymbol{\theta}) \,|\, \mathbf{x}_i^+)\right].
    \label{eq:ch2-empirical-risk}
\end{equation}
Minimizing $L_S$ is therefore equivalent to minimizing the probability of misclassification in a quantum state discrimination problem, for which the minimum achievable error probability is known exactly: the Helstrom bound~\cite{helstrom1976quantum}. Applying this bound, the empirical risk is lower bounded by:
\begin{equation}
    \boxed{L_S \geq \frac{1}{2} - D_{\mathrm{tr}}(p^- \rho^-,\, p^+ \rho^+),}
    \label{eq:ch2-lower-bound}
\end{equation}
where
\begin{equation}
    \rho^{\pm} = \frac{1}{N^{\pm}} \sum_{i=1}^{N^{\pm}} |\mathbf{x}_i^{\pm}\rangle \langle \mathbf{x}_i^{\pm}|
    \label{eq:ch2-data-ensemble}
\end{equation}
are the class-averaged density matrices (data ensembles), $p^{\pm} = N^{\pm}/N$ are the class priors, and $D_{\mathrm{tr}}(\rho, \sigma) = \frac{1}{2}\|\rho - \sigma\|_1$ denotes the trace distance.
For balanced binary datasets, where $p^+ = p^- = 1/2$, this bound can equivalently be written as
\begin{equation}
    L_S \geq \frac{1}{2}\left(1 - D_{\mathrm{tr}}(\rho^-,\, \rho^+)\right).
    \label{eq:ch2-balanced-lower-bound}
\end{equation}
This unweighted form is the one used for the balanced experiments and figures below.

This result has a direct implication: the best possible misclassification probability over all binary measurements is controlled by the trace distance between the two data ensembles $p^- \rho^-$ and $p^+ \rho^+$. This quantity is fixed once the embedded states have been prepared and is not improved by the subsequent parameterized unitary $U(\boldsymbol{\theta})$. Thus, even an expressive trainable circuit cannot overcome poor class distinguishability created at the embedding stage.

Furthermore, the minimum loss is achieved when the POVM $\{E_-(\boldsymbol{\theta}), E_+(\boldsymbol{\theta})\}$ forms a \emph{Helstrom measurement}---the measurement that optimally discriminates between the two quantum state ensembles. The training of a quantum neural network can therefore be viewed as a process of finding this optimal Helstrom measurement.

\paragraph{Contractive property of trace distance.}
The importance of the embedding is further underscored by the contractive property of the trace distance. For quantum channels, which are completely positive and trace preserving (CPTP), trace distance is contractive. More generally, the same contractivity holds for positive trace-preserving (PTP) maps on Hermitian operators~\cite{siudzinska2021entropy, wilde2017quantum}:
\begin{equation}
    D_{\mathrm{tr}}(\Lambda(\rho),\, \Lambda(\sigma)) \leq D_{\mathrm{tr}}(\rho,\, \sigma).
    \label{eq:ch2-contractive}
\end{equation}
Subsequent quantum operations after the embedding, including the parameterized circuit $U(\boldsymbol{\theta})$ and quantum noise channels, are described by CPTP maps. This is particularly relevant for NISQ devices, where hardware noise can further reduce the trace distance between quantum states.

\subsection{Limitations of Deterministic Embeddings}
\label{subsec:ch2-limitations}

We now examine commonly used deterministic quantum embedding schemes and explain why they do not guarantee a large trace distance for a given dataset.

\paragraph{Amplitude embedding} encodes a normalized classical vector $\mathbf{x} \in \mathbb{R}^{2^n}$ directly into the amplitudes of an $n$-qubit state:
\begin{equation}
    |\mathbf{x}\rangle = \sum_{i=0}^{2^n - 1} x_i |i\rangle.
    \label{eq:ch2-amplitude-embedding}
\end{equation}
This is maximally qubit-efficient ($2^n$ features with $n$ qubits) but requires circuit depth $\mathcal{O}(2^n)$~\cite{schuld2021effect}. The fidelity reduces to the classical inner product $|\langle \mathbf{x} | \mathbf{x}'\rangle|^2 = |\mathbf{x} \cdot \mathbf{x}'|^2$, so the quantum feature space inherits the geometry of the input space without additional structure.

\paragraph{Angle embedding} encodes each feature as a rotation angle on a dedicated qubit:
\begin{equation}
    |\mathbf{x}\rangle = \bigotimes_{j=1}^{n} R_P(x_j) |0\rangle,
    \label{eq:ch2-angle-embedding}
\end{equation}
where $R_P(\theta) = e^{-i\theta P/2}$ for $P \in \{X, Y, Z\}$. This produces only product states with no entanglement.

\paragraph{ZZ feature map} introduces entanglement through two-qubit $ZZ$ interactions:
\begin{equation}
    V(\boldsymbol{\phi}(\mathbf{x})) = \left[\exp\!\left(i \sum_{k} \phi_k(\mathbf{x})\, Z_k + i \sum_{k < l} \phi_{k,l}(\mathbf{x})\, Z_k Z_l \right) H^{\otimes n}\right]^L,
    \label{eq:ch2-zz-feature-map}
\end{equation}
where $L \geq 1$ is the number of repetitions~\cite{havlicek2019supervised}. The standard choices $\phi_k(\mathbf{x}) = x_k$ and $\phi_{k,l}(\mathbf{x}) = (\pi - x_k)(\pi - x_l)/2$ are made without data-dependent justification. Although \textcite{suzuki2020analysis} show that the choice of $\boldsymbol{\phi}$ significantly impacts performance, no guidelines exist for selecting it for a given problem.

\paragraph{The limitation of fixed embeddings.}
All three schemes are \emph{data-agnostic}: their structure is fixed independently of the training data, and there is no guarantee that they produce a large trace distance between data ensembles for a given classification task. A natural idea to address this is to introduce trainable parameters within the quantum circuit. In a \emph{trainable unitary embedding}~\cite{larose2020robust}, a parameterized unitary $U_{\mathrm{tra}}(\boldsymbol{\theta})$ is applied before the data encoding:
\begin{equation}
    |\mathbf{x}; \boldsymbol{\theta}\rangle = V(\mathbf{x})\, U_{\mathrm{tra}}(\boldsymbol{\theta}) |0\rangle^{\otimes n}.
    \label{eq:ch2-trainable-unitary}
\end{equation}
However, the maximum achievable trace distance is upper bounded by the diamond distance~\cite{hur2024nqe}:
\begin{equation}
    \max_{\boldsymbol{\theta}}\, D_{\mathrm{tr}}\!\left(p^+ \rho^+(\boldsymbol{\theta}),\, p^- \rho^-(\boldsymbol{\theta})\right) \leq D_{\diamond}\!\left(p^+ \mathcal{E}^+,\, p^- \mathcal{E}^-\right),
    \label{eq:ch2-diamond-bound}
\end{equation}
where $\mathcal{E}^{\pm}$ are quantum channels with Kraus operators $K_i^{\pm} = V(\mathbf{x}_i^{\pm})/\sqrt{N^{\pm}}$, determined entirely by the embedding circuit $V$. The trainable unitary $U_{\mathrm{tra}}(\boldsymbol{\theta})$ does not improve this upper bound. A related restriction applies to \emph{data re-uploading}~\cite{perez2020data}, since such circuits can be transformed into a form where the embedding component is separated from the trainable parameters by introducing ancilla qubits~\cite{jerbi2023quantum}. Thus, trainable quantum layers alone do not remove the need for an embedding that separates the classes well.
\paragraph{Experimental evidence.}

\begin{figure}[t]
    \centering
    \includegraphics[width=0.9\textwidth]{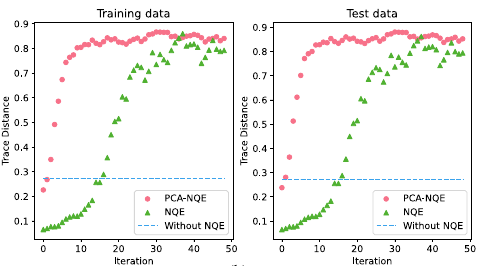}
    \caption{Trace distance between class-averaged data ensembles $D_{\mathrm{tr}}(\rho^-, \rho^+)$ for the conventional ZZ feature map and Neural Quantum Embedding on the balanced MNIST binary classification task (digits 0 vs.\ 1, 4 qubits). The blue dashed reference line indicates the trace distance obtained by the conventional ZZ feature map without NQE. Results obtained on IBM quantum hardware (ibmq\_toronto).}
    \label{fig:ch2-trace-distance-comparison}
\end{figure}

\Cref{fig:ch2-trace-distance-comparison} compares the trace distance achieved by the conventional ZZ feature map against a data-driven embedding on the balanced MNIST binary classification task (digits 0 vs.\ 1, 4 qubits); the embedding itself is developed in \Cref{sec:ch2-nqe} and we defer its details until then. The ZZ feature map achieves a trace distance of only $D_{\mathrm{tr}} \approx 0.27$, which by \Cref{eq:ch2-balanced-lower-bound} implies a lower bound on the empirical risk of $L_S \geq 0.36$. The data-driven embedding, by contrast, achieves $D_{\mathrm{tr}} \approx 0.84$, reducing the lower bound to $L_S \geq 0.08$. The same pattern holds on Fashion-MNIST across different qubit counts (4, 8, 12), and becomes even more pronounced under hardware noise, where the contractive property further degrades already-low trace distances.

These results motivate a \emph{data-driven} approach that increases the trace distance \emph{before} the data become quantum states, thereby sidestepping the PTP contractive constraint. We develop this next as Neural Quantum Embedding.

\section{Neural Quantum Embeddings}
\label{sec:ch2-nqe}

The preceding analysis reveals a bottleneck: the trace distance between data ensembles is fixed by the embedding circuit, and subsequent quantum processing cannot increase it. Neural Quantum Embedding (NQE)~\cite{hur2024nqe} addresses this limitation by operating \emph{before} the quantum domain, using a classical neural network to learn a data-dependent preprocessing that increases the trace distance.

\subsection{The NQE Framework}
\label{subsec:ch2-nqe-framework}

The core idea of NQE is to place a classical neural network $g : \mathbb{R}^m \times \mathbb{R}^r \to \mathbb{R}^{m'}$ in front of the quantum embedding, so that the data are mapped to the rotation angles that yield the most discriminative feature map. The NQE mapping is
\begin{equation}
    |\mathbf{x}\rangle = V(g(\mathbf{x}, \mathbf{w}))\, |0\rangle^{\otimes n},
    \label{eq:ch2-nqe-map}
\end{equation}
where $V$ is a fixed quantum embedding circuit (e.g., the ZZ feature map of \Cref{eq:ch2-zz-feature-map}), $g$ is a classical neural network with trainable parameters $\mathbf{w} \in \mathbb{R}^r$, and $m' \leq m$ allows for dimensionality reduction.

Because $g$ transforms the classical input parameters before any quantum state exists, NQE escapes the contractive bound that constrains the trainable-unitary embedding of \Cref{eq:ch2-trainable-unitary}. There, the trainable unitary acts on a fixed data-dependent channel ensemble and remains subject to the diamond-distance bound of \Cref{eq:ch2-diamond-bound}. Here, the achievable trace distance is determined by the chosen feature-map family, the Hilbert-space dimension, the class priors, and the capacity, optimization, and training data used to fit $g$.

For the ZZ feature map specifically, NQE supplies the guidance that \Cref{eq:ch2-zz-feature-map} lacks for choosing $\boldsymbol{\phi}$: the fixed angle functions $\phi_k$ and $\phi_{k,l}$ are replaced by learned functions $g_k(\mathbf{x}, \mathbf{w})$ and $g_{k,l}(\mathbf{x}, \mathbf{w})$. The NQE framework is not tied to this choice. It applies to any embedding circuit $V$, including amplitude and angle encoding. Whenever $g$ can represent the original preprocessing map, the fixed embedding is recovered as a special case, and any improvement beyond it reflects the added expressivity of the learned map.

\begin{figure}[t]
    \centering
    \includegraphics[width=\textwidth]{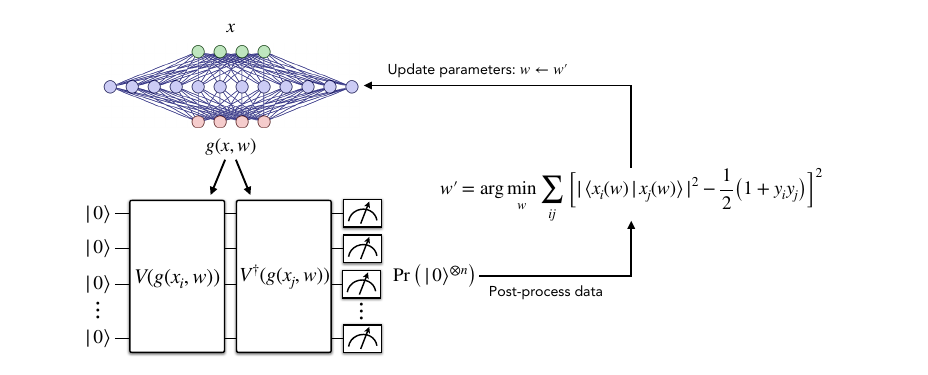}
    \caption{Overview of the NQE training procedure. A classical neural network $g(\mathbf{x}, \mathbf{w})$ transforms input data into rotation angles for the quantum embedding circuit $V$. The resulting quantum state $|\mathbf{x}\rangle = V(g(\mathbf{x}, \mathbf{w}))\,|0\rangle^{\otimes n}$ is used to compute the implicit fidelity loss via an overlap circuit $V^\dagger(g(\mathbf{x}_j, \mathbf{w}))\,V(g(\mathbf{x}_i, \mathbf{w}))\,|0\rangle^{\otimes n}$ by measuring the probability of the all-zero outcome. The classical parameters $\mathbf{w}$ are updated to maximize the distinguishability between classes.}
    \label{fig:ch2-nqe-schematic}
\end{figure}

\subsection{Training via Implicit Fidelity Loss}
\label{subsec:ch2-nqe-training}

The ideal training objective for NQE would directly maximize
$D_{\mathrm{tr}}(p^- \rho^-, p^+ \rho^+),$
since by \Cref{eq:ch2-lower-bound} this directly lowers the empirical risk bound. However, computing the trace distance requires full tomography of the class-averaged density matrices, which is computationally prohibitive. Instead, NQE employs an \emph{implicit fidelity loss} derived from pairwise state fidelities.

For a pair of training samples $(\mathbf{x}_i, y_i)$ and $(\mathbf{x}_j, y_j)$, the implicit fidelity loss is defined as:
\begin{equation}
    \ell_{\mathrm{fid}}\!\left((\mathbf{x}_i, y_i), (\mathbf{x}_j, y_j)\right) = \left[\left|\langle \mathbf{x}_i | \mathbf{x}_j \rangle\right|^2 - \frac{1}{2}\left(1 + y_i y_j\right)\right]^2,
    \label{eq:ch2-nqe-fidelity-loss}
\end{equation}
where $|\langle \mathbf{x}_i | \mathbf{x}_j \rangle|^2$ is the fidelity between the embedded quantum states. The target value $\frac{1}{2}(1 + y_i y_j)$ equals $1$ when the labels agree ($y_i = y_j$) and $0$ when they disagree ($y_i \neq y_j$). The NQE training objective is therefore:
\begin{equation}
    \mathbf{w}^* = \argmin_{\mathbf{w}} \sum_{i,j} \ell_{\mathrm{fid}}\!\left((\mathbf{x}_i, y_i), (\mathbf{x}_j, y_j)\right).
    \label{eq:ch2-nqe-objective}
\end{equation}

\paragraph{Train/test separation.}
The NQE preprocessing network is trained only from labeled pairs drawn from the training split.
After this embedding is fixed, the downstream quantum classifier is trained on the training split using the learned embedding.
Held-out test data are used only for final evaluation and for post hoc diagnostics such as test-set trace distance.

The connection between this pairwise fidelity loss and the trace distance operates through two complementary mechanisms:

\paragraph{Same-class pairs ($y_i = y_j$).}
The loss drives the fidelity $|\langle \mathbf{x}_i | \mathbf{x}_j \rangle|^2 \to 1$ for pairs within the same class. This directly increases the purity of the class-averaged density matrices:
\begin{equation}
    \mathrm{Tr}\!\left[(\rho^{\pm})^2\right] = \frac{1}{(N^{\pm})^2} \sum_{i,j=1}^{N^{\pm}} \left|\langle \mathbf{x}_i^{\pm} | \mathbf{x}_j^{\pm} \rangle\right|^2.
    \label{eq:ch2-purity}
\end{equation}
When $\mathrm{Tr}[(\rho^{\pm})^2] = 1$, the density matrix $\rho^{\pm}$ is a pure state, meaning all same-class data points are mapped to the same quantum state. Higher purity corresponds to tighter clustering within each class in the quantum feature space.

\paragraph{Cross-class pairs ($y_i \neq y_j$).}
The loss drives the fidelity $|\langle \mathbf{x}_i^+ | \mathbf{x}_j^- \rangle|^2 \to 0$ for pairs from different classes. For a balanced dataset with paired examples, convexity of the trace distance gives
\begin{equation}
    D_{\mathrm{tr}}(\rho^-,\, \rho^+) \leq \frac{1}{N}\sum_{i=1}^{N}\sqrt{1 - |\langle \mathbf{x}_i^+ | \mathbf{x}_i^- \rangle|^2},
    \label{eq:ch2-convex-trace}
\end{equation}
so reducing the cross-class fidelities relaxes this upper bound and, together with the same-class clustering that drives each ensemble toward a pure representative, makes a large trace distance attainable. 
Together, these two effects push the embedded quantum states toward a configuration where the two classes are represented by nearly pure states that are nearly orthogonal---precisely the condition that maximizes the trace distance and thus minimizes the lower bound on empirical risk. The full derivation is given in \Cref{subsec:app-nqe-fidelity-trace}.

\paragraph{Practical computation.}
The fidelity $|\langle \mathbf{x}_i | \mathbf{x}_j \rangle|^2$ between two embedded states can be measured on a quantum computer by preparing the overlap circuit
\begin{equation}
    V^\dagger(g(\mathbf{x}_j, \mathbf{w}))\, V(g(\mathbf{x}_i, \mathbf{w}))\, |0\rangle^{\otimes n}
\end{equation}
and measuring the probability of the all-zero outcome $\Pr(|0\rangle^{\otimes n})$, which equals the fidelity. This pure-state fidelity can also be computed via the swap test~\cite{buhrman2001quantum}. In \Cref{sec:ch2-dqc1}, we introduce a DQC1-compatible variant in which the training loss is written in terms of Hilbert--Schmidt inner products between feature-map unitaries.

\subsection{Experimental Results}
\label{subsec:ch2-nqe-results}

We now present experimental results demonstrating the effectiveness of NQE across multiple metrics. All experiments use binary classification of MNIST digits (0 vs.\ 1) with a 4-qubit QCNN (reviewed in \Cref{subsec:ch1-qcnn}), unless otherwise noted. We consider two NQE architectures that differ in how they handle high-dimensional inputs. In \textbf{PCA-NQE}, PCA first reduces the input to $n$ features, which are then passed through a fully connected neural network (two hidden layers, ReLU activations) that outputs $2n$ rotation angles for the ZZ feature map. In \textbf{NQE} (direct), a 2D convolutional neural network processes the full-resolution input (e.g., $28 \times 28$ images) with max pooling layers, bypassing PCA to learn nonlinear feature extraction end-to-end.

\paragraph{Convergence to theoretical bounds.}
\begin{figure}[t]
    \centering
    \includegraphics[width=\textwidth]{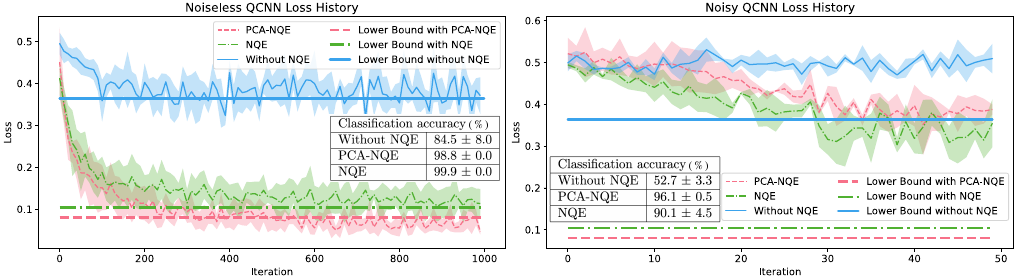}
    \caption{QCNN training loss histories for balanced 4-qubit MNIST binary classification (digits 0 vs.\ 1). Blue solid, red dashed, and green dash-dotted lines represent the conventional ZZ feature map, PCA-NQE, and NQE, respectively. Thick lines indicate the theoretical lower bounds from \Cref{eq:ch2-balanced-lower-bound}. Shaded regions represent one standard deviation over five (noiseless) or three (noisy) independent trials. \textbf{Left:} noiseless simulation. \textbf{Right:} IBM quantum hardware (ibmq\_jakarta, ibmq\_toronto, ibmq\_perth).}
    \label{fig:ch2-nqe-loss-history}
\end{figure}

\Cref{fig:ch2-nqe-loss-history}(a) shows the noiseless QCNN training loss histories for the conventional ZZ feature map, PCA-NQE, and NQE. For all three embedding methods, the QCNN training loss approaches the respective theoretical lower bounds predicted by \Cref{eq:ch2-lower-bound}, indicating that the trained QCNN approximates the corresponding Helstrom measurement in these experiments. The critical difference lies in the lower bounds themselves: NQE methods achieve substantially lower theoretical limits, translating into higher classification accuracy. \Cref{tab:ch2-nqe-results} summarizes the key results: the conventional ZZ feature map achieves $84.5 \pm 8.0\%$ noiseless accuracy, while PCA-NQE and NQE achieve $98.8 \pm 0.0\%$ and $99.9 \pm 0.0\%$, respectively.

\begin{table}[t]
    \centering
    \caption{Summary of NQE results for balanced 4-qubit MNIST binary classification (digits 0 vs.\ 1). Noiseless accuracy is from statevector simulation, while noisy accuracy is from the IBM quantum hardware runs. The lower bound on empirical risk is computed as $L_S \geq (1 - D_{\mathrm{tr}})/2$.}
    \label{tab:ch2-nqe-results}
    \begin{tabular}{lcccc}
        \toprule
        \textbf{Method} & $D_{\mathrm{tr}}$ & \textbf{Lower bound} & \textbf{Noiseless acc.\ (\%)} & \textbf{Noisy acc.\ (\%)} \\
        \midrule
        ZZ (no NQE) & 0.273 & 0.364 & $84.5 \pm 8.0$ & $52.7 \pm 3.3$ \\
        PCA-NQE     & 0.840 & 0.080 & $98.8 \pm 0.0$ & $96.1 \pm 0.5$ \\
        NQE         & 0.792 & 0.104 & $99.9 \pm 0.0$ & $90.1 \pm 4.5$ \\
        \bottomrule
    \end{tabular}
\end{table}

\begin{figure}[t]
    \centering
    \includegraphics[width=\textwidth]{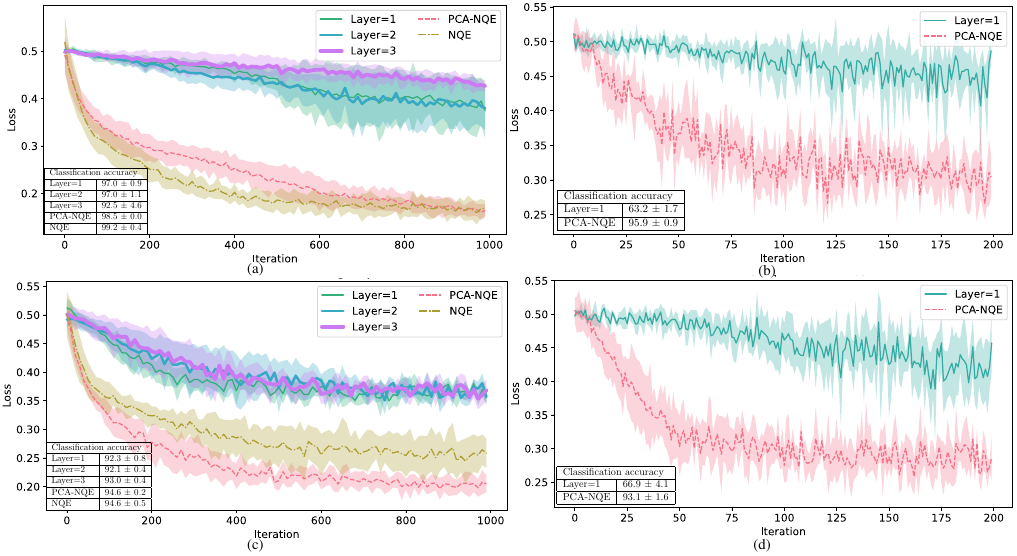}
    \caption{Comparison of NQE against trainable unitary embeddings with one, two, and three trainable layers, using 8-qubit circuits on MNIST (top row) and Fashion-MNIST (bottom row) under noiseless (left column) and noisy (right column) simulation. The noisy simulations use the IBM Qiskit FakeGuadalupe environment. Noiseless runs use $1000$ iterations, learning rate $0.01$, and batches of $128$ samples per iteration. Noisy runs use $200$ iterations, learning rate $0.05$, and batches of $15$ samples per iteration. Classification accuracies are evaluated on held-out sets of $2115$ (MNIST) and $2000$ (Fashion-MNIST) samples, and the loss histories show the mean and one standard deviation over five independent trials.}
    \label{fig:ch2-nqe-vs-tqe}
\end{figure}

\paragraph{NQE vs.\ trainable unitary embedding.}
We compare NQE against trainable unitary embeddings with $L = 1, 2, 3$ layers on both MNIST and Fashion-MNIST datasets using 8-qubit circuits (\Cref{fig:ch2-nqe-vs-tqe}). In both noiseless and noisy environments, NQE consistently achieves lower training loss and higher classification accuracy than all trainable unitary variants. Trainable unitary embedding adds parameterized quantum gates that increase the overall circuit depth, making the model more susceptible to hardware noise and barren plateaus~\cite{mcclean2018barren}. Under noisy simulation (IBM FakeGuadalupe), the advantage of NQE becomes even more pronounced: the trainable unitary embedding's deeper circuits suffer greater noise-induced degradation, while NQE's classical preprocessing adds no quantum circuit depth.

\paragraph{Robustness on real quantum hardware.}
\Cref{fig:ch2-nqe-loss-history}(b) presents the QCNN training loss histories obtained on IBM quantum hardware (ibmq\_jakarta, ibmq\_toronto, and ibmq\_perth). The contractive property (\Cref{eq:ch2-contractive}) provides a direct explanation for the dramatic difference in noise robustness between conventional and NQE-enhanced embeddings. Hardware noise constitutes a PTP map that can only \emph{reduce} trace distance:
\begin{itemize}
    \item \textbf{Small $D_{\mathrm{tr}}$ (ZZ $\approx 0.27$):} Even modest noise pushes the already-close quantum states past the decision boundary, collapsing classification accuracy to $52.7 \pm 3.3\%$---barely above random guessing.
    \item \textbf{Large $D_{\mathrm{tr}}$ (PCA-NQE $\approx 0.84$):} Noise reduces the trace distance, but the states remain well-separated and classification remains robust, achieving $96.1 \pm 0.5\%$ (PCA-NQE) and $90.1 \pm 4.5\%$ (NQE) on real hardware.
\end{itemize}
In these experiments, the noisy empirical risk achieved by NQE on real quantum hardware falls \emph{below} the noiseless theoretical lower bound of the conventional ZZ feature map. Thus, NQE with real hardware noise achieves a lower empirical risk than the best possible Helstrom limit associated with the unoptimized ZZ feature map in a noiseless setting. This result, summarized in \Cref{tab:ch2-nqe-results}, shows that improving the embedding can matter more than reducing moderate hardware noise for this task.

Beyond these training-loss improvements, NQE also affects two properties that are not visible in the training loss alone: the \emph{generalization} of the trained model to unseen data and its \emph{trainability}. We examine both in the following subsection.

\subsection{Generalization, Expressibility, and Trainability}
\label{subsec:ch2-generalization}

The trace-distance analysis of \Cref{subsec:ch2-optimization-trace} concerns only the \emph{training} loss. A low training loss does not by itself guarantee good performance on unseen inputs, nor does it indicate how readily the model can be trained. In this subsection we present numerical evidence that NQE improves two further, \emph{distinct} properties of the resulting models. The first is \emph{generalization}---the ability to predict well on unseen data---which we quantify through the local effective dimension of the quantum neural network and the weight-norm complexity of the quantum kernel model. The second is \emph{trainability}, which is governed by the \emph{expressibility} of the embedding and which we probe through the deviation from a unitary $2$-design and the variance of the quantum kernel entries. Here, a high expressibility is detrimental, as it induces barren plateaus and kernel concentration that impede optimization. These two axes are addressed separately below. Unless noted otherwise, the experiments use the same PCA-NQE and NQE networks introduced above (optimized on \texttt{ibmq\_toronto}; see \Cref{subsec:ch2-nqe-results}) and binary MNIST (digits 0 vs.\ 1).

\begin{figure}[t]
    \centering
    \includegraphics[width=\textwidth]{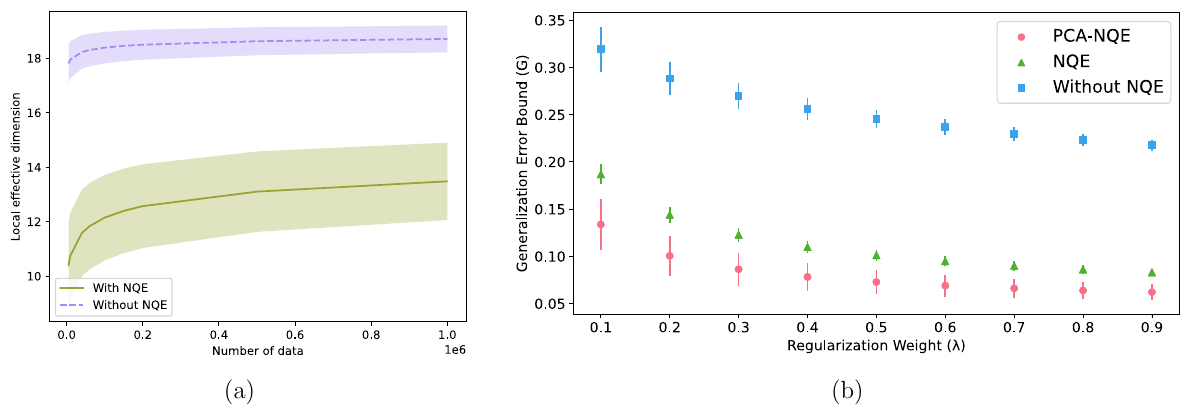}
    \caption{Generalization diagnostics for embeddings with and without NQE. \textbf{(a)}~Local effective dimension of a four-qubit QNN with (solid green) and without (dashed purple) NQE, as a function of the number of data. NQE yields a smaller effective model complexity, and hence a tighter generalization bound, across all dataset sizes (averaged over $200$ experiments---$10$ artificial datasets, each with $20$ random parameter initializations. Shaded regions denote one standard deviation). \textbf{(b)}~Weight-norm complexity $G = \|W^*\|_F/\sqrt{N}$, which controls the data-dependent term of the quantum-kernel generalization bound \Cref{eq:ch2-qkm-bound}, as a function of the regularization weight $\lambda$. PCA-NQE (red circles) and NQE (green triangles) both lower $G$ relative to the conventional ZZ feature map without NQE (blue squares) at every $\lambda$ (mean and one standard deviation over five independent draws of $1000$ MNIST samples). In both panels, lower values correspond to better expected generalization.}
    \label{fig:ch2-generalization}
\end{figure}

\paragraph{Effective dimension of the quantum neural network.}
The \emph{effective dimension} measures how many parameters of a model are ``active'' in the sense that they meaningfully influence its output, and it acts as a capacity measure that upper bounds the generalization error of a statistical model~\cite{abbas2021effective}. We use the \emph{local effective dimension} (LED), which is better matched to a specific learned model than the global effective dimension. The LED is positively correlated with the generalization error, so a smaller LED indicates better expected generalization. \Cref{fig:ch2-generalization}(a) reports the LED of a four-qubit QNN as a function of the number of data, averaged over $200$ instances ($10$ artificial datasets $\times$ $20$ random parameter initializations), with and without NQE. Across the entire range of dataset sizes, the NQE-enhanced model has a substantially lower effective dimension (rising from $\approx 10.4$ to $\approx 13.5$) than the conventional embedding (which plateaus near $\approx 18.5$), and the reduction holds in all $200$ instances. Because the effective dimension can also be read as the volume of solution space the model class occupies, the smaller value under NQE indicates a more constrained hypothesis class, consistent with the improved test accuracy reported above.

\paragraph{Generalization in the quantum kernel method.}
The benefit of NQE is not confined to quantum neural networks; it also tightens a generalization bound for the \emph{quantum kernel method} (QKM). Given the embedding, the quantum kernel is $k^Q(\mathbf{x}_i,\mathbf{x}_j)=|\langle \mathbf{x}_i|\mathbf{x}_j\rangle|^2$, and a kernel model predicts $f(\mathbf{x};W)=\mathrm{Tr}[W |\mathbf{x}\rangle\langle\mathbf{x}|]$, with $W$ obtained by minimizing a regularized least-squares cost,
\begin{equation}
    W^* = \argmin_{W \in \mathbb{C}^{2^n\times 2^n}} \frac{1}{N}\sum_{i=1}^{N}\left(f(\mathbf{x}_i;W)-h(\mathbf{x}_i)\right)^2 + \lambda \|W\|_F^2 ,
    \label{eq:ch2-qkm-objective}
\end{equation}
where $h$ is the target function, $\|\cdot\|_F$ is the Frobenius norm, and $\lambda$ is a regularization weight that trades training error for generalization. For this estimator, the generalization gap obeys~\cite{huang2021power}
\begin{equation}
    \left|R(W)-R_N(W)\right| \leq \mathcal{O}\!\left(\frac{\|W^*\|_F}{\sqrt{N}} + \sqrt{\frac{\log(1/\delta)}{N}}\right)
    \label{eq:ch2-qkm-bound}
\end{equation}
with probability at least $1-\delta$, where $R$ and $R_N$ denote the true and empirical risks. The data-dependent term is governed by the weight-norm complexity $G = \|W^*\|_F/\sqrt{N}$. Because NQE changes both the kernel matrix $K^Q$ and the embedded states $|\mathbf{x}\rangle$, it directly affects $G$. \Cref{fig:ch2-generalization}(b) plots $G$ for binary MNIST ($N = 1000$) across regularization weights $\lambda \in [0.1, 0.9]$. Both PCA-NQE and NQE yield a markedly smaller $G$ than the conventional embedding at every $\lambda$---that is, a tighter upper bound on the generalization error---with PCA-NQE giving the smallest values.

\paragraph{Expressibility and trainability.}
Both the QNN and QKM frameworks face a trade-off between expressibility and trainability: highly expressive circuits tend to exhibit barren plateaus---exponentially vanishing gradients that obstruct training~\cite{mcclean2018barren, larocca2025barren}---and, in the kernel setting, they induce an exponential concentration of the kernel entries, so that exponentially many measurements are needed to resolve $K^Q$~\cite{thanasilp2024exponential}. NQE mitigates both effects by deliberately \emph{limiting} the expressibility of the embedding, exploiting the prior that an embedding with large class distinguishability already suffices to approximate the target function.

\begin{figure}[t]
    \centering
    \includegraphics[width=0.9\textwidth]{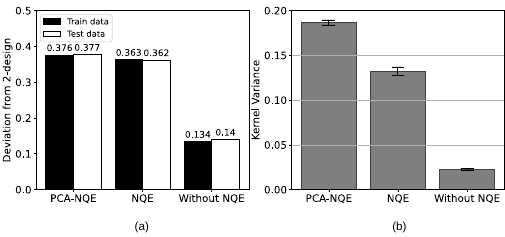}
    \caption{Expressibility and trainability diagnostics for embeddings with and without NQE. (a)~Deviation from a unitary $2$-design, $\epsilon = \sqrt{\mathrm{Tr}(A^\dagger A)}$ from \Cref{eq:ch2-2design-deviation}, on training (filled) and test (open) MNIST data. A larger deviation indicates \emph{lower} expressibility. Both NQE variants are markedly less expressive than the conventional embedding. (b)~Variance of the off-diagonal quantum-kernel elements, computed from $1000$ MNIST samples (mean and one standard deviation over five iterations). The larger variance under NQE indicates that the kernel entries are not exponentially concentrated, improving the trainability of the quantum kernel method.}
    \label{fig:ch2-expressibility}
\end{figure}

We quantify expressibility in two complementary ways. First, \Cref{fig:ch2-expressibility}(a) reports the Hilbert--Schmidt norm $\epsilon = \sqrt{\mathrm{Tr}(A^\dagger A)}$ of the deviation from a unitary $2$-design,
\begin{equation}
    A = \int_{\mathrm{Haar}} \left(|\psi\rangle\langle\psi|\right)^{\otimes 2} d\psi \;-\; \int_{\mathcal{E}} \left(|\phi\rangle\langle\phi|\right)^{\otimes 2} d\phi ,
    \label{eq:ch2-2design-deviation}
\end{equation}
where the first integral is over the Haar measure and the second is over the ensemble $\mathcal{E}$ of data-embedded states, evaluated on $12{,}665$ training and $2{,}115$ test MNIST samples. A small $\epsilon$ marks a highly expressive embedding (close to a $2$-design). Both NQE variants have a substantially larger deviation ($\epsilon \approx 0.37$) than the conventional embedding ($\epsilon \approx 0.13$), on both training and test data, confirming that NQE produces a \emph{less} expressive embedding. Second, \Cref{fig:ch2-expressibility}(b) shows the variance of the off-diagonal quantum-kernel entries (from $1000$ MNIST samples): the NQE variants exhibit far larger kernel variance ($\approx 0.19$ for PCA-NQE, $\approx 0.13$ for NQE) than the conventional embedding ($\approx 0.02$). A larger variance means the kernel entries are \emph{not} exponentially concentrated, so the kernel matrix can be estimated reliably with far fewer circuit executions. Together, the two panels show that NQE trades a controlled reduction in expressibility for improved trainability in both frameworks.
These diagnostics indicate that NQE improves not only the achievable training loss but also the generalization and trainability of the resulting models.

\section{Optimization via DQC1}
\label{sec:ch2-dqc1}

The NQE training procedure described in \Cref{subsec:ch2-nqe-training} relies on measuring the fidelity $|\langle \mathbf{x}_i | \mathbf{x}_j \rangle|^2$ between pairs of embedded quantum states, which requires preparing all $n$ qubits in the pure state $|0\rangle^{\otimes n}$. While this is straightforward on gate-based quantum computers such as superconducting and trapped-ion platforms, it is not naturally matched to \emph{ensemble quantum systems}---most notably nuclear magnetic resonance (NMR) quantum processors~\cite{cory1997ensemble}---where the thermal equilibrium state is highly mixed and preparing a global pure state across all qubits is impractical. In this section, we show that NQE can be reformulated using the \emph{Hilbert--Schmidt inner product} in place of fidelity, and that this quantity can be computed efficiently using the \emph{deterministic quantum computation with one qubit} (DQC1) model~\cite{knill1998dqc1}. This reformulation extends NQE to ensemble quantum platforms and demonstrates its experimental viability on an NMR quantum processor.

\subsection{The DQC1 Model}
\label{subsec:ch2-dqc1-model}

The DQC1 model, introduced by \textcite{knill1998dqc1}, is a restricted model of quantum computation that uses one probe qubit with nonzero polarization alongside $n$ qubits in the maximally mixed state. In DQC1 framework, the initial state of the $(n+1)$-qubit system is:
\begin{equation}
    \rho_0 = |0\rangle\langle 0| \otimes \frac{I}{2^n},
    \label{eq:ch2-dqc1-initial}
\end{equation}
where the first qubit (the \emph{probe}) is prepared as a clean qubit and the remaining $n$ qubits are in the maximally mixed state $I/2^n$.

The DQC1 circuit proceeds as follows: a Hadamard gate $H$ is applied to the probe qubit, followed by a controlled-$U$ gate (with the probe as control and the $n$ mixed qubits as targets), and finally another Hadamard gate on the probe. Measuring the Pauli $\sigma_z$ observable on the probe qubit yields:
\begin{equation}
    \langle \sigma_z \rangle = \frac{\mathrm{Re}\{\mathrm{Tr}(U)\}}{2^n}.
    \label{eq:ch2-dqc1-trace}
\end{equation}
By replacing the final Hadamard with a phase gate $S^\dagger H$, one can similarly extract $\mathrm{Im}\{\mathrm{Tr}(U)\}/2^n$, thereby obtaining the full normalized trace $\mathrm{Tr}(U)/2^n$.

\begin{figure}[t]
    \centering
    \includegraphics[width=0.7\textwidth]{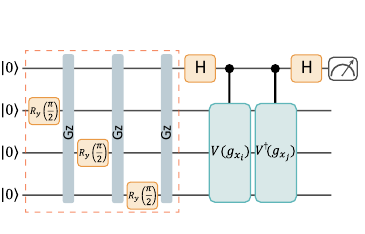}
    \caption{Experimental NQE-DQC1 circuit. A single probe qubit is acted upon by Hadamard gates, while the remaining register implements the controlled feature-map unitary and its Hermitian conjugate. Measurement of $\sigma_z$ on the probe yields the Hilbert--Schmidt inner product required for the NQE-DQC1 loss.}
    \label{fig:ch2-nqe-dqc1-schematic}
\end{figure}

Despite its limited initial resources, the DQC1 model is believed to offer computational power beyond classical simulation. \textcite{shor2008estimating} showed that estimating the Jones polynomial---a problem related to knot invariants---is complete for the DQC1 complexity class. Moreover, \textcite{poulin2004exponential} demonstrated an exponential quantum speedup for estimating fidelity decay in quantum chaos using DQC1. The quantum correlations responsible for this computational power arise not from entanglement in the conventional sense, but from quantum discord between the probe and the mixed register~\cite{datta2005entanglement}.

Crucially for our purposes, the DQC1 model is a natural fit for NMR quantum processors. In NMR, the quantum state at thermal equilibrium is close to the maximally mixed state, with only a small polarization on each spin. Through spatial averaging techniques, one can prepare an effective DQC1 state with a polarized probe and a maximally mixed encoding register, which realizes the structure idealized in \Cref{eq:ch2-dqc1-initial}.

\subsection{NQE Loss Function via Hilbert--Schmidt Inner Product}
\label{subsec:ch2-nqe-dqc1-loss}

We now reformulate the NQE training objective in terms of a quantity that DQC1 can directly measure.
The \emph{Hilbert--Schmidt (HS) inner product} between two unitary operators $U_1$ and $U_2$ is defined as:
\begin{equation}
    \langle U_1, U_2 \rangle_{\mathrm{HS}} = \frac{\mathrm{Tr}(U_1^\dagger U_2)}{2^n}.
    \label{eq:ch2-hs-inner-product}
\end{equation}
This quantity is directly related to the normalized \emph{Frobenius distance} between the two unitaries:
\begin{equation}
    \frac{1}{2^n}\|V(g(\mathbf{x}_i, \mathbf{w})) - V(g(\mathbf{x}_j, \mathbf{w}))\|_F^2 = 2 - 2\,\mathrm{Re}\!\left\{\mathrm{Tr}\!\left[V(g(\mathbf{x}_i, \mathbf{w}))\, V^\dagger(g(\mathbf{x}_j, \mathbf{w}))\right]\right\}/2^n,
    \label{eq:ch2-frobenius}
\end{equation}
where $V(g(\mathbf{x}, \mathbf{w}))$ is the quantum embedding circuit parameterized by the neural network output, as in \Cref{eq:ch2-nqe-map}. Minimizing this normalized Frobenius distance for same-class pairs and increasing it for cross-class pairs is analogous to the fidelity-based objective of \Cref{eq:ch2-nqe-fidelity-loss}, but now expressed entirely in terms of the HS inner product.

The NQE-DQC1 loss function is defined as:
\begin{equation}
    L_{\mathrm{NQE}} = \sum_{i,j} \left[\frac{1}{2^n}\mathrm{Re}\!\left\{\mathrm{Tr}\!\left[V(g(\mathbf{x}_i, \mathbf{w}))\, V^\dagger(g(\mathbf{x}_j, \mathbf{w}))\right]\right\} - \frac{1 + y_i y_j}{2}\right]^2.
    \label{eq:ch2-nqe-dqc1-loss}
\end{equation}
The target value $(1 + y_i y_j)/2$ is identical to the fidelity-based loss: it equals $1$ for same-class pairs and $0$ for different-class pairs. The key difference is that the fidelity $|\langle \mathbf{x}_i | \mathbf{x}_j \rangle|^2$ has been replaced by the measured real part of the HS inner product
\begin{equation}
    \frac{\mathrm{Tr}[V(g(\mathbf{x}_i, \mathbf{w}))\, V^\dagger(g(\mathbf{x}_j, \mathbf{w}))]}{2^n}.
\end{equation}

The crucial observation is that this HS inner product is precisely what DQC1 measures. Setting
\begin{equation}
    U = V(g(\mathbf{x}_i, \mathbf{w}))\, V^\dagger(g(\mathbf{x}_j, \mathbf{w}))
\end{equation}
in \Cref{eq:ch2-dqc1-trace}, the DQC1 measurement on the probe qubit yields:
\begin{equation}
    \langle \sigma_z \rangle = \frac{\mathrm{Re}\!\left\{\mathrm{Tr}\!\left[V(g(\mathbf{x}_i, \mathbf{w}))\, V^\dagger(g(\mathbf{x}_j, \mathbf{w}))\right]\right\}}{2^n},
    \label{eq:ch2-dqc1-measurement}
\end{equation}
which directly provides the real part of the HS inner product needed for the loss function (\Cref{eq:ch2-nqe-dqc1-loss}). The loss drives same-class pairs toward an HS inner product of $1$ (similar unitaries) and different-class pairs toward low HS overlap, providing an HS-compatible class-separation surrogate.

The classical neural network $g(\mathbf{x}, \mathbf{w})$ is optimized via standard gradient descent. Each loss evaluation is estimated from repeated DQC1 probe-qubit measurements. As in the original NQE framework, the embedding circuit $V$ can be any fixed quantum circuit, such as the ZZ feature map (\Cref{eq:ch2-zz-feature-map}) with $\boldsymbol{\phi}(\mathbf{x})$ replaced by the learned function $g(\mathbf{x}, \mathbf{w})$.

\subsection{Experimental Demonstration}
\label{subsec:ch2-dqc1-experiment}

\paragraph{NMR platform and initialization.}
The NQE-DQC1 protocol was experimentally demonstrated on a Bruker 300 MHz NMR spectrometer using ${}^{13}$C-labeled \emph{trans}-crotonic acid dissolved in d6-acetone~\cite{liu2025nqe_dqc1}. The molecule provides four carbon nuclear spins: C1 serves as the probe qubit, while C2--C4 serve as the three encoding qubits ($n = 3$). The initial state is prepared via spatial averaging as an effective DQC1 state $\rho_0 = |0\rangle\langle 0| \otimes I/2^3$, matching the structure of \Cref{eq:ch2-dqc1-initial}. The quantum embedding circuit uses the ZZ feature map (\Cref{eq:ch2-zz-feature-map}) with $M = 1$ repetition, and the classical input data is preprocessed via PCA to reduce the dimensionality to five principal components before being fed to the neural network $g$.

\begin{figure}[t]
    \centering
    \includegraphics[width=0.85\textwidth]{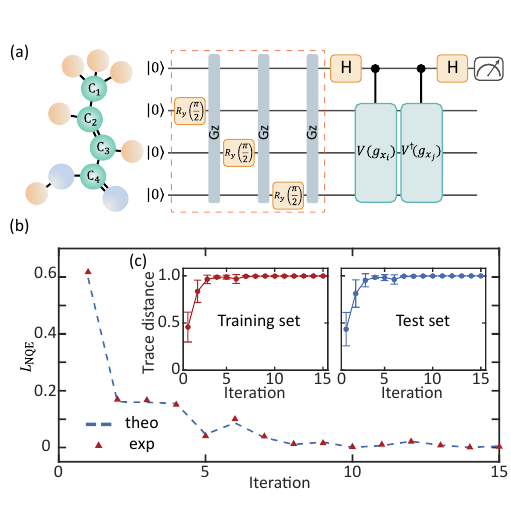}
    \caption{NQE-DQC1 training on the NMR platform. (a)~Schematic of the NMR DQC1 circuit: the probe qubit C1 controls the application of $V^\dagger(g(\mathbf{x}_j))\,V(g(\mathbf{x}_i))$ on qubits C2--C4. (b)~Training loss versus iteration for NMR experiments (markers) and numerical simulation (solid line). (c)~Trace distance between class ensembles for the training set (circles) and test set (triangles) across NQE training iterations.}
    \label{fig:ch2-dqc1-training}
\end{figure}

\paragraph{NQE-DQC1 training results.}
The training was performed on 500 MNIST images (binary classification of digits 0 vs.\ 1), with 15 training iterations and 10 randomly sampled data pairs per iteration. For each pair $(\mathbf{x}_i, \mathbf{x}_j)$, the DQC1 circuit implements the unitary $U = V^\dagger(g(\mathbf{x}_j, \mathbf{w}))\, V(g(\mathbf{x}_i, \mathbf{w}))$ controlled by the probe qubit to measure the HS inner product via \Cref{eq:ch2-dqc1-measurement}. \Cref{fig:ch2-dqc1-training}(b) shows the training loss as a function of iteration: the loss converges to approximately zero by iteration 10, and the NMR experimental results closely match the numerical simulations, confirming the practical feasibility of the DQC1-based approach. \Cref{fig:ch2-dqc1-training}(c) demonstrates that the trace distance between class ensembles increases for both the training and test sets across iterations, consistent with the HS surrogate being aligned with improved distinguishability in this experiment.

\FloatBarrier

\begin{figure}[ht]
    \centering
    \includegraphics[width=0.8\textwidth]{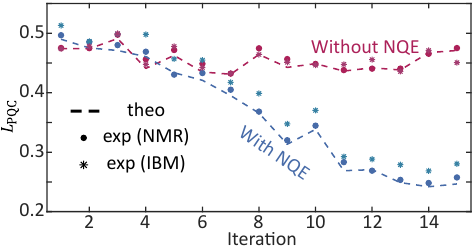}
    \caption{Cross-platform training loss of the downstream PQC classifier with and without NQE. The classifier is optimized on the NQE-enhanced embedding (blue) and on the conventional ZZ feature map (red, ``without NQE''). Dashed lines denote noiseless simulation, filled circles the NMR experiment, and asterisks the IBM superconducting hardware.}
    \label{fig:ch2-dqc1-crossplatform}
\end{figure}
\FloatBarrier

\paragraph{Classification results.}
After NQE training, a parameterized quantum circuit (PQC) classifier is deployed on the NQE-enhanced embedding. The classifier uses 2 circuit layers with 4 trainable rotation parameters and 2 CNOT gates, and classification is performed by measuring $\langle \sigma_z \rangle$ on the third qubit. Without NQE (using the raw ZZ feature map), the classifier achieves 54\% accuracy on the 500-sample dataset, which is close to random guessing and consistent with the low trace distance of the unoptimized embedding. With NQE preprocessing, the accuracy reaches 98\%.

\begin{figure}[h]
    \centering
    \includegraphics[width=\textwidth]{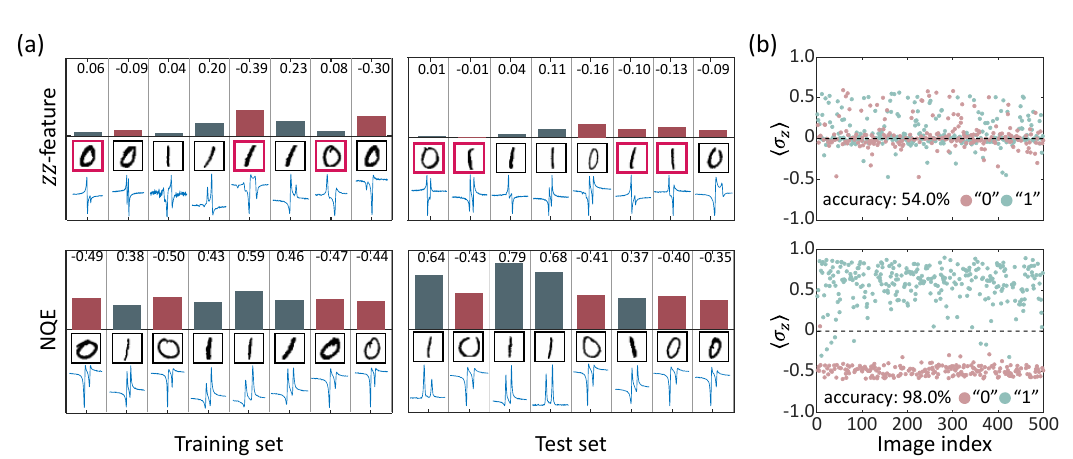}
    \caption{Classification results with and without NQE. (a)~Per-sample classification outputs using the ZZ feature map alone (top) versus the NQE-enhanced embedding (bottom). (b)~Scatter plot of predicted labels for all 500 MNIST test images (digits 0 vs.\ 1).}
    \label{fig:ch2-classification-results}
\end{figure}
\FloatBarrier

\paragraph{Cross-platform extendability.}
An important feature of the NQE-DQC1 approach is cross-platform transferability. The neural network $g(\mathbf{x}, \mathbf{w})$ trained through the NMR-compatible DQC1 protocol was deployed on IBM cloud superconducting quantum processors without retraining, and the experimental trends agreed with the numerical simulations (\Cref{fig:ch2-dqc1-crossplatform})~\cite{liu2025nqe_dqc1}. This is possible because NQE training produces a set of classical neural network parameters $\mathbf{w}^*$ that define a data preprocessing function $g(\mathbf{x}, \mathbf{w}^*)$, which is independent of the hardware used to evaluate the training loss. The DQC1 protocol is used during the NQE training phase to evaluate the loss function. Once training is complete, the optimized preprocessing can be paired with another platform that implements the same embedding circuit $V$. Additional validation on Fashion-MNIST and satellite image datasets supports the applicability of the approach across the tested data domains~\cite{liu2025nqe_dqc1}.
Further supplementary analyses on feature-map ansatz robustness and multi-class extensions, together with reproducibility details, are collected in \Cref{sec:app-nqe-details}.

\section{Summary}
\label{sec:ch2-conclusions}

In this chapter, we addressed a central question in quantum machine learning: how does the data embedding constrain the achievable performance of a quantum classifier? We established three main results.

First, we showed that the empirical risk of the binary quantum classifier under the linear loss is lower bounded by the trace distance between embedded data ensembles (\Cref{eq:ch2-lower-bound}), a quantity determined by the data embedding. The contractive property of trace distance under quantum channels (\Cref{eq:ch2-contractive}) further implies that subsequent noisy quantum processing cannot increase this distinguishability. We demonstrated that conventional deterministic embeddings (amplitude encoding, angle encoding, and the ZZ feature map) are data-agnostic and offer no guarantee of producing a large trace distance for a given classification task, and that trainable quantum embedding strategies remain constrained by the distinguishability allowed by the embedding construction.

Second, we introduced Neural Quantum Embedding (NQE), which avoids the contractive constraint by inserting a classical neural network between the raw data and the quantum embedding circuit. Because the neural network operates on classical parameters rather than quantum states, it is not subject to the quantum-state contractive inequality at this preprocessing stage. The implicit fidelity loss (\Cref{eq:ch2-nqe-fidelity-loss}) provides a tractable training objective that simultaneously clusters same-class states and separates cross-class states in the quantum feature space. In the reported IBM hardware experiments on binary MNIST, PCA-NQE raises the trace distance from $\approx 0.27$ to $\approx 0.84$, improving classification accuracy from $52.7\%$ to $96.1\%$ under real hardware noise and achieving a lower empirical risk than the noiseless Helstrom limit associated with the conventional ZZ feature map.

Third, we showed that NQE can be extended to ensemble quantum systems by replacing the fidelity-based loss with the Hilbert--Schmidt inner product, which is directly measurable via the DQC1 model using a single probe qubit and a mixed encoding register. This reformulation was experimentally validated on an NMR quantum processor, where it achieved 98\% classification accuracy on MNIST compared to 54\% without NQE. The deployment of the learned neural network parameters on superconducting quantum processors further shows that the preprocessing learned through the DQC1 protocol can be transferred across platforms that implement the same embedding circuit.

Throughout this chapter, we focused on a representation-induced source of training error in quantum machine learning: the achievable training loss can be reduced only if the embedding makes the class ensembles sufficiently distinguishable. However, low training error does not automatically guarantee good performance on unseen data. In the next chapter, we turn to the complementary question of \emph{generalization}: under what conditions does a quantum classifier trained on a finite dataset maintain its performance on new, previously unseen inputs? We will show that the trace-distance perspective also connects naturally to margin-based generalization analysis.


\chapter{Generalization in Quantum Machine Learning}
\label{chap:generalization}


\chaptersource{The results in this chapter are based on \textit{Understanding Generalization in Quantum Machine Learning with Margins}~\cite{hur2024margin}. Code is available at \url{https://github.com/takh04/Q-margin}.}
\bigskip

In the previous chapter, we studied a representation-induced limit on quantum supervised learning. Neural Quantum Embedding increases the trace distance between embedded data ensembles and thereby lowers the achievable training loss. However, achieving low training error is only part of the challenge: a model that memorizes its training data without learning the underlying pattern will fail on new, unseen inputs. This chapter turns to the complementary question of \emph{generalization}---the ability of a model to maintain its performance beyond the training set.

As discussed in \Cref{subsubsec:error-decomposition}, the generalization error is the discrepancy between the true risk and the empirical risk, arising because the model is trained on a finite sample rather than the full data distribution. Controlling this error requires understanding the complexity of the hypothesis class. In classical machine learning, uniform generalization bounds based on complexity measures such as VC dimension and Rademacher complexity have been the primary theoretical tools. However, recent work has shown that such bounds are often too vacuous to explain empirical comparisons between models that generalize and models that memorize data, both in classical deep learning~\cite{zhang2021understanding} and in quantum machine learning~\cite{gilfuster2024understanding}.

In this chapter, we develop a \emph{margin-based} generalization framework for quantum neural networks that responds to these limitations. Drawing inspiration from the classical margin theory of \textcite{bartlett2017spectrally}, we establish a high-probability bound for multiclass QNN classifiers that remains a function-class bound, but is margin-sensitive through the empirical margin loss. We experimentally show that margin-based metrics are stronger predictors of generalization performance than parameter-based metrics in the studied QPR setting. Furthermore, we connect the margin to quantum state discrimination, showing that quantum embeddings with large trace distances---such as those produced by NQE (\Cref{chap:nqe})---can support larger margins and thus tighter margin-based generalization bounds when the other bound parameters are comparable.

\section{Theoretical Background}
\label{sec:ch3-background}

This section reviews the key concepts from classical generalization theory that form the foundation for our quantum margin bounds. We build the theory progressively: starting from the simplest case of finite hypothesis classes, moving to the tools needed for infinite classes, and finally arriving at margin theory. We assume a supervised learning setting as established in \Cref{subsubsec:error-decomposition}: given a hypothesis class $\mathcal{H}$ and a training set $S = \{(x_i, y_i)\}_{i=1}^m$ drawn i.i.d.\ from an unknown distribution $\mathcal{D}$, we seek to bound the difference between the true risk $R(h) = \mathbb{E}_{(x,y) \sim \mathcal{D}}[\mathds{1}(\argmax_j h(x)_j \neq y)]$ and the empirical risk $\hat{R}(h) = m^{-1} \sum_{i=1}^m \mathds{1}(\argmax_j h(x_i)_j \neq y_i)$ for any hypothesis $h \in \mathcal{H}$.

\subsection{Generalization Bounds for Finite Hypothesis Classes}
\label{subsec:ch3-finite}

We begin with the simplest setting: a hypothesis class $\mathcal{H}$ containing finitely many hypotheses. Even this restricted case reveals the core tension in generalization theory---the gap between controlling error for a \emph{single fixed} hypothesis and controlling error for the \emph{best hypothesis selected from data}.

\paragraph{Concentration for a single hypothesis.}
Consider a single, fixed hypothesis $h$. Since the training samples $(x_1, y_1), \ldots, (x_m, y_m)$ are drawn i.i.d., the empirical risk $\hat{R}(h)$ is an average of $m$ independent bounded random variables, each taking values in $[0,1]$. Hoeffding's inequality~\cite{shalev2014understanding} gives a sharp concentration bound:
\begin{equation}
    \Pr\!\left[|R(h) - \hat{R}(h)| > \epsilon\right] \leq 2\exp(-2m\epsilon^2).
    \label{eq:ch3-hoeffding}
\end{equation}
Thus, for any \emph{fixed} hypothesis, the empirical risk converges exponentially fast to the true risk as the sample size $m$ grows. If we could specify which hypothesis to evaluate before seeing the data, generalization would be straightforward.

\paragraph{The selection problem.}
In practice, however, we do not fix $h$ in advance. Instead, we use the training data to select the hypothesis with the lowest empirical risk: $\hat{h} = \argmin_{h \in \mathcal{H}} \hat{R}(h)$. This data-dependent selection invalidates the direct application of \Cref{eq:ch3-hoeffding}, because the selected hypothesis is correlated with the training sample. A hypothesis that happens to perform well on the training set may do so by chance rather than by capturing the true pattern.

\paragraph{Union bound over $\mathcal{H}$.}
When $\mathcal{H}$ is finite, we can resolve this by applying Hoeffding's inequality to \emph{every} $h \in \mathcal{H}$ simultaneously and taking a union bound. The probability that \emph{any} hypothesis deviates by more than $\epsilon$ is at most:
\begin{equation}
    \Pr\!\left[\exists\, h \in \mathcal{H}: |R(h) - \hat{R}(h)| > \epsilon\right] \leq \sum_{h \in \mathcal{H}} 2\exp(-2m\epsilon^2) = 2|\mathcal{H}|\exp(-2m\epsilon^2).
    \label{eq:ch3-union-bound}
\end{equation}
Setting this probability to $\delta$ and solving for $\epsilon$, we obtain the \emph{finite hypothesis class bound}: with probability at least $1 - \delta$, for all $h \in \mathcal{H}$ simultaneously,
\begin{equation}
    R(h) \leq \hat{R}(h) + \sqrt{\frac{\ln|\mathcal{H}| + \ln(2/\delta)}{2m}}.
    \label{eq:ch3-finite-bound}
\end{equation}
This result has an appealing interpretation: the ``price'' of selecting the best hypothesis from $|\mathcal{H}|$ candidates is only logarithmic in the class size. Even if $|\mathcal{H}|$ is astronomically large, the generalization penalty grows slowly.

\paragraph{Limitation.}
The finite hypothesis class bound breaks down when $|\mathcal{H}| = \infty$---as is the case for hypothesis classes parameterized by continuous parameters. Neural networks, support vector machines, and quantum circuits all define infinite hypothesis classes, rendering \Cref{eq:ch3-finite-bound} vacuous. This motivates the development of complexity measures that can handle infinite classes, which we turn to next.

\subsection{Complexity Measures for Infinite Hypothesis Classes}
\label{subsec:ch3-complexity}

When the hypothesis class is infinite, we can no longer enumerate its elements. Instead, we need complexity measures that capture the ``effective size'' of $\mathcal{H}$---how richly it can fit arbitrary patterns---without requiring finiteness.

\subsubsection{VC Dimension}

The \emph{Vapnik--Chervonenkis (VC) dimension}~\cite{vapnik1995nature} was the first such measure. A hypothesis class $\mathcal{H}$ \emph{shatters} a set of $m$ points if, for every possible labeling of those points, there exists an $h \in \mathcal{H}$ that realizes that labeling. The VC dimension $d_{\mathrm{VC}}$ is the largest $m$ such that some set of $m$ points can be shattered by $\mathcal{H}$.

The fundamental theorem of PAC learning~\cite{vapnik1995nature, shalev2014understanding} establishes that $d_{\mathrm{VC}}$ characterizes learnability: a binary hypothesis class is PAC-learnable if and only if its VC dimension is finite. Moreover, the VC generalization bound states that with high probability,
\begin{equation}
    R(h) \leq \hat{R}(h) + O\!\left(\sqrt{\frac{d_{\mathrm{VC}}}{m}}\right).
    \label{eq:ch3-vc-bound}
\end{equation}
While the VC dimension elegantly extends the finite class bound to infinite classes (note the structural similarity between \Cref{eq:ch3-finite-bound,eq:ch3-vc-bound}, with $\ln|\mathcal{H}|$ replaced by $d_{\mathrm{VC}}$), it has a significant limitation for modern models: the VC dimension of a neural network scales with its number of parameters, producing bounds that are vacuous for overparameterized models that nonetheless generalize well in practice.

\subsubsection{Rademacher Complexity}

\emph{Rademacher complexity}~\cite{mohri2018foundations} provides a data-dependent measure of hypothesis class complexity. For a sample $S = \{z_i\}_{i=1}^m$ and a function class $\mathcal{F}$, the \emph{sample Rademacher complexity} is defined as:
\begin{equation}
    \mathfrak{R}(\mathcal{F}|_S) = \mathbb{E}_{\boldsymbol{\sigma}}\left[\sup_{f \in \mathcal{F}} \frac{1}{m} \sum_{i=1}^m \sigma_i f(z_i)\right],
    \label{eq:ch3-rademacher}
\end{equation}
where $\sigma_1, \ldots, \sigma_m$ are independent Rademacher random variables taking values $\pm 1$ with equal probability. Intuitively, the Rademacher complexity measures how well functions in $\mathcal{F}$ can correlate with random noise---a class that can fit random labels has high Rademacher complexity and is prone to overfitting.

A fundamental result connects Rademacher complexity to generalization~\cite{mohri2018foundations, shalev2014understanding}: for any $\delta > 0$, with probability at least $1 - \delta$ over the random draw of an i.i.d.\ sample $S$ of size $m$, the following holds for all $h \in \mathcal{H}$:
\begin{equation}
    R(h) \leq \hat{R}(h) + 2\mathfrak{R}(\mathcal{H}|_S) + 3\sqrt{\frac{\ln(2/\delta)}{2m}}.
    \label{eq:ch3-rademacher-bound}
\end{equation}
This bound is \emph{uniform}---it holds simultaneously for all hypotheses in $\mathcal{H}$---and becomes tighter as the sample size $m$ increases or the Rademacher complexity decreases. Compared to the VC bound, the Rademacher bound has the advantage of being \emph{data-dependent}: the sample Rademacher complexity depends on the specific data points, not just the abstract hypothesis class.

\subsubsection{Covering Numbers and Dudley's Entropy Integral}

In practice, computing the Rademacher complexity directly can be challenging. A powerful technique for bounding it uses \emph{covering numbers}, which bridge the gap between infinite and finite by discretizing function classes at progressively finer resolutions.

The $\epsilon$-covering number of a set $A$ with respect to a norm $\|\cdot\|$, denoted $\mathcal{N}(A, \epsilon, \|\cdot\|)$, is the minimum number of balls of radius $\epsilon$ needed to cover $A$. The key idea is that even an infinite function class can be approximated to precision $\epsilon$ by a finite set of size $\mathcal{N}(A, \epsilon, \|\cdot\|)$---effectively reducing the problem to the finite case at each resolution scale.

The connection between covering numbers and Rademacher complexity is established through \emph{Dudley's entropy integral}~\cite{shalev2014understanding}:
\begin{equation}
    \mathfrak{R}(A) \leq \inf_{\alpha \geq 0} \left(\frac{4\alpha}{\sqrt{m}} + \frac{12}{m} \int_{\alpha}^{\sqrt{m}} \sqrt{\ln \mathcal{N}(A, \beta, \|\cdot\|_2)}\, d\beta\right).
    \label{eq:ch3-dudley}
\end{equation}
This integral aggregates the covering numbers across all scales $\beta$. Fine scales (small $\beta$) capture local complexity, while coarse scales (large $\beta$) capture global structure. Dudley's entropy integral is the key technical tool we will use in \Cref{subsec:ch3-margin-bound} to derive our margin generalization bound for QNNs.

\subsection{Margin Theory}
\label{subsec:ch3-margin}

The complexity measures discussed above---VC dimension, Rademacher complexity, covering numbers---all provide \emph{uniform} bounds that depend on the hypothesis class $\mathcal{H}$ as a whole. \emph{Margin theory} introduces a complementary, \emph{hypothesis-dependent} perspective: the generalization bound depends on how confidently the specific learned hypothesis classifies the training data. The concept of margin has been central to generalization theory since the early days of support vector machines~\cite{cortes1995support}.

\subsubsection{Classical Margin Bounds}

For binary classification, the margin of a classifier $f$ on a sample $(x, y)$ is defined as $y \cdot f(x)$, where $y \in \{-1, +1\}$. The sample is correctly classified with margin $\gamma$ if $y \cdot f(x) \geq \gamma > 0$.

For multiclass classification with $k$ classes, the margin is generalized through the \emph{margin operator}:
\begin{equation}
    \mathcal{M}(v, y) = v_y - \max_{i \neq y} v_i,
    \label{eq:ch3-margin-operator}
\end{equation}
where $v = (v_1, \ldots, v_k)$ is the output vector of the classifier and $y$ is the true label. A positive margin $\mathcal{M}(v, y) > 0$ indicates correct classification, and a larger margin corresponds to a more confident prediction.

To incorporate margins into generalization bounds, one replaces the hard $0$-$1$ loss with a Lipschitz surrogate such as the \emph{ramp loss} $l_\gamma : \mathbb{R} \to \mathbb{R}^+$, parameterized by the margin threshold $\gamma > 0$:
\begin{equation}
    l_\gamma(x) = \begin{cases}
        0, & \text{if } x > \gamma, \\
        1 - x/\gamma, & \text{if } 0 \leq x \leq \gamma, \\
        1, & \text{if } x < 0.
    \end{cases}
    \label{eq:ch3-ramp-loss}
\end{equation}
The ramp loss upper bounds the $0$-$1$ loss while remaining Lipschitz, which makes it suitable for Rademacher-complexity arguments. The theorem below is stated in terms of the empirical margin error,
\begin{equation}
    \hat{R}_\gamma(h) = \frac{1}{m} \sum_{i=1}^m \mathds{1}\!\left(\mathcal{M}(h(x_i), y_i) \leq \gamma\right),
    \label{eq:ch3-empirical-margin-loss}
\end{equation}
which counts both misclassified samples and correctly classified samples whose margin is at most $\gamma$.

\subsubsection{Margin Bounds for Deep Networks}

A landmark result by \textcite{bartlett2017spectrally} established spectrally-normalized margin bounds for deep neural networks. Their key insight was that the margin distribution, normalized by the spectral norms of the weight matrices, is strongly correlated with generalization performance. This represented a significant advance over classical uniform bounds, which depend on the number of parameters or the VC dimension and often provide vacuous estimates for overparameterized networks.

Specifically, for a deep network with weight matrices $W_1, \ldots, W_L$, the generalization bound scales as:
\begin{equation}
    R(h) \leq \hat{R}_\gamma(h) + \tilde{O}\left(\frac{\prod_{\ell=1}^L \|W_\ell\|_\sigma}{\gamma \sqrt{m}}\right),
    \label{eq:ch3-bartlett-bound}
\end{equation}
where $\|W_\ell\|_\sigma$ denotes the spectral norm of the $\ell$-th weight matrix. Unlike parameter-count bounds, this bound becomes tighter when the model achieves larger margins. Subsequent work confirmed that margin-based metrics are among the strongest predictors of generalization in deep learning~\cite{neyshabur2017pac, jiang2018predicting, jiang2019fantastic, dziugaite2020search}.

Our contribution in this chapter is to extend this margin-based framework from classical neural networks to quantum neural networks, adapting the covering number techniques to the quantum setting.

\section{Generalization Bounds for Quantum Models}
\label{sec:ch3-qml-generalization}

We now review the existing approaches to understanding generalization in quantum machine learning. These approaches predominantly rely on \emph{uniform bounds}---bounds that hold for all hypotheses within the function class---and we discuss their strengths and limitations.

\subsection{Uniform Generalization Bounds}
\label{subsec:ch3-uniform-bounds}

A growing body of work has established uniform generalization bounds for quantum models using various complexity measures.

\paragraph{Parameter-based bounds.}
The most widely adopted approach is based on the number of trainable parameters. \textcite{caro2022generalization} proved that the generalization gap of a QNN with $T$ trainable gates, trained on $m$ samples, is bounded by:
\begin{equation}
    |R(h) - \hat{R}(h)| \leq O\left(\sqrt{\frac{T}{m}}\right),
    \label{eq:ch3-caro-bound}
\end{equation}
with high probability. When only a subset of parameters undergo substantial change during training, the bound improves to scale with the number of \emph{effective parameters} rather than the total parameter count. This result has become a standard tool in the QML community, providing a simple and interpretable estimate of generalization.

\paragraph{Effective dimension.}
\textcite{abbas2021effective} introduced the \emph{effective dimension} of a QML model as a measure of its complexity, connecting it to the Fisher information matrix. Models with smaller effective dimension are expected to generalize better, as they occupy a smaller effective volume in the hypothesis space.

\paragraph{Quantum Rademacher and statistical complexity.}
\textcite{bu2021rademacher} derived bounds on the Rademacher complexity of quantum circuits, relating the generalization capability to properties of the quantum circuit architecture. Subsequent work~\cite{bu2022statistical, bu2023effects} extended this analysis to study the effects of quantum resources and noise on the statistical complexity of quantum models.

\paragraph{Information-theoretic approaches.}
\textcite{banchi2021generalization} analyzed QML generalization from a quantum information standpoint, connecting the generalization gap to information-theoretic quantities such as the mutual information between the training data and the learned hypothesis. More recently, \textcite{caro2023information} established information-theoretic generalization bounds for learning from quantum data, providing a complementary perspective to the complexity-based approaches.

\subsection{Limitations of Uniform Bounds}
\label{subsec:ch3-uniform-limitations}

Despite the theoretical elegance of the bounds reviewed above, they share a limitation for the comparisons studied here: they are \emph{uniform bounds} that hold for all hypotheses in the function class simultaneously. In classical deep learning, a seminal study by \textcite{zhang2021understanding} demonstrated that modern neural networks can easily memorize datasets with completely randomized labels, achieving zero training error on random noise. Since uniform bounds must hold for \emph{all} functions in the hypothesis class---including those that memorize random labels---they can be too vacuous to distinguish a model that has learned meaningful patterns from one that has merely memorized the training data in many practical settings~\cite{nagarajan2019uniform, dziugaite2017computing}.

Building on this observation, \textcite{gilfuster2024understanding} demonstrated that the same phenomenon occurs in QML: quantum neural networks, including QCNNs, can overfit randomized labels on quantum datasets. Despite the small number of qubits and parameters, QNNs are expressive enough to fit random labels, showing that uniform bounds can also be too vacuous for QML memorization regimes.

These findings motivate a shift from uniform bounds toward \emph{data-dependent} and \emph{hypothesis-dependent} generalization measures. In classical deep learning, this shift led to the discovery that margin-based metrics are among the strongest predictors of generalization~\cite{jiang2018predicting, jiang2019fantastic, dziugaite2020search}. In the next section, we extend this margin-based perspective to quantum neural networks.

\section{Margin-Based Generalization for Quantum Neural Networks}
\label{sec:ch3-margin-qnn}

We now present our main contribution, a margin-based generalization bound for multiclass classification with quantum neural networks. We begin by formalizing the QNN classification setup, derive the margin bound, and validate it experimentally.

\subsection{Multiclass Classification with Quantum Neural Networks}
\label{subsec:ch3-qnn-classification}

Consider a $k$-class classification task where the input is an $n$-qubit quantum state $\rho \in \mathbb{C}^{N \times N}$ with $N = 2^n$, and the label $y \in [k] = \{1, \ldots, k\}$. A QNN performs classification using a parameterized unitary circuit $U(\theta) \in \mathbb{U}_{\text{QNN}}$ and a set of positive operator-valued measurements (POVMs) $\{E_i\}_{i=1}^k$ satisfying $\sum_{i=1}^k E_i = I$ and $E_i \geq 0$. The QNN maps an input quantum state to a $k$-dimensional probability vector:
\begin{equation}
    h_\theta(\rho) = \left\{\mathrm{Tr}\!\left(U(\theta)\rho U^\dagger(\theta) E_i\right)\right\}_{i=1}^k,
    \label{eq:ch3-qnn-output}
\end{equation}
where each component $h_\theta(\rho)_i$ represents the probability of assigning the input to class $i$.

Given $m$ training samples $S = \{(\rho_i, y_i)\}_{i=1}^m$ drawn i.i.d.\ from an unknown distribution $\mathcal{D}$, the goal is to find optimal parameters $\theta^*$ that minimize the true error:
\begin{equation}
    R(h^*) = \mathbb{E}_{(\rho, y) \sim \mathcal{D}}\left[\mathds{1}\!\left(\argmax_j h^*(\rho)_j \neq y\right)\right].
    \label{eq:ch3-true-error}
\end{equation}
The generalization gap $g(h) = R(h) - \hat{R}(h)$ measures the discrepancy between the true error and the empirical error on the training set.

\subsection{Margin Generalization Bound}
\label{subsec:ch3-margin-bound}

We now derive the margin-based generalization bound for QNNs. The key idea is to bound the Rademacher complexity of the margin loss function class in terms of the quantum channel components.

For a margin threshold $\gamma > 0$, the margin loss function class is:
\begin{equation}
    \mathcal{F}_\gamma = \left\{(\rho, y) \mapsto l_\gamma(\mathcal{M}(h(\rho), y)) : h \in \mathcal{H}\right\},
    \label{eq:ch3-margin-function-class}
\end{equation}
where $\mathcal{M}$ is the margin operator (\Cref{eq:ch3-margin-operator}) and $l_\gamma$ is the ramp loss (\Cref{eq:ch3-ramp-loss}). Applying the Rademacher generalization bound (\Cref{eq:ch3-rademacher-bound}) to $\mathcal{F}_\gamma$, we obtain:
\begin{equation}
    \Pr_{(\rho,y)\sim\mathcal{D}}\!\left[\argmax_i h(\rho)_i \neq y\right] \leq \hat{R}_\gamma(h) + 2\mathfrak{R}(\mathcal{F}_\gamma|_S) + 3\sqrt{\frac{\ln(2/\delta)}{2m}},
    \label{eq:ch3-margin-rademacher}
\end{equation}
for all $h \in \mathcal{H}$, with probability at least $1 - \delta$.

The central technical contribution is an analytic bound on the Rademacher complexity $\mathfrak{R}(\mathcal{F}_\gamma|_S)$ in terms of the quantum channel components. This derivation proceeds in three steps:

\paragraph{Step 1: Lipschitz continuity.}
We first establish the Lipschitz properties of the function class. The margin loss $l_\gamma(\mathcal{M}(\cdot, y))$ is $2/\gamma$-Lipschitz with respect to the $l_p$ norm for any $p \geq 1$. The quantum measurement function $g(x) = \{x^\dagger E_i x\}_{i=1}^k$ (for pure state inputs $\rho = |x\rangle\langle x|$ with $x \in \mathbb{C}^N$) is $2E$-Lipschitz, where $E = \sqrt{\sum_i \|E_i\|_\sigma^2}$ and $\|\cdot\|_\sigma$ denotes the spectral norm. This result follows from each measurement function $g_i$ being $2\|E_i\|_\sigma$-Lipschitz for normalized quantum states.

\paragraph{Step 2: Covering number bound.}
Using the Lipschitz continuity, we reduce the covering number of $\mathcal{F}_\gamma$ restricted to the sample $S$ to the covering number of the set $\{UX : U \in \mathbb{U}_{\text{QNN}}\}$, where $X \in \mathbb{C}^{N \times m}$ is the data matrix whose columns are the quantum state vectors:
\begin{equation}
    \ln \mathcal{N}\!\left((\mathcal{F}_\gamma)|_S, \epsilon, \|\cdot\|_2\right) \leq \left\lceil \frac{32mb^2 E^2}{\epsilon^2 \gamma^2} \right\rceil \ln 4N^2,
    \label{eq:ch3-covering-bound}
\end{equation}
where $b$ is a distance bound such that $\|U - U_{\mathrm{ref}}\|_{2,1} \leq b$ for all $U \in \mathbb{U}_{\text{QNN}}$, with $U_{\mathrm{ref}}$ serving as a reference unitary matrix. This bound is derived using matrix covering techniques originally introduced by \textcite{bartlett2017spectrally} and \textcite{zhang2002covering}, adapted to the complex-valued setting of quantum circuits.
Here $\|A\|_{2,1}$ denotes the sum of Euclidean norms of the columns of $A$. The parameter $b$ measures the radius of the accessible unitary family around the reference unitary in this matrix norm.

\paragraph{Step 3: Dudley's entropy integral.}
Finally, we bound the Rademacher complexity using Dudley's entropy integral (\Cref{eq:ch3-dudley}) with the covering number bound from Step 2. This yields our main result:

\begin{theorem}[Margin Generalization Bound for QNNs]
\label{thm:ch3-margin-bound}
Consider an $n$-qubit QNN with unitary $U \in \mathbb{U}_{\emph{QNN}}$ and POVMs $\{E_i\}_{i=1}^k$ for $k$-class classification. Let $b$ be a distance bound such that $\|U - U_{\emph{ref}}\|_{2,1} \leq b$ for any $U \in \mathbb{U}_{\emph{QNN}}$. Then, for any $\delta > 0$ and $\gamma > 0$, with probability at least $1 - \delta$ over the random draw of an i.i.d.\ sample $S$ of size $m$, the following holds for all $h \in \mathcal{H}$:
\begin{equation}
    R(h) \leq \hat{R}_\gamma(h) + \tilde{O}\!\left(\frac{b}{\gamma}\sqrt{\frac{n}{m}\sum_{i=1}^k \|E_i\|_\sigma^2} + \sqrt{\frac{\ln(1/\delta)}{m}}\right),
    \label{eq:ch3-main-bound}
\end{equation}
where $\tilde{O}$ hides logarithmic factors in $n$, $b$, $E$, $\gamma$, and $m$.
\end{theorem}

\paragraph{Interpreting the quantities in the bound.}
The parameter $n$ is the number of qubits, so $N=2^n$ sets the Hilbert-space dimension entering the covering argument.
The number of classes is $k$, the sample size is $m$, and $\delta$ is the failure probability of the high-probability statement.
The margin threshold $\gamma$ controls the trade-off between empirical margin loss and complexity: larger $\gamma$ is more demanding on the training margins but reduces sensitivity to small confidence gaps.
The term $b=\|U-U_{\mathrm{ref}}\|_{2,1}$ is a radius of the reachable unitary family around a fixed reference unitary.
Finally, $\sum_i\|E_i\|_\sigma^2$ measures how the POVM can amplify perturbations of the quantum state into perturbations of class probabilities.

This bound has several important properties that distinguish it from the uniform bounds discussed in \Cref{sec:ch3-qml-generalization}:

\paragraph{Dependence on margin.}
Unlike parameter-count or norm-only uniform bounds, the margin bound contains an empirical margin term evaluated on the learned classifier. The first term $\hat{R}_\gamma(h)$ measures the fraction of training samples with margin at most $\gamma$, and the second term scales as $1/\gamma$. A model that achieves large margins on the training data will have a small $\hat{R}_\gamma(h)$ for a reasonably large $\gamma$, resulting in a tighter bound. In the randomized-label experiments below, memorizing models are reflected by smaller margins and looser margin bounds, which is precisely the behavior missing from parameter-count bounds.

\paragraph{Role of measurement operators.}
The bound depends on the spectral norms of the POVMs $\sum_i \|E_i\|_\sigma^2$, which quantifies the influence of the measurement choice on generalization. This is a uniquely quantum feature: the generalization performance depends not only on the circuit architecture but also on how information is extracted from the quantum state.

\paragraph{Corollary for projective measurements.}\mbox{}\\
When the POVMs are projective measurements---a common choice in QML models---the bound simplifies significantly:

\begin{corollary}
\label{cor:ch3-projective}
Under the conditions of \Cref{thm:ch3-margin-bound}, suppose the POVMs $\{E_i\}_{i=1}^k$ are projective measurements, i.e., $E_i^2 = E_i$ and $E_i E_j = 0$ for all $i \neq j$. Then:
\begin{equation}
    R(h) \leq \hat{R}_\gamma(h) + \tilde{O}\!\left(\frac{b}{\gamma}\sqrt{\frac{nk}{m}} + \sqrt{\frac{\ln(1/\delta)}{m}}\right).
    \label{eq:ch3-projective-bound}
\end{equation}
\end{corollary}

For projective measurements, $E_i^2 = E_i$ implies $\|E_i\|_\sigma = 1$, so $\sum_i \|E_i\|_\sigma^2 = k$. The bound is therefore independent of the specific measurement operators, depending only on the number of classes $k$.

\subsection{Experimental Validation}
\label{subsec:ch3-experiments}

We now validate our theoretical framework through extensive experiments on the quantum phase recognition (QPR) task---a quantum data classification problem of significant physical relevance.

\subsubsection{Quantum Phase Recognition}
\label{subsubsec:ch3-qpr}

Quantum phase recognition (QPR)~\cite{sachdev1999quantum, sachdev2023quantum} is a classification task aimed at identifying quantum phases of matter, a problem of fundamental importance in condensed matter physics~\cite{broecker2017machine, ebadi2021quantum, carrasquilla2017machine}. We consider the generalized cluster Hamiltonian:
\begin{equation}
    H(J_1,J_2)
    =
    \sum_{j=1}^{n}
    \left(
        Z_j
        - J_1 X_j X_{j+1}
        - J_2 X_{j-1} Z_j X_{j+1}
    \right),
    \label{eq:ch3-cluster-hamiltonian}
\end{equation}
where $X_j$ and $Z_j$ are Pauli operators acting on site $j$, and $J_1$, $J_2$ are tunable parameters controlling the interaction strengths. Depending on the values of $J_1$ and $J_2$, the ground state falls into one of four distinct phases: (1) ferromagnetic, (2) antiferromagnetic, (3) symmetry-protected topological (SPT), or (4) trivial. Determining the phase of a given ground state with unknown interaction parameters constitutes a four-class classification problem.

We use 8-qubit QCNNs~\cite{cong2019quantum, hur2022quantum} for this task, training on 20 data points evenly split across the four classes and evaluating test accuracy on 1,000 held-out samples. The small training set is deliberately chosen to explore the overfitting regime, following the methodology of \textcite{gilfuster2024understanding}. To test the models' ability to distinguish genuine patterns from noise, we introduce label randomization at varying levels: 0\% (pure labels), 50\% (half random), and 100\% (fully random). The QCNNs are trained using Adam with learning rate $0.001$ and full-batch updates, train for up to 5,000 iterations with early stopping based on convergence of interval-averaged losses, and average the margin-distribution plots over 15 repetitions with different training samples.

\subsubsection{Margin Distribution and Generalization Performance}

\begin{figure}[t]
    \centering
    \includegraphics[width=\textwidth]{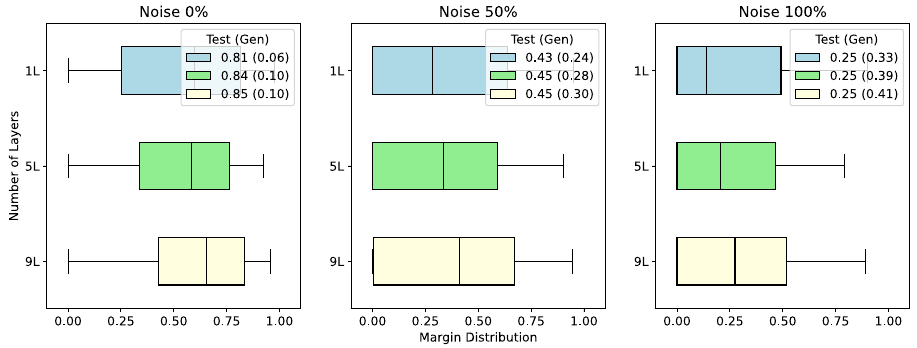}
    \caption{Tukey box-and-whisker plots of margin distributions for optimized 8-qubit QCNNs on the QPR task. Results are shown for QCNNs with one, five, and nine layers, under varying degrees of label randomization: 0\% (left), 50\% (middle), and 100\% (right). Each legend entry reports the test accuracy with the corresponding generalization gap in parentheses. As randomization increases, the margin distributions shift to the left and the generalization gap widens, consistent with poorer generalization under the margin-sensitive bound in \Cref{eq:ch3-main-bound}.}
    \label{fig:ch3-margin-distribution}
\end{figure}

\Cref{fig:ch3-margin-distribution} presents the margin distributions of optimized QCNNs using Tukey box-and-whisker plots, along with the corresponding test accuracies and generalization gaps, for models with one, five, and nine QCNN layers.

The results reveal a clear and consistent pattern across all configurations:
\begin{itemize}
    \item A right-skewed margin distribution (larger margins) is associated with higher test accuracy and a tighter generalization bound, as indicated by the smaller right-hand side of \Cref{eq:ch3-main-bound}.
    \item Increasing label randomization shifts the margin distributions to the left, reflecting that the model assigns lower confidence to randomly labeled data. This leftward shift is accompanied by a larger measured generalization gap (reported in the legend) and a larger generalization bound, correctly capturing the deteriorating generalization under label noise.
    \item When labels are not fully randomized, deeper QCNNs (more layers) tend to achieve higher test accuracy and rightward-shifted margin distributions, suggesting that increased expressibility helps the model find hypotheses closer to the optimal one.
\end{itemize}

These observations support the margin distribution as an informative indicator of generalization performance in the studied settings, even in scenarios where the model can overfit random labels---precisely the regime where uniform bounds fail.

\subsubsection{Predicting the Generalization Gap: Margins vs.\ Parameters}

A key question is whether margin-based metrics offer a practical advantage over parameter-based metrics for predicting the generalization gap. To address this, we compare three margin-based metrics (lower quartile, median, and mean of the margin distribution) against three parameter-based metrics (total parameter count, and effective parameters at thresholds $10^{-1}$ and $10^{-2}$).

\begin{figure}[t]
    \centering
    \includegraphics[width=\textwidth]{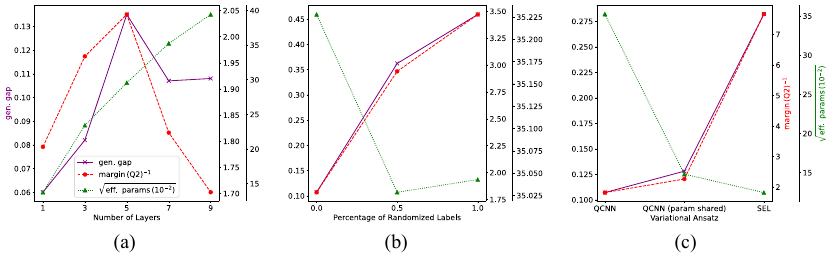}
    \caption{Comparison of the generalization gap, median margin (a margin-based metric), and effective parameters with threshold $10^{-2}$ (a parameter-based metric) as a function of: (a) the number of QCNN layers, (b) the percentage of randomized labels, and (c) the choice of variational ansatz. Since margins are inversely correlated with the generalization gap (\Cref{eq:ch3-main-bound}), the inverse median margin is plotted. The margin more closely tracks the generalization gap in these experiments, while effective parameters show inconsistent or opposite trends.}
    \label{fig:ch3-margin-vs-params}
\end{figure}

\Cref{fig:ch3-margin-vs-params} compares these metrics across three experimental axes: (a) varying the number of QCNN layers, (b) varying the percentage of randomized labels, and (c) varying the variational ansatz (QCNN, QCNN with shared parameters~\cite{cong2019quantum, hur2022quantum}, and Strongly Entangling Layers~\cite{bergholm2020pennylane}). The results show that:
\begin{itemize}
    \item The margin reliably captures variations in the generalization gap across all three hyperparameters. For example, the generalization gap peaks at five layers before slightly decreasing, and the inverse median margin accurately tracks this non-monotonic behavior.
    \item Effective parameters fail to capture the generalization gap in several settings and sometimes show the opposite trend. For instance, the effective parameters increase monotonically with the number of layers, failing to capture the peak in generalization gap at five layers.
\end{itemize}

For a more comprehensive comparison, we compute both the \emph{mutual information} and the \emph{Kendall rank correlation coefficient} between each metric and the generalization gap across all hyperparameter configurations simultaneously.

The mutual information $I(g; \mu)$ quantifies the reduction in uncertainty about the generalization gap $g$ given the metric $\mu$. The Kendall rank correlation coefficient $\tau$ measures the ordinal association between the generalization gap and the metric---specifically, whether the ranking of models by the metric agrees with their ranking by generalization performance:
\begin{equation}
    \tau_{G,M} = \frac{1}{n(n-1)} \sum_{i < j} \left[1 + \mathrm{sgn}(g_i - g_j)\,\mathrm{sgn}(\mu_i - \mu_j)\right],
    \label{eq:ch3-kendall-tau}
\end{equation}
where $G = [g_1, \ldots, g_n]$ and $M = [\mu_1, \ldots, \mu_n]$ are lists of generalization gaps and corresponding metrics. This coefficient ranges from $0$ to $1$, where $\tau = 1$ indicates perfect agreement between the two rankings (all pairs concordant), and $\tau = 0$ perfect disagreement (all pairs discordant).

\begin{figure}[t]
    \centering
    \includegraphics[width=0.95\textwidth]{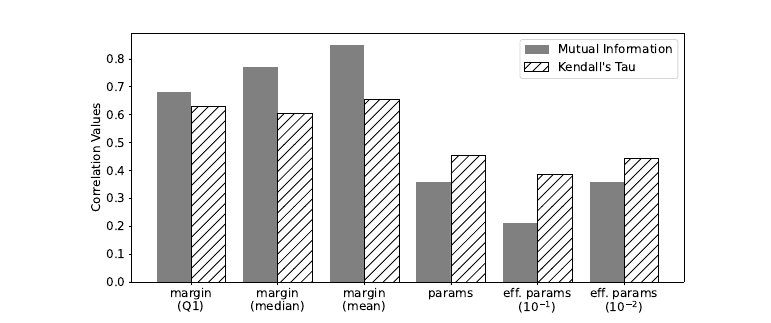}
    \caption{Mutual information (solid) and Kendall rank correlation coefficient (shaded) between the generalization gap and various metrics. The first three columns represent margin-based metrics (lower quartile, median, and mean of the margin distribution), while the last three represent parameter-based metrics (total parameters, effective parameters at thresholds $10^{-1}$ and $10^{-2}$). Margin-based metrics show stronger correlations with the generalization gap than parameter-based metrics in these experiments.}
    \label{fig:ch3-correlation}
\end{figure}

\Cref{fig:ch3-correlation} shows that margin-based metrics are more strongly correlated with the generalization gap than parameter-based metrics in both evaluation methods. This supports the use of margin-based quantities not only as theoretically motivated terms in \Cref{thm:ch3-margin-bound}, but also as practical diagnostics for evaluating the generalization performance of QML models.

\subsection{Connection to Quantum State Discrimination}
\label{subsec:ch3-quantum-embedding}

We now establish a connection between the margin and quantum state discrimination, bridging our generalization analysis with the quantum embedding theory developed in \Cref{chap:nqe}.

\paragraph{Margin mean and trace distance.}
For binary classification, the margin mean can be expressed as:
\begin{equation}
    \mu_{\mathrm{mean}} = \frac{1}{m}\sum_i 2\,\mathrm{Tr}(U\rho(x_i)U^\dagger E_{y_i}) - 1,
    \label{eq:ch3-margin-mean}
\end{equation}
where $U$ is the optimized unitary and $E_{y_i}$ is the POVM element corresponding to class $y_i$. Defining the effective POVM $E_{\pm 1}^* = U^\dagger E_{\pm 1} U$ and the class-averaged density matrices $\rho^\pm = \frac{1}{m^\pm} \sum_i \rho(x_i^\pm)$, the margin mean becomes:
\begin{equation}
    \mu_{\mathrm{mean}} = 2\,\mathrm{Tr}(p^+ \rho^+ E_{+1}^*) + 2\,\mathrm{Tr}(p^- \rho^- E_{-1}^*) - 1,
    \label{eq:ch3-margin-trace}
\end{equation}
where $p^\pm$ are the class priors. Using $E_{+1}^* + E_{-1}^* = I$, the success-probability term symmetrizes as:
\begin{equation}
    \mathrm{Tr}(p^+ \rho^+ E_{+1}^*) + \mathrm{Tr}(p^- \rho^- E_{-1}^*) = \frac{1}{2} + \frac{1}{2}\,\mathrm{Tr}\!\left[(p^+ \rho^+ - p^- \rho^-)(E_{+1}^* - E_{-1}^*)\right].
    \label{eq:ch3-margin-sym}
\end{equation}
Maximizing over binary POVMs $\{E_{\pm 1}^*\}$ recovers the \emph{Helstrom measurement} for discriminating $p^+ \rho^+$ from $p^- \rho^-$~\cite{helstrom1976quantum}, which yields:
\begin{equation}
    \mu_{\mathrm{mean}} \leq 2D_{\mathrm{tr}}(p^+ \rho^+, p^- \rho^-),
    \label{eq:ch3-margin-trace-bound}
\end{equation}
where $D_{\mathrm{tr}}$ is the trace distance, as in \Cref{chap:nqe}, and the inequality becomes an equality if and only if $\{E_{\pm 1}^*\}$ is the Helstrom measurement.

Thus, the trace distance serves as an upper bound on the achievable margin mean. Quantum embeddings that produce a large trace distance between class ensembles---such as those obtained through NQE (\Cref{chap:nqe})---raise the ceiling on the margin, which can lead to tighter bounds in \Cref{thm:ch3-margin-bound}. This provides a theoretical explanation for the empirically observed relationship between large trace distances and improved generalization~\cite{hur2024nqe, lloyd2020quantum, hubregtsen2022training}: these embeddings make larger classification margins attainable.

\paragraph{Experimental validation.}
We empirically examine this connection by comparing QCNNs with three different quantum embedding schemes on classical datasets (MNIST~\cite{lecun2010mnist}, Fashion-MNIST~\cite{xiao2017fashion}, and Kuzushiji-MNIST~\cite{clanuwat2018deep}): (1) a fixed ZZ feature map, (2) trainable quantum embedding (TQE)~\cite{lloyd2020quantum}, and (3) Neural Quantum Embedding (NQE)~\cite{hur2024nqe}. These three schemes produce progressively increasing initial trace distances.

\begin{figure}[t]
    \centering
    \includegraphics[width=\textwidth]{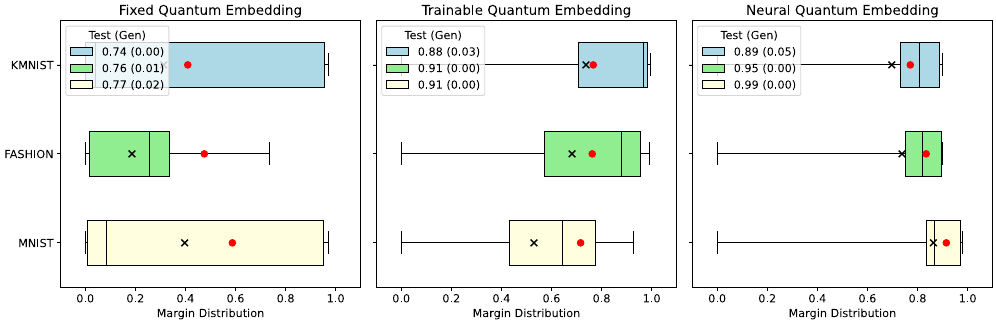}
    \caption{Tukey box-and-whisker plots of margin distributions for optimized 8-qubit QCNNs on binary classification tasks using three quantum embedding schemes: fixed ZZ feature map (left), trainable quantum embedding (middle), and Neural Quantum Embedding (right). Results are shown for MNIST (bottom), Fashion-MNIST (middle), and Kuzushiji-MNIST (top). Each legend entry reports the test accuracy with the corresponding generalization gap in parentheses. The black cross indicates the margin mean, and the red circle indicates the unweighted trace distance between class ensembles. Higher trace distances correspond to larger attainable margins and higher test accuracies---while the generalization gap remains small throughout---in these experiments, consistent with the connection established in \Cref{eq:ch3-margin-trace}.}
    \label{fig:ch3-embedding-comparison}
\end{figure}

\Cref{fig:ch3-embedding-comparison} supports the theoretical prediction: the trace distance (red dots) increases progressively from the fixed embedding to TQE to NQE, accompanied by a corresponding increase in the margin mean (black crosses) and rightward shift of the margin distribution. Larger margin means, coupled with right-skewed margin distributions, are associated with higher test accuracies while the generalization gap stays small across all three embeddings, aligning with the margin-based generalization framework.

This result connects the training-loss and generalization perspectives: NQE (\Cref{chap:nqe}) reduces the achievable training loss by increasing the trace distance, and the same increased trace distance can support larger classification margins. The margin-based perspective thus links state distinguishability and generalization through the quantum embedding.

\section{Summary}
\label{sec:ch3-summary}

In this chapter, we developed a margin-based framework for understanding generalization in quantum machine learning. Our main contributions are:

\begin{enumerate}
    \item \textbf{Margin generalization bound for QNNs} (\Cref{thm:ch3-margin-bound}): We established a generalization bound for multiclass classification with QNNs that depends on the margin distribution. Unlike parameter-count bounds that are often too vacuous for the empirical memorization comparisons studied here, the margin bound yields tighter estimates when the model classifies training data with large margins.

    \item \textbf{Empirical comparison of margin-based metrics}: Through experiments on the quantum phase recognition task, we found that margin-based metrics (lower quartile, median, and mean of the margin distribution) are stronger predictors of generalization performance than parameter-based metrics in the tested settings, as measured by both mutual information and the Kendall rank correlation coefficient.

    \item \textbf{Connection to quantum state discrimination}: We showed that the margin mean is upper bounded by the trace distance between class ensembles (\Cref{eq:ch3-margin-trace-bound}). This establishes a theoretical link between the generalization framework of this chapter and the trace-distance framework of \Cref{chap:nqe}: embeddings with large trace distances make larger margins attainable.
\end{enumerate}

In Part~II, we reverse the direction and explore how artificial intelligence techniques can address fundamental challenges in quantum computing, beginning with neural decoders for quantum error correction in \Cref{chap:decoder}.


\part{AI for Quantum}


\chapter{Neural Decoders for Quantum Error Correction}
\label{chap:decoder}


\chaptersource{The results in this chapter are based on \textit{Scalable Neural Decoders for Practical Real-Time Quantum Error Correction}~\cite{lee2025mamba}. The numerical data underlying the real-time decoding results for the Transformer and Mamba decoders are publicly available in the GitHub repository at \url{https://github.com/qDNA-yonsei/NeuralDecoder_v1}.}
\bigskip

In Part~I of this thesis, we explored how quantum computing can enhance machine learning---from neural quantum embeddings (\Cref{chap:nqe}) to generalization theory for quantum neural networks (\Cref{chap:generalization}).
In Part~II, we reverse this direction and ask: \emph{how can artificial intelligence help solve quantum problems?}
The two Part~II chapters examine this question at complementary levels: \Cref{chap:decoder} studies real-time quantum error correction, while \Cref{chap:nqs} studies neural representations and optimizers for quantum many-body ground states.

A central bottleneck on the path to fault-tolerant quantum computation is \emph{quantum error correction decoding}---the classical inference task of identifying and correcting errors from noisy syndrome measurements.
As quantum processors scale to hundreds and eventually thousands of physical qubits, decoders must provide throughput comparable to the syndrome-generation rate while maintaining high accuracy, so that the classical backlog remains bounded~\cite{terhal2015qec}.
This chapter presents a neural decoder based on the Mamba architecture~\cite{gu2023mamba}, a state-space model that replaces the attention mechanism of Transformer-based decoders with a selective scan of $\mathcal{O}(d^2)$ complexity, where $d$ is the code distance.
The experiments show that this architectural choice preserves comparable decoding accuracy relative to the reproduced Transformer baseline in the tested settings while improving the asymptotic scaling relative to the $\mathcal{O}(d^4)$ attention mechanism used by Transformer decoders such as AlphaQubit~\cite{bausch2024alphaqubit}. Under the simulated latency model introduced in this chapter, this improved scaling yields a higher finite-size effective threshold.

The chapter is organized as follows.
\Cref{sec:ch4-qec} introduces the basics of quantum error correction, focusing on surface codes and the computational challenges of decoding.
\Cref{sec:ch4-neural-decoders} surveys neural network approaches to decoding, with particular attention to AlphaQubit and its limitations.
\Cref{sec:ch4-mamba} presents our Mamba-based decoder---its architecture, training procedure, and experimental results on both hardware data and simulated real-time scenarios.
\Cref{sec:ch4-summary} summarizes the chapter.

\section{Introduction to Quantum Error Correction}
\label{sec:ch4-qec}

Quantum information is inherently fragile: interactions with the environment, imperfect gate operations, and faulty measurements introduce errors that accumulate rapidly during a computation.
Unlike classical bits, which can be copied and checked, quantum states cannot be cloned~\cite{nielsen2010quantum}, making direct error detection impossible without carefully designed encoding schemes.

Quantum error correction (QEC) overcomes this obstacle by encoding a \emph{logical} qubit redundantly across many \emph{physical} qubits, so that errors can be detected and corrected without destroying the encoded quantum information~\cite{shor1995scheme,calderbank1996good,steane1996error,gottesman1997stabilizer}.
The field has matured from early theoretical proposals to experimental demonstrations of error suppression in regimes relevant to fault tolerance~\cite{googlequantum2023suppressing}, yet a critical classical bottleneck remains: the \emph{decoder} that interprets syndrome measurements and prescribes corrections must keep pace with syndrome production---a challenge that grows with code size.

\subsection{Surface Codes}
\label{subsec:ch4-surface-codes}

Among the many families of quantum error-correcting codes, the \emph{surface code}~\cite{kitaev2003fault,dennis2002topological,fowler2012surface} stands out as the leading candidate for near-term fault-tolerant quantum computing.
Its appeal rests on three properties: (i) it requires only nearest-neighbor interactions on a two-dimensional lattice; (ii) under standard circuit-noise assumptions it has one of the highest known fault-tolerance thresholds, approximately $1\%$ per physical gate~\cite{dennis2002topological,fowler2012surface}; and (iii) syndrome extraction requires only local measurements.

\paragraph{Stabilizer formalism.}
The surface code is defined within the \emph{stabilizer formalism}~\cite{gottesman1997stabilizer}.
An $\llbracket n, k, d \rrbracket$ stabilizer code encodes $k$ logical qubits into $n$ physical qubits and can detect any error acting on fewer than $d$ qubits, where $d$ is the \emph{code distance}.
The code is specified by an Abelian \emph{stabilizer group} $\mathcal{S} = \langle g_1, g_2, \ldots, g_{n-k} \rangle \subset \mathcal{P}_n$, where $\mathcal{P}_n$ is the $n$-qubit Pauli group.
The \emph{codespace} is the simultaneous $+1$ eigenspace of all stabilizer generators:
\begin{equation}
    \mathcal{C} = \bigl\{ \ket{\psi} \in (\mathbb{C}^2)^{\otimes n} : g_i \ket{\psi} = \ket{\psi} \;\; \forall\, i \bigr\}.
    \label{eq:ch4-codespace}
\end{equation}
An error $E \in \mathcal{P}_n$ is detectable if it anticommutes with at least one stabilizer generator, producing a $-1$ measurement outcome that flags the error without revealing the encoded information.

\paragraph{Rotated surface code.}
Throughout this chapter we focus on the \emph{rotated surface code}~\cite{tomita2014low}, the variant used in the experimental benchmarks below and in current superconducting-qubit demonstrations.
A distance-$d$ rotated patch encodes a single logical qubit using $2d^2-1$ physical qubits: $d^2$ data qubits and $d^2-1$ measurement (ancilla) qubits that mediate stabilizer readout.
Its stabilizers come in two types, $X$-type and $Z$-type.
\emph{Logical operators} are Pauli strings that commute with all stabilizers but are not themselves in $\mathcal{S}$.
On the surface code, logical $\bar{X}$ ($\bar{Z}$) is a chain of $X$ ($Z$) operators spanning the lattice from one boundary to the opposite boundary.

\paragraph{Syndrome measurement.}
In a QEC cycle, every stabilizer generator is measured once, yielding a binary outcome $s_i \in \{0,1\}$ for each generator.
The collection $\mathbf{s} = (s_1, s_2, \ldots, s_{n-k})$ is called the \emph{syndrome}.
In an ideal setting, $s_i = 0$ for all $i$ indicates no detectable error.

In practice, syndrome measurements are themselves noisy---ancilla qubits and measurement gates are imperfect.
To reliably extract the syndrome, the measurement is repeated $d$ times (matching the code distance), creating a three-dimensional \emph{spacetime} volume of syndrome data.
A \emph{detection event} is defined as a change in a stabilizer outcome between consecutive rounds:
\begin{equation}
    \delta_{i,t} = s_{i,t} \oplus s_{i,t-1},
    \label{eq:ch4-detection-event}
\end{equation}
where $s_{i,t}$ is the outcome of stabilizer $i$ at round $t$ and $\oplus$ denotes addition modulo 2.
Detection events, rather than raw syndromes, form the input to modern decoders because they are invariant to the initial syndrome configuration and directly encode the locations of faults in spacetime.

\begin{figure}[t]
    \centering
    \includegraphics[width=\textwidth]{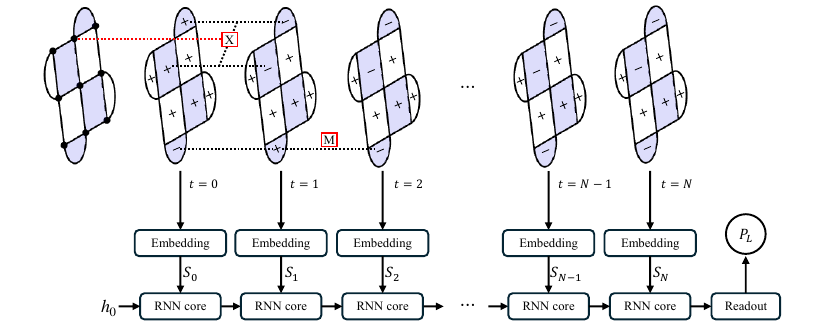}
    \caption{Spacetime view of surface-code syndrome extraction for recurrent neural decoding. Each quantum error-correction (QEC) cycle produces stabilizer measurements and detection events, which are embedded as spatial features, processed recurrently over time, and decoded into a logical-error probability.}
    \label{fig:ch4-surface-code}
\end{figure}

\subsection{The Decoding Problem}
\label{subsec:ch4-decoding-problem}

Given a syndrome history $\mathbf{s} = \{s_{i,t}\}$, the task of the \emph{decoder} is to infer a correction operator $\mathcal{R}$ that returns the system to the codespace.
Because many different physical errors can produce the same syndrome (they differ by elements of $\mathcal{S}$), the decoder only needs to identify the correct \emph{equivalence class} of the error, not the exact physical error.

\paragraph{Maximum likelihood decoding.}
The optimal decoding strategy is \emph{maximum likelihood} (ML) decoding, which selects the equivalence class with the highest total probability:
\begin{equation}
    \hat{c} = \argmax_{c \in \{0,1\}} \sum_{E \in c} \Pr(E \mid \mathbf{s}),
    \label{eq:ch4-ml-decoding}
\end{equation}
where $c$ labels the equivalence class (trivial or logical error).
ML decoding is \#P-hard in general~\cite{bravyi2014efficient}, motivating the development of efficient approximate decoders.

\paragraph{Decoder taxonomy.}
A variety of decoding algorithms have been developed, broadly categorized as follows:
\begin{itemize}
    \item \textbf{Matching-based decoders.} Minimum-weight perfect matching (MWPM)~\cite{higgott2022pymatching,higgott2023improved} maps the syndrome to a graph problem and finds the minimum-weight set of edges connecting detection events. MWPM runs in $\mathcal{O}(n^3)$ worst-case time and has been the workhorse decoder for surface codes. Standard matching formulations are most direct for independent graphlike error models. Extensions such as correlated MWPM~\cite{fowler2013optimal} and belief matching~\cite{higgott2023improved} incorporate correlations between $X$ and $Z$ errors at the cost of increased complexity.
    \item \textbf{Tensor network decoders.} By contracting a tensor network representation of the error model, these decoders can approximate ML decoding with tunable accuracy~\cite{ferris2014tensor,bravyi2014efficient}. However, the bond dimension required for accuracy grows with code distance, limiting scalability.
    \item \textbf{Neural decoders.} Machine learning models trained on syndrome--error pairs can learn complex noise correlations directly from data~\cite{torlai2017neural,krastanov2017deep,bausch2024alphaqubit}. Their accuracy can match or exceed traditional decoders, but inference latency is a concern for real-time operation (discussed in \Cref{sec:ch4-neural-decoders}).
\end{itemize}

\paragraph{Real-time constraints.}
In a fault-tolerant quantum computer, error correction operates continuously as each QEC cycle produces new syndrome information.
For memory experiments and Clifford operations, Pauli-frame updates can often be deferred, so the decoder does not always need to finish strictly before the next cycle.
Nevertheless, a scalable processor still requires decoding throughput comparable to the syndrome-generation rate so that the backlog remains bounded~\cite{terhal2015qec}.
Latency becomes especially consequential when classical feedforward is needed, and delayed corrections can be modeled as an additional effective noise source.
This latency constraint makes decoding speed an essential companion to accuracy for practical QEC.

\paragraph{Data-driven decoding.}
A recent paradigm for decoder design is the \emph{data-driven} approach~\cite{bausch2024alphaqubit,lange2025data}: a neural decoder is pretrained on large synthetic datasets generated from a calibrated noise model, then fine-tuned on experimental hardware data.
This two-stage pipeline allows the decoder to learn the specific noise characteristics of the target device---including cross-talk, leakage, and measurement-induced errors---without requiring an explicit analytical noise model.

\paragraph{Performance metrics.}
The quality of a decoder is measured by the \emph{logical error rate} (LER), denoted $\varepsilon$, which is the probability of a logical error per QEC cycle.
The LER is estimated by fitting the measured logical fidelity $F(n)$ after $n$ cycles to the model
\begin{equation}
    \log F(n) = \log F_0 + n \log(1 - 2\varepsilon),
    \label{eq:ch4-ler-fit}
\end{equation}
where $F_0$ accounts for state preparation and measurement errors~\cite{bausch2024alphaqubit}.
A good code--decoder pair is usually characterized by an \emph{error threshold} $p_\text{th}$: the maximum physical error rate below which increasing the code distance $d$ exponentially suppresses the LER in an asymptotic setting.
The \emph{error suppression ratio} $\Lambda = \varepsilon(d) / \varepsilon(d+2)$ quantifies how much benefit each increase in code distance provides.

\section{Neural Decoders}
\label{sec:ch4-neural-decoders}

The idea of using neural networks for QEC decoding dates back to 2017, when the earliest decoders ranged from generative Boltzmann machines~\cite{torlai2017neural} to feedforward networks for small codes~\cite{varsamopoulos2017decoding}.
Since then, the field has grown rapidly, with architectures ranging from simple multilayer perceptrons to recurrent transformers.
For a broader review of machine learning approaches to decoding topological quantum codes, see \textcite{lee2026machine}.
This section surveys the neural decoder literature (\Cref{subsec:ch4-neural-overview}) and then discusses AlphaQubit (\Cref{subsec:ch4-alphaqubit}), a high-performing Transformer-based decoder that motivates our Mamba-based approach.

\subsection{Overview of Neural Decoder Literature}
\label{subsec:ch4-neural-overview}

\paragraph{Early feedforward approaches.}
\textcite{torlai2017neural} introduced one of the first neural decoders, using a restricted Boltzmann machine to decode the toric code, demonstrating that unsupervised generative models could learn the error distribution from syndrome data.
Concurrently, \textcite{krastanov2017deep} proposed a deep neural network probabilistic decoder for stabilizer codes that could handle depolarizing noise.
\textcite{varsamopoulos2017decoding} systematically studied feedforward networks for small surface codes, followed by comparisons across architectures~\cite{varsamopoulos2019comparing} and distributed decoder designs~\cite{varsamopoulos2020decoding}.

\paragraph{Recurrent and specialized architectures.}
\textcite{baireuther2018machine} applied recurrent neural networks to decode correlated qubit errors in a topological code, showing that temporal correlations across QEC cycles could be exploited.
This was extended to color codes with circuit-level noise~\cite{baireuther2019neural}.
\textcite{chamberland2018deep} designed deep neural decoders specifically for near-term fault-tolerant experiments, while \textcite{maskara2019advantages} demonstrated the versatility of neural-network decoding for various topological codes.
\textcite{ni2020neural} scaled neural decoders to large-distance 2D toric codes, and \textcite{liu2019neural} introduced neural belief-propagation decoders that combined neural networks with message-passing algorithms.

\paragraph{Modern approaches.}
More recently, \textcite{gicev2023scalable} developed a scalable artificial neural network syndrome decoder optimized for speed.
\textcite{lange2025data} applied graph neural networks (GNNs) to QEC decoding, exploiting the natural graph structure of stabilizer codes.
\textcite{sweke2020reinforcement} formulated decoding as a reinforcement learning problem, and \textcite{cao2023qecgpt} used generative pre-trained transformers for decoding.
\textcite{egorov2023equivariant} proposed equivariant neural decoders that respect the symmetries of the code, improving sample efficiency.
Further work has explored hardware cost-performance tradeoffs~\cite{overwater2022neural}, symmetry-aware architectures~\cite{wagner2020symmetries}, and deep reinforcement-learning decoders~\cite{fitzek2020deep}.

Despite this progress, a persistent tension remains between \emph{accuracy} and \emph{latency}.
Highly accurate neural decoders tend to use large models with expensive inference, while lightweight decoders sacrifice accuracy for speed.
The breakthrough of AlphaQubit showed that neural decoders can be highly competitive with strong traditional decoders on important benchmarks, but at a computational cost that challenges real-time operation.

\subsection{AlphaQubit}
\label{subsec:ch4-alphaqubit}

AlphaQubit~\cite{bausch2024alphaqubit} is a recurrent Transformer-based neural decoder developed by Google DeepMind and Google Quantum AI.
It represents a landmark in neural decoding: it demonstrated that a data-driven neural decoder can achieve highly competitive performance on real hardware data from Google's Sycamore processor.

\paragraph{Architecture.}
AlphaQubit follows a recurrent encoder architecture with three components (cf.\ \Cref{fig:ch4-decoder-architecture}):
\begin{enumerate}
    \item \textbf{Stabilizer Embedder:} At each QEC cycle $n$, raw stabilizer measurements, detection events, and optionally analog I/Q readout values and leakage flags are embedded into $d_\text{model}$-dimensional feature vectors $\mathbf{s}_{n,i}$ for each stabilizer $i$, via linear projections combined with positional encodings and processed through a small ResNet.
    \item \textbf{RNN Core:} The sequence of stabilizer embeddings $S_n = \{\mathbf{s}_{n,i}\}_{i=1}^{|\mathcal{S}|}$ is processed recurrently across QEC cycles. At each cycle, the core updates a hidden state:
    \begin{equation}
        h_{n+1} = f_\text{core}(h_n, S_n),
        \label{eq:ch4-rnn-update}
    \end{equation}
    where $f_\text{core}$ consists of multiple \emph{Syndrome Mixer} layers. In AlphaQubit, each Syndrome Mixer uses \emph{multi-head attention} (MHA)~\cite{vaswani2017attention} as the core mixing operation, followed by gated dense layers and dilated 2D convolutions.
    \item \textbf{Readout Network:} After the final cycle, the hidden state $h_N$ is scattered onto a 2D grid, converted to the data qubit lattice via convolutions, and passed through a deep ResNet to produce the logical error probability $P_L$.
\end{enumerate}

The multi-head attention in the Syndrome Mixer computes:
\begin{equation}
    \text{Attention}(Q, K, V) = \text{softmax}\!\left(\frac{QK^\top}{\sqrt{d_k}}\right) V,
    \label{eq:ch4-attention}
\end{equation}
where $Q$, $K$, $V \in \mathbb{R}^{n_s \times d_k}$ are the query, key, and value matrices derived from the $n_s = |\mathcal{S}|$ stabilizer embeddings, and $d_k$ is the head dimension.
This allows the model to learn arbitrary pairwise correlations between stabilizers.

\paragraph{Training.}
AlphaQubit employs a two-stage data-driven training pipeline. In \emph{pretraining}, the model is trained on large synthetic syndrome datasets generated from a detector error model (DEM) calibrated using cross-entropy benchmarking (XEB) data from the Sycamore processor. In \emph{fine-tuning}, the pretrained model is then adapted on real experimental data from Sycamore memory experiments, allowing it to learn hardware-specific noise patterns including cross-talk, leakage, and correlated errors.

\paragraph{Results on Sycamore data.}
On distance-3 and distance-5 surface codes, AlphaQubit achieved logical error rates competitive with the strongest tested decoders: $\varepsilon = 2.901 \times 10^{-2}$ at distance 3 and $\varepsilon = 2.748 \times 10^{-2}$ at distance 5, corresponding to an error suppression ratio of $\Lambda = 1.056$.
The full AlphaQubit result outperformed both matching-based and tensor-network decoders on the Sycamore data.

\paragraph{The latency bottleneck.}
Despite its accuracy, AlphaQubit has a fundamental scalability limitation rooted in its attention mechanism.
The self-attention operation in \Cref{eq:ch4-attention} computes pairwise interactions between all $n_s$ stabilizers, with complexity $\mathcal{O}(n_s^2)$ per layer.
For a distance-$d$ surface code, the number of stabilizers scales as $n_s \propto d^2$, giving an overall per-cycle complexity of $\mathcal{O}(d^4)$.

AlphaQubit's latency is already approximately $40\,\mu$s at distance 9---an order of magnitude slower than the $\sim 1\,\mu$s QEC cycle time of superconducting qubits.
This latency bottleneck motivates the central contribution of this chapter: replacing the attention mechanism with a Mamba-based selective state-space model that achieves $\mathcal{O}(d^2)$ complexity---a quadratic improvement---while preserving comparable accuracy.

\section{Mamba Decoder}
\label{sec:ch4-mamba}

We now present the Mamba decoder~\cite{lee2025mamba}, a neural decoder that replaces the multi-head attention in AlphaQubit's Syndrome Mixer with a Mamba module---a selective state-space model with linear complexity.
We first introduce state-space models and the Mamba architecture (\Cref{subsec:ch4-ssm}), then describe the decoder architecture and training (\Cref{subsec:ch4-mamba-architecture}), and finally present experimental results (\Cref{subsec:ch4-results}).

\subsection{Mamba and State Space Models}
\label{subsec:ch4-ssm}

State-space models (SSMs) are a class of sequence models inspired by continuous-time dynamical systems.
They have recently emerged as a compelling alternative to Transformers for modeling long sequences, offering linear-time complexity while maintaining competitive performance across language, audio, and genomics tasks~\cite{gu2021efficiently,gu2023mamba}.

\paragraph{Continuous-time SSM.}
A linear time-invariant (LTI) state-space model maps an input signal $x(t) \in \mathbb{R}$ to an output $y(t) \in \mathbb{R}$ through a latent state $h(t) \in \mathbb{R}^N$:
\begin{align}
    h'(t) &= A\, h(t) + B\, x(t), \label{eq:ch4-ssm-continuous-h} \\
    y(t)  &= C\, h(t), \label{eq:ch4-ssm-continuous-y}
\end{align}
where $A \in \mathbb{R}^{N \times N}$ is the state transition matrix, $B \in \mathbb{R}^{N \times 1}$ is the input projection, and $C \in \mathbb{R}^{1 \times N}$ is the output projection.

\paragraph{Discretization.}
To process discrete sequences, the continuous system is discretized using a step size $\Delta > 0$.
Under the zero-order hold (ZOH) assumption:
\begin{align}
    \bar{A} &= \exp(\Delta A), \label{eq:ch4-zoh-a} \\
    \bar{B} &= (\Delta A)^{-1}(\exp(\Delta A) - I) \cdot \Delta B, \label{eq:ch4-zoh-b}
\end{align}
yielding the discrete recurrence:
\begin{align}
    h_k &= \bar{A}\, h_{k-1} + \bar{B}\, x_k, \label{eq:ch4-ssm-discrete-h} \\
    y_k &= C\, h_k. \label{eq:ch4-ssm-discrete-y}
\end{align}
This recurrence can be computed in $\mathcal{O}(L)$ time for a sequence of length $L$.
Alternatively, unrolling the recurrence yields a convolution $y = \bar{K} * x$ with kernel $\bar{K} = (C\bar{B}, C\bar{A}\bar{B}, C\bar{A}^2\bar{B}, \ldots)$, which can be computed in $\mathcal{O}(L \log L)$ via the Fast Fourier Transform.
This \emph{dual view}---recurrence for inference, convolution for training---is a key advantage of SSMs.

\paragraph{S4: Structured State Spaces.}
The S4 model~\cite{gu2021efficiently} introduced structured initialization of the state matrix $A$ using the HiPPO (High-order Polynomial Projection Operator) framework, enabling SSMs to capture long-range dependencies that challenge RNNs and even Transformers.
S4 demonstrated state-of-the-art performance on the Long Range Arena benchmark, but its LTI nature means that the same dynamics are applied regardless of input content---the model cannot selectively attend to or ignore parts of the sequence.

\paragraph{Mamba: Selective State Spaces.}
Mamba~\cite{gu2023mamba} addresses this limitation by making the SSM parameters \emph{input-dependent}.
Specifically, the matrices $B$, $C$, and the step size $\Delta$ are computed as functions of the input:
\begin{align}
    B_k &= \text{Linear}_B(x_k), \qquad
    C_k = \text{Linear}_C(x_k), \qquad
    \Delta_k = \text{softplus}(\text{Linear}_\Delta(x_k)).
    \label{eq:ch4-mamba-selective}
\end{align}
This \emph{selective} mechanism allows the model to perform content-aware filtering: it can dynamically decide which information to propagate through the hidden state and which to discard, analogous to the gating mechanisms in LSTMs~\cite{hochreiter1997long} but within the SSM framework.

The input-dependent parameters break the LTI structure, precluding the use of the convolution mode during training.
Mamba compensates with a \emph{hardware-aware parallel scan}.
The key observation is that the linear recurrence $h_k = \bar{A}_k h_{k-1} + \bar{B}_k x_k$ is \emph{associative}: each step can be written as a pair $(\bar{A}_k,\, \bar{B}_k x_k)$ composed under the rule $(A_2, b_2) \circ (A_1, b_1) = (A_2 A_1,\, A_2 b_1 + b_2)$, so the entire sequence can be evaluated with a parallel prefix scan~\cite{blelloch1990prefix} in $\mathcal{O}(\log L)$ parallel depth rather than $L$ sequential steps.
This recovers parallelism across the time dimension even though the parameters are now time-varying.
The implementation is \emph{hardware-aware}: because the selective SSM expands each input channel to an $N$-dimensional state, na\"ively storing all intermediate states in the GPU's high-bandwidth memory (HBM) would make the operation memory-bandwidth bound.
Mamba instead fuses discretization, the scan, and the output projection into a single kernel that keeps the expanded states in fast on-chip SRAM and writes only the final outputs back to HBM---similar in spirit to FlashAttention~\cite{dao2022flashattention}---yielding near-linear scaling in sequence length in practice.

\paragraph{Mamba block.}
The Mamba block processes its input through two parallel paths:
\begin{enumerate}
    \item \textbf{SSM path:} The input is linearly projected to an expanded dimension, passed through a 1D convolution for local feature extraction, activated by SiLU (Sigmoid Linear Unit), and then processed by the selective SSM.
    \item \textbf{Gating path:} A parallel linear projection with SiLU activation provides a multiplicative gate.
\end{enumerate}
The two paths are combined via element-wise multiplication and projected back to the original dimension:
\begin{equation}
    \text{MambaBlock}(x) = \text{Linear}_\text{out}\!\bigl(\text{SSM}(\text{Conv1D}(\text{Linear}_1(x))) \odot \text{SiLU}(\text{Linear}_2(x))\bigr).
    \label{eq:ch4-mamba-block}
\end{equation}

\subsection{Architecture and Training}
\label{subsec:ch4-mamba-architecture}

\begin{figure}[t]
    \centering
    \includegraphics[width=\textwidth]{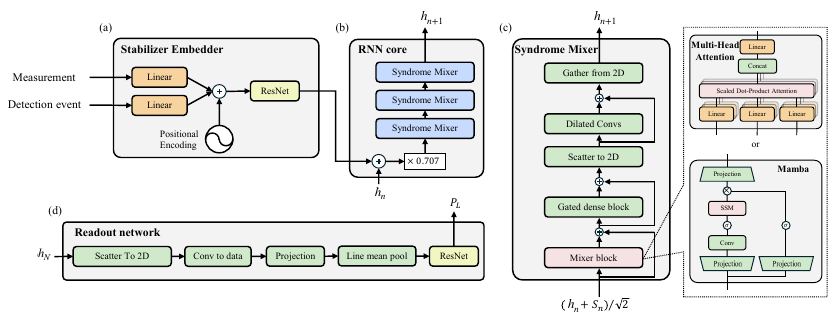}
    \caption{Architecture of the Mamba decoder. (a) The Stabilizer Embedder converts raw measurements and detection events into $d_\text{model}$-dimensional embeddings via linear projections, positional encodings, and a ResNet. (b) The RNN Core processes each QEC cycle through $L=3$ Syndrome Mixer layers with scaled skip connections. (c) Each Syndrome Mixer contains a Mamba-based mixer block, gated dense block, and dilated 2D convolutions. (d) The Readout Network maps the final hidden state to the logical-error probability $P_L$.}
    \label{fig:ch4-decoder-architecture}
\end{figure}

The Mamba decoder shares the same overall recurrent structure as AlphaQubit (\Cref{fig:ch4-decoder-architecture})---a Stabilizer Embedder, an RNN Core with Syndrome Mixer layers, and a Readout Network (described in \Cref{subsec:ch4-alphaqubit}).
The key difference is the replacement of multi-head attention with a Mamba module in each Syndrome Mixer.

\paragraph{Hyperparameters.}
\Cref{tab:ch4-hyperparameters} lists the hyperparameters for both the Transformer and Mamba decoder variants across the two experimental settings (Sycamore memory experiments and real-time decoding simulations).

\begin{table}[t]
    \centering
    \caption{Decoder hyperparameters. Top: model-specific parameters for Transformer and Mamba variants in Sycamore memory experiments (Syc.) and real-time decoding simulations (RT). Bottom: shared training and architecture parameters.}
    \label{tab:ch4-hyperparameters}
    \begin{tabular}{lcc|lcc}
        \toprule
        \multicolumn{3}{c|}{\textbf{Transformer-specific}} & \multicolumn{3}{c}{\textbf{Mamba-specific}} \\
        \textbf{Param.} & \textbf{Syc.} & \textbf{RT} & \textbf{Param.} & \textbf{Syc.} & \textbf{RT} \\
        \midrule
        $d_\text{model}$ & 320 & 256 & $d_\text{model}$ & 320 & 256 \\
        $H$ & 4 & 4 & $d_\text{state}$ & 16 & 16 \\
        $d_b$ & 48 & 48 & $d_\text{conv}$ & 4 & 4 \\
        $d_\text{attn}$ & 32 & 32 & $w_\text{mamba}$ & 1 & 1 \\
        $d_\text{mid}$ & 32 & 32 & & & \\
        \midrule
        \multicolumn{6}{c}{\textbf{Shared Hyperparameters}} \\
        \midrule
        \textbf{Param.} & \multicolumn{2}{c|}{\textbf{Sycamore}} & \textbf{Param.} & \multicolumn{2}{c}{\textbf{Real-time}} \\
        \midrule
        $L_\text{stab}$ & \multicolumn{2}{c|}{2} & $L_\text{stab}$ & \multicolumn{2}{c}{2} \\
        $L_\text{res}$ & \multicolumn{2}{c|}{16} & $L_\text{res}$ & \multicolumn{2}{c}{16} \\
        $d_\text{read}$ & \multicolumn{2}{c|}{64} & $d_\text{read}$ & \multicolumn{2}{c}{48} \\
        $w_\text{gate}$ & \multicolumn{2}{c|}{5} & $w_\text{gate}$ & \multicolumn{2}{c}{5} \\
        $D_\text{conv}$ ($d\!=\!3$) & \multicolumn{2}{c|}{$[1,1,1]$} & $D_\text{conv}$ ($d\!=\!3$) & \multicolumn{2}{c}{$[1,1,1]$} \\
        $D_\text{conv}$ ($d\!=\!5$) & \multicolumn{2}{c|}{$[1,1,2]$} & $D_\text{conv}$ ($d\!=\!5$) & \multicolumn{2}{c}{$[1,1,2]$} \\
         & \multicolumn{2}{c|}{} & $D_\text{conv}$ ($d\!=\!7$) & \multicolumn{2}{c}{$[1,2,4]$} \\
        \bottomrule
    \end{tabular}
\end{table}

\paragraph{Training procedure.}
The training procedure differs between the two experimental settings:

\subparagraph{Sycamore memory experiments.}
Following AlphaQubit, we employ a two-stage training pipeline:
\begin{enumerate}
    \item \textbf{Pretraining:} We generate up to 100 million synthetic syndrome samples from a detector error model (DEM) whose parameters are derived from a Pauli noise model calibrated using cross-entropy benchmarking (XEB) data from the Sycamore processor. The pretraining dataset includes QEC sequences of varying cycle counts $r \in \{1, 3, \ldots, 25\}$, covering both $X$- and $Z$-basis memory experiments. We employ a \emph{curriculum learning} strategy: the model initially trains on shorter sequences $r \in \{1, 3, 5, 7, 9\}$, and every 150,000 iterations the training set is expanded to include four additional cycle counts, eventually covering the full range up to $r = 25$.

    The Lion optimizer~\cite{chen2023lion} is used with an initial learning rate of $5 \times 10^{-6}$ and weight decay of $1 \times 10^{-5}$. The learning rate follows a cosine annealing schedule. We apply gradient clipping with a maximum norm of 1 and maintain an exponential moving average (EMA) of model weights with a decay rate of 0.9999.

    \item \textbf{Fine-tuning:} The pretrained decoder is adapted on the actual Sycamore experimental dataset (50\% train / 50\% evaluation split) for 10 epochs with a reduced learning rate of $2 \times 10^{-6}$ and increased weight decay of $7 \times 10^{-5}$.
\end{enumerate}

\subparagraph{Real-time decoding experiments.}
For the simulated real-time setting, we train decoders from scratch on synthetic data generated on-the-fly using the Superconducting-Inspired Circuit Depolarizing Noise (SI1000) model~\cite{gidney2021fault} via the Stim circuit simulator~\cite{gidney2021stim}.
Separate models are trained for code distances $d \in \{3, 5, 7\}$ at a base physical error rate of $p = 0.002$.
Training runs for 500,000 iterations with a batch size of 256, using the Lion optimizer with cosine annealing.

For finite-size effective threshold analysis under the specified latency model, we fine-tune the baseline models (trained at $p = 0.002$) at higher physical error rates $p \in \{0.006,$ $0.008,$ $0.010,$ $0.012\}$ for 250,000 iterations each, using a transfer learning approach that significantly reduces the computational cost of training models at multiple noise levels.

\subsection{Experimental Results}
\label{subsec:ch4-results}

We evaluate the Mamba decoder across two experimental settings: memory experiments on real hardware data (\Cref{subsubsec:ch4-memory}) and simulated real-time decoding with decoder-induced noise (\Cref{subsubsec:ch4-realtime}).

\subsubsection{Memory Experiments on Sycamore Data}
\label{subsubsec:ch4-memory}

\begin{figure}[t]
    \centering
    \begin{subfigure}[b]{0.66\textwidth}
        \centering
        \includegraphics[width=\textwidth]{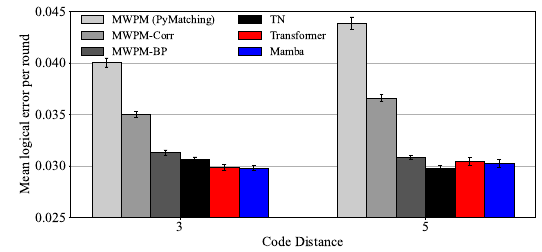}
        \caption{}
        \label{fig:ch4-memory-results}
    \end{subfigure}
    \hfill
    \begin{subfigure}[b]{0.29\textwidth}
        \centering
        \includegraphics[width=\textwidth]{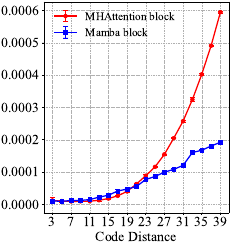}
        \caption{}
        \label{fig:ch4-inference-time}
    \end{subfigure}
    \caption[Mamba decoder accuracy and inference speed]{Accuracy and speed of the Mamba decoder. \textbf{(a)}~Logical error per round on the Sycamore memory experiment dataset for various decoders at code distances $d=3$ and $d=5$; Mamba tracks the reproduced Transformer baseline while replacing multi-head attention with a Mamba block. \textbf{(b)}~Wall-clock inference time for a single Mamba block versus a Multi-Head Attention (MHA) block as a function of code distance, measured on a local RTX 4090 GPU.}
    \label{fig:ch4-benchmarks}
\end{figure}

We benchmark our Mamba decoder against Transformer-based, matching-based, and tensor network decoders using the Sycamore memory experiment dataset from AlphaQubit~\cite{bausch2024alphaqubit,googlequantum2023suppressing}.
This publicly available dataset was obtained on the 72-qubit Sycamore superconducting processor comprising four distance-3 surface code blocks and a single distance-5 block.
Both $X$- and $Z$-basis memory experiments were conducted for up to 25 QEC cycles, with 50,000 experiments performed for each odd-numbered cycle count $n \in \{1, 3, \ldots, 25\}$.
The logical error rate is estimated by fitting the measured fidelity to \Cref{eq:ch4-ler-fit}.
\Cref{fig:ch4-memory-results} summarizes the results.

Our Mamba decoder achieves LERs of approximately $2.98 \times 10^{-2}$ at distance 3 and $3.03 \times 10^{-2}$ at distance 5, closely tracking the reproduced Transformer baseline.
At distance 3, the reproduced neural baselines are competitive with the strongest listed decoders.
At distance 5, both reproduced neural baselines underperform the tensor network decoder ($\varepsilon = 2.915 \times 10^{-2}$), while outperforming the matching-based decoders.
We attribute this gap in part to the size of the pretraining dataset: while AlphaQubit is described as being pretrained on up to 1 billion samples, our models were pretrained on 100 million samples due to computational constraints.
With a larger pretraining dataset, the neural baselines might reduce this remaining gap.

The key takeaway from the memory experiments is that \emph{replacing multi-head attention with a Mamba module preserves comparable decoding accuracy}.
This is a necessary condition for the Mamba decoder to be useful: its speed advantage matters only if the accuracy remains close to the reproduced Transformer baseline.

\subsubsection{Inference Speed}
\label{subsubsec:ch4-speed}

To quantify the computational advantage of the Mamba block over multi-head attention, we measure wall-clock inference time on an NVIDIA RTX 4090 GPU as a function of code distance (\Cref{fig:ch4-inference-time}).

The measured inference times track the complexity of the two blocks. The curves deviate near $d \approx 21$, beyond which the Mamba block becomes increasingly faster, reaching roughly $3\times$ the speed of the MHA block at $d = 39$ and widening further as $d$ grows.
This is highly relevant as fault-tolerant algorithms targeting practical applications are expected to require substantially larger code distances~\cite{fowler2012surface}, precisely the regime in which the quadratic vs.\ quartic scaling gap translates into large latency reductions.
The Mamba block is thus better positioned than attention for the long-sequence decoding demanded by future large-distance processors.

\subsubsection{Real-Time Decoding with Decoder-Induced Noise}
\label{subsubsec:ch4-realtime}

\paragraph{Experimental setup.}
We simulate real-time decoding for surface codes of distance $d \in \{3, 5, 7\}$ using the SI1000 noise model~\cite{gidney2021fault} with a base physical error rate of $p = 0.002$.
Each model is trained on sequences of $2d + 1$ QEC cycles and evaluated over $8d + 4$ cycles (four repetitions of the training-length block).
To simulate the effect of decoder latency, \emph{decoder-induced noise} is injected after every $2d + 1$ cycles.

\paragraph{Decoder-induced noise model.}
The strength of the decoder-induced noise is calibrated to reflect each architecture's computational complexity.
For large code distances ($d > 20$), the overall inference time is dominated by the most computationally intensive component: the MHA block (for the Transformer) or the Mamba block.
We model the decoder-induced error probability as:
\begin{equation}
    p_\text{dec} =
    \begin{cases}
        \alpha \cdot d^4 & \text{(Transformer)}, \\
        \alpha \cdot d^2 & \text{(Mamba)},
    \end{cases}
    \label{eq:ch4-decoder-noise}
\end{equation}
where $\alpha = 7.623 \times 10^{-6}$ is a scaling constant based on AlphaQubit's reported $\sim 40\,\mu$s latency at $d = 9$.
At this distance, $\alpha d^4 \approx 25p$ for the base physical error rate $p=0.002$.
Since the SI1000 model assigns measurement error probability $5p$, this corresponds to a Transformer-induced error rate five times larger than the physical measurement-error rate.

\begin{figure}[t]
    \centering
    \includegraphics[width=0.75\textwidth]{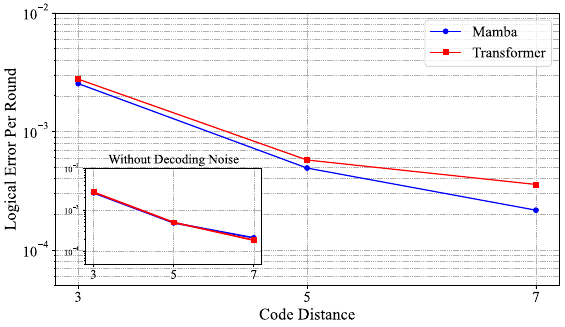}
    \caption{Real-time decoding performance. \textbf{Main:} Logical error per round (LER) for Mamba and Transformer decoders under simulated real-time conditions with decoder-induced noise proportional to computational complexity. \textbf{Inset:} LER without decoder-induced noise, showing comparable baseline accuracy.}
    \label{fig:ch4-realtime-results}
\end{figure}

\paragraph{Results.}
As a baseline, we first compare the two decoders \emph{without} decoder-induced noise (\Cref{fig:ch4-realtime-results}, inset).
In this setting, the Mamba and Transformer decoders exhibit nearly identical logical error rates at all distances, confirming that the Mamba architecture does not sacrifice accuracy when latency is not a factor.
When decoder-induced noise is included (\Cref{fig:ch4-realtime-results}, main panel), the two architectures diverge.
At $d = 3$, both decoders experience mild noise penalties, with comparable LERs.
At $d = 5$, the Transformer's LER begins to degrade more rapidly.
At $d = 7$, the Transformer's LER rises to approximately $10^{-1}$, while the Mamba decoder maintains an LER of approximately $10^{-2}$ in this simulation.

\paragraph{Finite-size effective threshold analysis.}
Here, we estimate a \emph{finite-size effective threshold under the specified decoder-induced-noise/latency model}, rather than the asymptotic surface-code threshold.
Decoder-induced noise can shift this finite-size crossover by adding an additional noise channel proportional to the decoder's modeled latency.

To estimate the finite-size effective threshold under this latency model, we fine-tune both decoders at physical error rates $p \in \{0.006,$ $0.008,$ $0.010,$ $0.012\}$ and compare the LERs at distances $d = 3$ and $d = 5$.
The estimate is the crossover point where increasing code distance no longer helps in this finite-size comparison under the assumed latency-induced noise.

\begin{figure}[t]
    \centering
    \includegraphics[width=\textwidth]{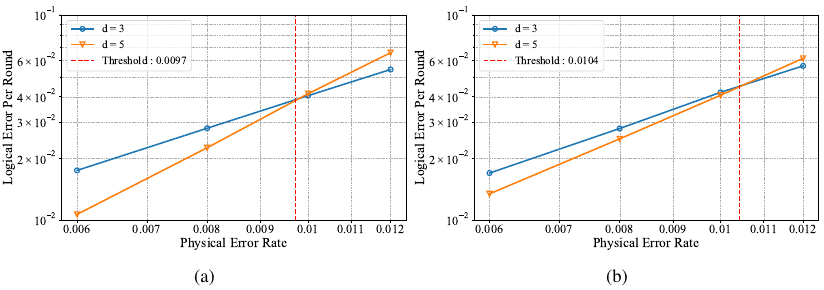}
    \caption{Finite-size effective threshold under this latency model. Each panel shows logical error per round versus physical error rate for code distances $d=3$ and $d=5$. The effective threshold is identified where the $d=5$ curve crosses above the $d=3$ curve. Under the assumed decoder-induced-noise model, the Transformer decoder yields $p_\text{th}\approx0.0097$, while the Mamba decoder yields $p_\text{th}\approx0.0104$.}
    \label{fig:ch4-error-threshold}
\end{figure}

As shown in (\Cref{fig:ch4-error-threshold}), the Transformer reaches $p_\text{th} \approx 0.0097$, while the Mamba decoder reaches $p_\text{th} \approx 0.0104$.
A higher finite-size effective threshold means that, under the assumed decoder-induced noise scaling, the decoder can tolerate a larger physical error rate before increasing the code distance ceases to help.
Moreover, as code distance increases, the Transformer's $\mathcal{O}(d^4)$ decoder-induced noise would degrade the model-dependent effective threshold faster than the Mamba decoder's $\mathcal{O}(d^2)$ noise, making the scaling advantage increasingly important at larger distances.

\section{Summary}
\label{sec:ch4-summary}

In this chapter, we introduced a neural decoder for quantum error correction based on the Mamba architecture---a selective state-space model that replaces the attention mechanism of Transformer-based decoders with a linear-complexity alternative.
Our investigation established three key findings:

\begin{enumerate}
    \item \textbf{Comparable accuracy to the reproduced Transformer baseline.} In memory experiments on real hardware data from Google's Sycamore processor, the Mamba decoder achieved logical error rates of $\varepsilon \approx 2.98 \times 10^{-2}$ (distance 3) and $\varepsilon \approx 3.03 \times 10^{-2}$ (distance 5), closely tracking the reproduced Transformer-based decoder in this comparison.

    \item \textbf{Faster inference with growing advantage.} Wall-clock benchmarks on an RTX 4090 GPU showed that inference time tracks the $\mathcal{O}(d^2)$ vs.\ $\mathcal{O}(d^4)$ complexity of the two blocks: the curves deviate near $d \approx 21$, beyond which the Mamba block pulls increasingly ahead, reaching roughly $3\times$ faster at $d = 39$ and widening further as $d$ grows. Constant-factor overhead masks this advantage at the small distances accessible today ($d \leq 5$), so its significance is asymptotic, favoring the Mamba architecture at the larger code distances relevant for future processors.

    \item \textbf{Higher finite-size effective threshold under this latency model.} Under simulated real-time decoding conditions with decoder-induced noise, the Mamba decoder achieved a finite-size effective threshold of $p_\text{th} \approx 0.0104$, compared to $p_\text{th} \approx 0.0097$ for the Transformer---a 7\% improvement under the assumed latency model. This higher threshold reflects the benefit of lower decoder-induced noise in that model and suggests that latency scaling could become more important as code distance increases.
\end{enumerate}

In the next chapter, we turn from classical neural networks that \emph{assist} quantum hardware to neural networks that \emph{represent} quantum states directly.


\chapter{Neural Quantum States}
\label{chap:nqs}

\chaptersource{The results in this chapter are based on Tak Hur, \textit{Stochastic Reconfiguration as Statistical Filtering for Overparameterized Neural Quantum States}~\cite{hur2026srfilter}. Code is available at \url{https://github.com/takh04/sr_filter}.}
\bigskip

In \Cref{chap:decoder}, neural networks were used to process syndrome data for quantum error correction.
This chapter studies how neural networks can serve as variational representations of quantum many-body wave functions.
Neural quantum states (NQS) combine expressive neural-network ansatzes with variational Monte Carlo (VMC), making it possible to search for ground states in Hilbert spaces far beyond exact diagonalization.

The chapter focuses on stochastic reconfiguration (SR), the standard optimizer for NQS\@.
For modern NQS, the number of parameters can greatly exceed the number of Monte Carlo samples used in one optimization step.
In this parameter-rich regime, we show that the diagonal shift in SR is more than a numerical safeguard: it plays a statistical role, controlling how well the optimizer generalizes from a finite number of Monte Carlo samples.
This perspective motivates a multi-shift variant of SR (MS-SR), which combines several differently shifted updates to make better use of the available samples.

The chapter is organized as follows.
\Cref{sec:ch5-nqs-vmc} introduces the NQS-VMC workflow and SR.
\Cref{sec:ch5-statistical-filtering} develops the fixed-checkpoint regression view and the bias-variance role of the shift.
\Cref{sec:ch5-experiments} summarizes the exact and large-scale diagnostics supporting this view.
\Cref{sec:ch5-ms-sr} presents MS-SR and the main empirical evidence for the method.
\Cref{sec:ch5-summary} summarizes the chapter.

\section{Neural Quantum States and Stochastic Reconfiguration}
\label{sec:ch5-nqs-vmc}

The central obstacle in quantum many-body physics is the exponential size of the Hilbert space.
A system of $N$ spin-$\tfrac{1}{2}$ degrees of freedom has basis states $\ket{x}$ indexed by bit strings $x\in\{\pm1\}^N$, so a generic state requires $2^N$ complex amplitudes.
Tensor-network methods such as density matrix renormalization group exploit low entanglement and are highly effective in many one-dimensional systems~\cite{white1992density,schollwoeck2011density}.
However, frustrated and higher-dimensional systems often require more flexible representations.
NQS address this by parameterizing the wave-function amplitude $\psi_\theta(x)$ with a neural network.
Since the RBM construction of \textcite{carleo2017solving}, NQS have expanded to recurrent, fermionic, transformer-based, and foundation-style wave functions~\cite{hibatallah2020recurrent,pfau2020ferminet,hermann2020deep,viteritti2023transformer,rende2024simple,rende2025foundation}.

VMC approximates the ground state of a Hamiltonian $\hat{H}$ by minimizing the Rayleigh quotient
\begin{equation}
    E(\theta)
    =
    \frac{\braket{\psi_\theta | \hat{H} | \psi_\theta}}
         {\braket{\psi_\theta | \psi_\theta}},
    \qquad
    \ket{\psi_\theta}=\sum_x\psi_\theta(x)\ket{x}.
    \label{eq:ch5-variational-principle}
\end{equation}
The energy is estimated by sampling configurations from the Born distribution
\begin{equation}
    \pi_\theta(x)
    =
    \frac{|\psi_\theta(x)|^2}{\sum_{x'}|\psi_\theta(x')|^2}
\end{equation}
and averaging the local energy
\begin{equation}
    H_{\mathrm{loc}}(x)
    =
    \frac{\bra{x}\hat{H}\ket{\psi_\theta}}{\braket{x|\psi_\theta}}
    =
    \sum_{x'} \braket{x|\hat{H}|x'}\frac{\psi_\theta(x')}{\psi_\theta(x)}.
    \label{eq:ch5-local-energy}
\end{equation}
The resulting training loop is conceptually simple: sample from $|\psi_\theta|^2$, evaluate local energies and wave-function derivatives, and update the parameters.

Ordinary gradient descent treats the parameter vector as Euclidean, but the same Euclidean movement can correspond to very different changes in the represented quantum state.
SR corrects this mismatch by using the geometry induced by the variational state manifold~\cite{amari1998natural,stokes2020quantum,hackl2020geometry}.
Let
\begin{equation}
    O_c(x)
    =
    \nabla_\theta\log\psi_\theta(x)
    -
    \mathbb{E}_{\pi_\theta}[\nabla_\theta\log\psi_\theta]
    \label{eq:ch5-centered-features}
\end{equation}
denote centered tangent features.
For real wave functions and real parameters, the quantum geometric tensor (QGT) and SR force are
\begin{equation}
    S=\mathbb{E}_{\pi_\theta}\!\left[O_c(x)O_c(x)^T\right],
    \qquad
    g=\mathbb{E}_{\pi_\theta}\!\left[O_c(x)H_{\mathrm{loc},c}(x)\right],
    \label{eq:ch5-population-s-g}
\end{equation}
where $H_{\mathrm{loc},c}(x)=H_{\mathrm{loc}}(x)-E(\theta)$.
SR updates parameters by solving
\begin{equation}
    S \delta = g,
    \qquad
    \theta^+ = \theta-\eta\delta.
    \label{eq:ch5-sr-linear-system}
\end{equation}
Geometrically, this is a natural-gradient step; physically, it is the projection of imaginary-time evolution onto the tangent space of $\ket{\psi_\theta}$~\cite{sorella1998green,sorella2001generalized}.

In practice, the exact expectations defining $S$ and $g$ are unavailable and are replaced by Monte Carlo averages over a batch of $N_s$ samples $\{x_j\}$.
Stacking the centered features as rows of $O\in\mathbb{R}^{N_s\times P}$, with row $j$ equal to $O_c(x_j)^T$, and collecting the centered local energies in $\mathbf{h}\in\mathbb{R}^{N_s}$ with $h_j=H_{\mathrm{loc},c}(x_j)$, the empirical QGT and force are
\begin{equation}
    \hat{S}
    =
    \frac{1}{N_s}\sum_{j=1}^{N_s}O_c(x_j)O_c(x_j)^T
    =
    \frac{1}{N_s}O^{T}O,
    \qquad
    \hat{g}
    =
    \frac{1}{N_s}\sum_{j=1}^{N_s}O_c(x_j)H_{\mathrm{loc},c}(x_j)
    =
    \frac{1}{N_s}O^{T}\mathbf{h}.
    \label{eq:ch5-empirical-s-g}
\end{equation}
Because $\hat{S}$ is rank-deficient and ill-conditioned when estimated from a finite batch, SR regularizes the linear system with a diagonal shift $\lambda$, giving the ridge solution
\begin{equation}
    \hat{\delta}_\lambda
    =
    (\hat{S}+\lambda I)^{-1}\hat{g}.
    \label{eq:ch5-empirical-ridge}
\end{equation}

\paragraph{MinSR.}
The computational regime has changed with modern NQS\@.
Classical VMC and early RBM calculations often had fewer parameters than samples.
Transformer and foundation-style NQS can instead have $P$ larger than the sample count $N_s$ by one to three orders of magnitude~\cite{chen2024empowering,rende2024simple,rende2025foundation}.
In this regime, forming and inverting the $P\times P$ matrix $\hat{S}+\lambda I$ of \cref{eq:ch5-empirical-ridge} becomes prohibitive.
Kernel-form SR, often called minSR, instead solves in sample space using the empirical neural tangent kernel
\begin{equation}
    \hat{T}
    =
    \frac{1}{N_s}O O^{T}
    \in\mathbb{R}^{N_s\times N_s}.
    \label{eq:ch5-ntk}
\end{equation}
By the push-through identity, the ridge solution of \cref{eq:ch5-empirical-ridge} is obtained equivalently as
\begin{equation}
    \hat{\delta}_\lambda
    =
    \frac{1}{N_s}O^{T}(\hat{T}+\lambda I)^{-1}\mathbf{h},
    \label{eq:ch5-minsr-update}
\end{equation}
which inverts an $N_s\times N_s$ system with the same shift $\lambda$ rather than the full $P\times P$ QGT\@.
This reduces the linear-algebra dimension from $P\times P$ to $N_s\times N_s$, but forming the dense kernel still costs $\mathcal{O}(N_s^2P)$ and solving it costs $\mathcal{O}(N_s^3)$.
Increasing the batch size therefore remains expensive, and each SR update is still estimated from a small random batch relative to a very expressive tangent space.

\section{SR as Statistical Spectral Filtering}
\label{sec:ch5-statistical-filtering}

The key simplification is to analyze one SR step at a fixed checkpoint $\theta$.
At this checkpoint, the wave function, sampling distribution, tangent features, and local energies are fixed.
For notational clarity, we present the case of real wave functions and Hamiltonians in the sampling basis.
For complex wave functions, the regression uses modulus-squared residuals, with $Q=\mathbb{E}_{\pi_\theta}[\overline{O_c}O_c^T]$ and $f=\mathbb{E}_{\pi_\theta}[\overline{O_c}H_{\mathrm{loc},c}]$.
Real parameter increments then satisfy $\operatorname{Re}(Q)\delta=\operatorname{Re}(f)$, equivalently a regression with stacked real and imaginary residuals.

At a fixed checkpoint, SR fits tangent predictions $O_c(x)^T\delta$ to centered local energies.

The ideal population SR direction is the least-squares solution
\begin{equation}
    \delta^*
    =
    S^\dagger g
    \in
    \argmin_\delta
    \mathbb{E}_{\pi_\theta}
    \left[
        \left(O_c(x)^T\delta-H_{\mathrm{loc},c}(x)\right)^2
    \right].
    \label{eq:ch5-population-sr}
\end{equation}

The centered local energy is not generally representable exactly by the current tangent features.
The population residual of the best tangent-space fit is the \emph{expressivity gap}
\begin{equation}
    \epsilon(x)
    =
    H_{\mathrm{loc},c}(x)-O_c(x)^T\delta^*.
    \label{eq:ch5-expressivity-gap}
\end{equation}
The normal equations imply
\begin{equation}
    \mathbb{E}_{\pi_\theta}[O_c(x)\epsilon(x)] = 0,
    \qquad
    \sigma_{\mathrm{gap}}^2
    =
    \mathbb{E}_{\pi_\theta}[\epsilon(x)^2]
    =
    \mathrm{Var}_{\pi_\theta}(H_{\mathrm{loc}})-g^TS^\dagger g.
    \label{eq:ch5-gap-variance}
\end{equation}
Physically, this gap is the component of imaginary-time evolution that the current tangent space cannot express.
Statistically, it behaves like residual noise: it is orthogonal to the tangent features in population, but finite Monte Carlo batches contain sampled values of $\epsilon(x)$ that an overparameterized empirical solve can fit.

This regression view clarifies the role of the diagonal shift.
The target is $\delta^*$, while the empirical ridge update $\hat{\delta}_\lambda$ of \cref{eq:ch5-empirical-ridge} depends on a random batch.
The relevant excess risk is
\begin{equation}
    \mathcal{E}_\lambda
    =
    \mathbb{E}_D
    \left[
        \|\hat{\delta}_\lambda-\delta^*\|_S^2
    \right].
    \label{eq:ch5-excess-risk}
\end{equation}
This $S$-seminorm excess risk is physically meaningful because it is directly related to the \emph{infidelity}, a standard distance between quantum states. For normalized states, bounded updates, and a small step size $\eta$,
\begin{equation}
    I_\theta(\delta,\delta')
    =
    1-|\langle\psi_{\theta-\eta\delta}|\psi_{\theta-\eta\delta'}\rangle|^2
    =
    \eta^2\|\delta-\delta'\|_S^2
    +
    O(\eta^3),
    \label{eq:ch5-infidelity}
\end{equation}
so $\mathcal{E}_\lambda$ controls the expected local infidelity between the ideal population update and the empirical SR update, up to the factor $\eta^2$ and higher-order terms.

The excess risk has the exact decomposition
\begin{equation}
    \mathcal{E}_\lambda
    =\left\|\mathbb{E}_D[\hat{\delta}_\lambda]-\delta^*\right\|_S^2
    +\mathbb{E}_D\!\left[\left\|\hat{\delta}_\lambda-\mathbb{E}_D[\hat{\delta}_\lambda]\right\|_S^2\right].
    \label{eq:ch5-exact-bias-variance}
\end{equation}
Writing $S=V\mathrm{diag}(s_i)V^T$ on the non-null QGT subspace and $\beta_i^*=(V^T\delta^*)_i$, a spectral approximation for random-design ridge regression~\cite{hsu2014random} gives the bias--variance proxy
\begin{equation}
    \mathcal{E}_\lambda
    \approx
    \underbrace{
    \sum_i
    s_i
    \left(\frac{\lambda}{s_i+\lambda}\right)^2
    (\beta_i^*)^2
    }_{\text{shrinkage bias}}
    +
    \underbrace{
    \frac{\sigma_{\mathrm{gap}}^2}{N_s}
    \sum_i
    \left(\frac{s_i}{s_i+\lambda}\right)^2
    }_{\text{finite-sample variance}}.
    \label{eq:ch5-spectral-risk}
\end{equation}
This approximation replaces the empirical inverse by $(S+\lambda I)^{-1}$ and approximates the residual-noise covariance $\mathbb{E}_{\pi_\theta}[O_cO_c^T\epsilon^2]$ by $\sigma_{\mathrm{gap}}^2S$.
It therefore isolates a noise mechanism rather than giving an exact finite-sample risk formula; the held-out identities below do not require this scalar-noise approximation.
The same scalar filter
\begin{equation}
    f_\lambda(s)=\frac{s}{s+\lambda}
    \label{eq:ch5-single-shift-filter}
\end{equation}
controls both terms.
Increasing $\lambda$ damps noisy small-eigenvalue directions and reduces variance, but it also shrinks useful components of $\delta^*$ and increases bias.
The diagonal shift is therefore a statistical regularizer, not only a numerical tolerance.

For small Hilbert spaces, $S$, $\delta^*$, and $\sigma_{\mathrm{gap}}^2$ can be computed by exact summation.
For large NQS, the population quantities are unavailable, so we introduce held-out tangent predictions.
For an independent validation set $D_{\mathrm{val}}$,
\begin{equation}
    R_{\mathrm{val}}(\delta)
    =
    \frac{1}{|D_{\mathrm{val}}|}
    \sum_{x\in D_{\mathrm{val}}}
    \left(O_c(x)^T\delta-H_{\mathrm{loc},c}(x)\right)^2,
    \label{eq:ch5-validation-residual}
\end{equation}
with expectation
\begin{equation}
    \mathbb{E}_{D_{\mathrm{val}}}[R_{\mathrm{val}}(\delta)]
    =
    \|\delta-\delta^*\|_S^2+\sigma_{\mathrm{gap}}^2.
    \label{eq:ch5-validation-identity}
\end{equation}
Thus validation residual tracks excess risk plus the irreducible expressivity-gap term.
A complementary diagnostic measures how much the update varies across independent training batches. Given $m$ SR updates $\hat{\delta}_{\lambda,j}$ obtained from independently resampled batches at the same checkpoint, with mean $\bar{\delta}_\lambda = m^{-1}\sum_j \hat{\delta}_{\lambda,j}$, the multi-batch variance is
\begin{equation}
    \mathcal{V}_{\mathrm{mb}}(\lambda)
    =
    \frac{1}{m-1}
    \sum_{j=1}^m
    \frac{1}{|D_{\mathrm{val}}|}
    \sum_{x\in D_{\mathrm{val}}}
    \left(
        O_c(x)^T(\hat{\delta}_{\lambda,j}-\bar{\delta}_\lambda)
    \right)^2.
    \label{eq:ch5-multi-batch-variance}
\end{equation}
This is the sample variance of the tangent predictions $O_c(x)^T\hat{\delta}_{\lambda,j}$ across the independent updates, evaluated on held-out configurations. Because the average over $x\in D_{\mathrm{val}}$ approximates the QGT seminorm, $\mathcal{V}_{\mathrm{mb}}$ isolates the batch-to-batch fluctuation of the update and, as $m$ and $|D_{\mathrm{val}}|$ grow, converges to the variance component of the excess risk in \Cref{eq:ch5-excess-risk}. Like the validation residual, it requires only held-out tangent-feature predictions, so it remains practical at large scale.
The filtering view predicts a U-shaped validation residual as $\lambda$ varies and a monotone decrease in $\mathcal{V}_{\mathrm{mb}}$ as finite-batch noise is suppressed.

\section{Evidence Across System Scales}
\label{sec:ch5-experiments}

\paragraph{Small-scale test.}
The first test uses an exactly tractable $4\times4$ spin-$\tfrac12$ Heisenberg graph,
\begin{equation}
    \hat{H}
    =
    J_1\sum_{(i,j)\in\mathcal{B}_1}\hat{S}_i\cdot\hat{S}_j
    +
    J_2\sum_{(i,j)\in\mathcal{B}_2}\hat{S}_i\cdot\hat{S}_j,
    \qquad
    J_1=1,\quad J_2=0.5.
    \label{eq:ch5-j1j2}
\end{equation}
Here $\hat{\mathbf S}_i=\boldsymbol\sigma_i/2$.
The graph has $32$ periodic nearest-neighbor bonds, but only $24$ diagonal bonds: the historical constructor omits the eight diagonals across one periodic seam.
The reported data therefore concern this graph, rather than the fully periodic $J_1$--$J_2$ lattice.
All $2^{16}=65{,}536$ basis states are included without a magnetization restriction, allowing exact Born-weighted population sums.
The experiment compares an underparameterized RBM with $P/N_s=0.27$ against an overparameterized Vision Transformer with $P/N_s=3.28$, using $N_s=4096$ and a near-ridgeless shift $\lambda=10^{-9}$.

The networks represent complex wave functions, but these offline diagnostics score only the real-amplitude regression: tangent features are $\nabla_\theta\operatorname{Re}\log\psi_\theta$, and targets are $\operatorname{Re}H_{\mathrm{loc}}$, each Born-centered.
The complex state is retained in the Born distribution and local energy.
Phase derivatives and the imaginary local-energy target are excluded, so the following gap and risk values describe this restricted regression, rather than the complete complex-SR error or local infidelity.
The excess-risk estimates are means with standard errors over $100$ independent Born-resampled batches for matched noise and $10$ for matched state.

Two protocols separate the two effects of overparameterization.
In the matched-noise comparison, RBM and ViT checkpoints are selected with comparable gap variance, $\sigma_{\mathrm{gap}}^2\approx0.8$.
The RBM remains close to the population-optimal amplitude direction, with excess risk approximately $0.53\pm0.08$.
The ViT has comparable gap variance but much larger excess risk, approximately $3.39\pm0.73$, because its larger tangent space gives the empirical solve more directions in which to fit finite-sample residuals.
In the matched-state comparison, a ViT is fit to represent the same wave function as a partially converged RBM checkpoint, reaching fidelity $F\geq0.98$.
Now the represented state is nearly fixed while the tangent space changes.
The ViT reduces the gap variance from $1.95$ to $0.816$ and lowers excess risk from approximately $2.53\pm0.19$ to $1.59\pm0.08$.
Thus overparameterization helps when it reduces the expressivity gap, but it can hurt when a comparable gap remains and finite-sample SR overfits it.

\begin{figure}[t]
    \centering
    \includegraphics[width=0.9\textwidth]{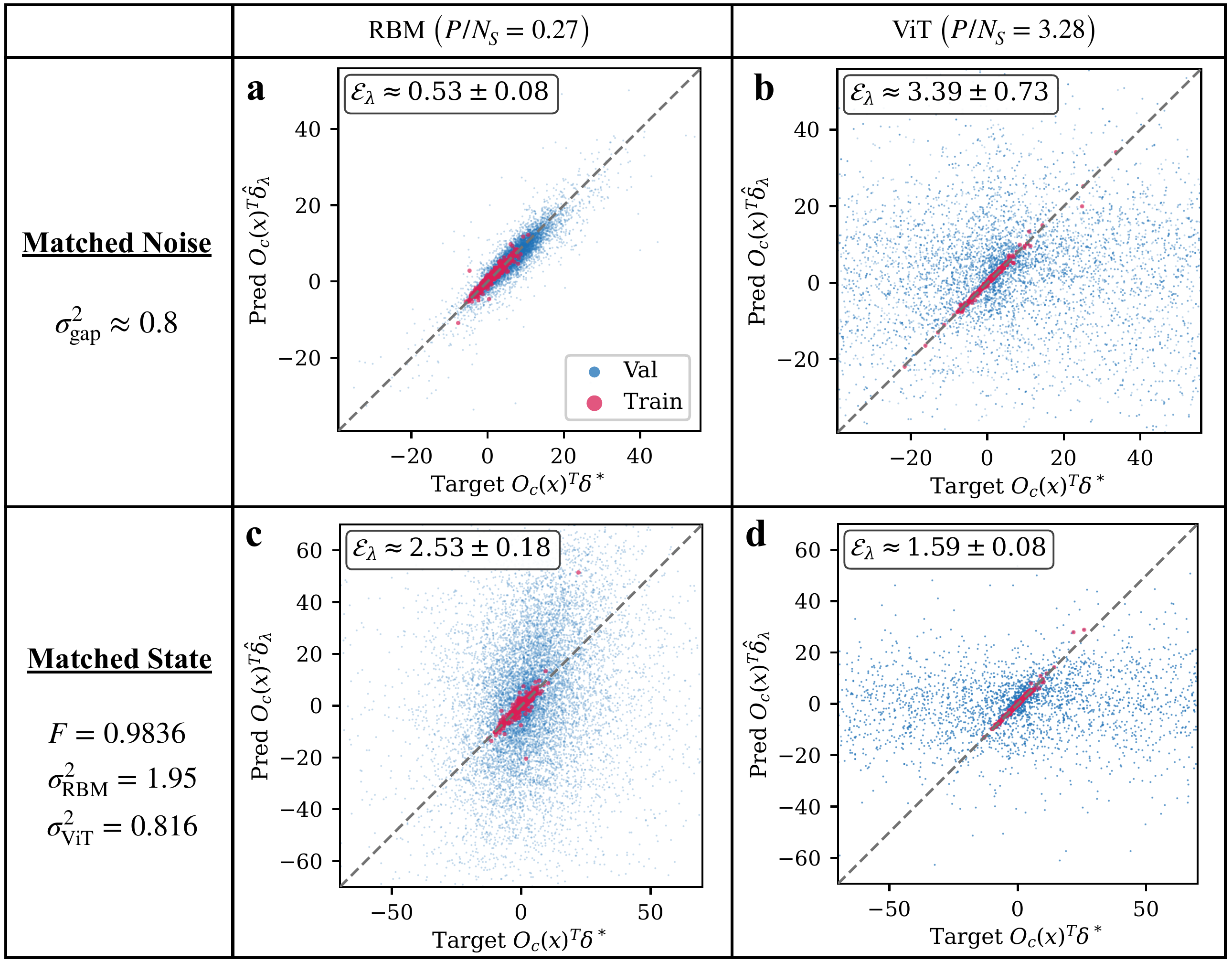}
    \caption{Exact amplitude-regression diagnostics for the $4\times4$ Heisenberg graph of \Cref{eq:ch5-j1j2}, with $32$ nearest-neighbor and $24$ diagonal bonds. At $N_s=4096$ and $\lambda=10^{-9}$, matched noise exposes ViT overfitting at comparable amplitude-gap variance, while matched state shows lower gap variance and excess risk for the ViT. The plotted predictions and risks exclude the phase channel.}
    \label{fig:ch5-small-scale-risk}
\end{figure}

\paragraph{Large-scale test.}
The large-scale test asks whether the same noisy-ridge mechanism appears in a modern NQS setting.
The system is a periodic one-dimensional transverse-field Ising family,
\begin{equation}
    \hat{H}(h)
    =
    -h\sum_{i=1}^{L}\hat{X}_i
    -
    \sum_{i=1}^{L}\hat{Z}_i\hat{Z}_{i+1},
    \qquad
    L=100,\quad h\in[0.8,1.2].
    \label{eq:ch5-tfim}
\end{equation}
A single foundation neural quantum state represents the conditional wave function $\log\psi_\theta(x;h)$ across the Hamiltonian family~\cite{rende2025foundation}.
The model has $P=198{,}144$ trainable parameters, while each SR step uses $N_s=12{,}000$ samples, organized as $6000$ training fields with two spin configurations per field.

\begin{figure}[t]
    \centering
    \includegraphics[width=\textwidth]{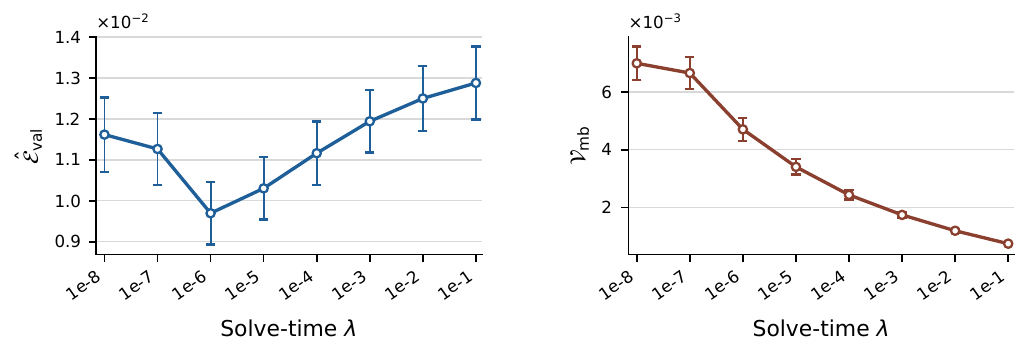}
    \caption{Large-scale fixed-checkpoint diagnostics for the $L=100$ transverse-field Ising model family using a foundation NQS\@. The validation residual exhibits a U-shaped dependence on the SR diagonal shift, while the multi-batch variance decreases with stronger regularization, as predicted by the noisy-ridge interpretation in \Cref{eq:ch5-spectral-risk}.}
    \label{fig:ch5-tfim-shift-sweep}
\end{figure}

The diagnostic protocol holds the wave function fixed and varies only the solve-time shift.
Ten late-training checkpoints from a run trained with $\lambda_{\mathrm{train}}=10^{-4}$ are reused.
For each checkpoint and each $\lambda\in\{10^{-8},10^{-7},\ldots,10^{-1}\}$, $m=100$ independent standard-SR updates are recomputed from fresh batches.
Validation uses $120{,}000$ held-out samples, arranged as $6000$ held-out fields disjoint from the training fields with $20$ spin configurations per field.
Predictions and local energies are centered separately within each field.
\Cref{fig:ch5-tfim-shift-sweep} shows the predicted pattern: the validation residual is U-shaped in the solve-time shift, while the multi-batch variance decreases as $\lambda$ increases.
Because all shifts are evaluated at the same fixed checkpoints, this isolates the statistical effect of the ridge filter rather than comparing different training trajectories.
The large-scale implementation uses the energy-gradient target $2H_{\mathrm{loc},c}$ and doubled update directions, so the raw residual and variance values in the large-scale figures are four times those under the normalization of \Cref{eq:ch5-validation-residual,eq:ch5-multi-batch-variance}.
This common factor leaves relative comparisons and shift selection unchanged.

\section{Multi-Shift Stochastic Reconfiguration}
\label{sec:ch5-ms-sr}

The spectral-filtering view suggests the importance of the diagonal shift in SR.
Single-shift SR applies the one-parameter filter $f_\lambda(s)=s/(s+\lambda)$ to every QGT eigendirection.
MS-SR replaces this with a mixture of shifted solves computed on independent batches.

\FloatBarrier
\begin{algorithm}[t]
\caption{Multi-Shift Stochastic Reconfiguration (MS-SR)}
\label{alg:ch5-ms-sr}
\begin{algorithmic}[1]
\Require Parameters $\theta_t$, batch size $N_s$, number of shifts $K$, NTK quantiles $\{q_k\}_{k=1}^K$, learning rate $\eta$
\State Draw independent batches $D_1,\ldots,D_K\sim\pi_{\theta_t}$, each of size $N_s$
\State Use the positive NTK spectrum of $D_1$ to choose shifts $\lambda_k$ from quantiles $\{q_k\}_{k=1}^K$
\For{$k=1,\ldots,K$}
    \State Solve the shifted minSR system on $D_k$ to obtain $\hat{\delta}_k\leftarrow\mathrm{minSR}(D_k,\lambda_k)$
\EndFor
\State Evaluate leave-one-batch-out tangent-space residuals for the candidate updates
\State Fit simplex-constrained stacking weights $w\in\Delta^{K-1}$ from the held-out residuals
\State Apply $\theta_{t+1}\leftarrow\theta_t-\eta\sum_{k=1}^K w_k\hat{\delta}_k$
\end{algorithmic}
\end{algorithm}

For fixed shifts and weights, the population spectral approximation of MS-SR has the filter
\begin{equation}
    f_{w,\lambda}(s)
    =
    \sum_{k=1}^K w_k\frac{s}{s+\lambda_k}.
    \label{eq:ch5-ms-filter}
\end{equation}
This filter family is richer than any single shifted SR solve.
Independent-batch candidates have different empirical QGTs, so their average is not exactly a filter of one empirical matrix.
Different shifts can suppress noisy low-eigenvalue directions while preserving useful high-eigenvalue directions more flexibly than a single $\lambda$.
Independent batches provide a complementary variance reduction: for fixed shifts and weights, independent candidate errors contribute $\sum_k w_k^2 V_k$ to the mixture variance, where $V_k$ is the variance of candidate $k$ in the QGT seminorm.
Uniform weights give a factor-$K$ reduction relative to the average candidate variance in this ideal independent-batch limit.
Data-dependent shifts and weights, as well as sampling correlations, require empirical assessment.
Computationally, $K$ independent minSR solves of size $N_s$ cost $\mathcal{O}(K N_s^3)$ and can be parallelized, whereas one solve on a batch of size $K N_s$ costs $\mathcal{O}(K^3 N_s^3)$.
MS-SR is most worthwhile when finite-batch variance is a visible bottleneck and multiple solve-time shifts cover distinct useful parts of the empirical spectrum.

In the experiments, $K=4$ and the shifts are chosen from positive empirical NTK-spectrum quantiles $\{0.9,0.7,0.4,0.1\}$.
The weights are fit by leave-one-batch-out stacking.
Writing $O_j$ and $\mathbf h_j$ for the centered feature matrix and local-energy vector on batch $D_j$, the objective is
\begin{equation}
    \min_{w\in\Delta^{K-1},\,w_j<1}
    \sum_{j=1}^K
    \left\|
        \sum_{i\ne j}\frac{w_i}{1-w_j}O_j\hat\delta_i-\mathbf h_j
    \right\|^2.
    \label{eq:ch5-stacking-objective}
\end{equation}
Each held-out prediction excludes the candidate fitted on that batch and renormalizes the remaining weights.
The resulting mixture is convex, but this objective is not generally a convex function of the weights; numerical fitting from uniform weights does not guarantee a global minimum.

\paragraph{Checkpoint-local ablations.}
We evaluate the ingredients of MS-SR on the $L=100$ TFIM foundation NQS of \Cref{sec:ch5-experiments} and an $8\times8$ periodic $J_1$--$J_2$ ViT at $J_2/J_1=0.5$.
The latter uses Pauli interactions $\boldsymbol\sigma_i\cdot\boldsymbol\sigma_j$, $128$ bonds of each type, and the zero-magnetization sector.
Its raw energies are four times those of the spin-operator convention in \Cref{eq:ch5-j1j2}; squared diagnostics retain the corresponding Pauli units.
The ViT has $P=147{,}216$ real parameters, and its source training progressively adds translations and $C_4$ rotations.
Ten fixed source checkpoints per system and $20$ resampling repeats per checkpoint are used for the ablations.
The NTK quantile grid is estimated on a separately seeded reference batch once per checkpoint and reused across repeats.

Standard SR uses one batch and the training shift $\lambda_{\mathrm{train}}=10^{-4}$.
\emph{Bagged SR} uniformly averages four independent candidates at one common shift, isolating independent-batch averaging.
\emph{Shared-batch multi-shift SR} fits four shifts on shared data and uses uniform weights; its \emph{stacked} counterpart fits weights on held-out samples using the same candidates.
\emph{Independent-batch multi-shift SR} uses four independent fitting batches with uniform weights, and full \emph{MS-SR} learns the weights by \Cref{eq:ch5-stacking-objective}.
The shared-batch TFIM comparison uses an independent weight-fit batch; the $J_1$--$J_2$ shared-batch variants reserve one quarter of their sampled batch for weight fitting.
All multi-shift methods use the same quantile grid.

Bagging is evaluated both at $\lambda_{\mathrm{train}}$ and at a checkpoint-specific $\lambda_{\mathrm{tuned}}$ selected by an eight-point sweep over $\{10^{-8},\ldots,10^{-1}\}$.
The latter minimizes the bagged residual on the reporting data and is therefore an oracle reference requiring additional solves and validation, rather than an independently tuned baseline.
Relative to standard SR at the training shift, MS-SR reduces the validation residual and multi-batch variance by approximately $39\%$ and $43\%$ on TFIM, and by $14\%$ and $66\%$ on $J_1$--$J_2$.
Oracle-tuned bagging reduces them by $45\%$ and $49\%$, and by $18\%$ and $80\%$, respectively, giving lower mean values than MS-SR in both systems.
Shared-batch multi-shift SR reduces the TFIM residual by $17\%$, but increases the $J_1$--$J_2$ residual by $2\%$.

On TFIM, MS-SR has $27\%$ lower residual than bagging at the training shift, with the same ordering on all ten checkpoints.
Training-shift bagging has lower variance but higher total residual, illustrating the cost of shrinkage bias.
Bagging near $10^{-6}$--$10^{-7}$ achieves a lower residual than MS-SR, showing that a well-chosen single shift can suffice.
On $J_1$--$J_2$, the mean residuals are $3.98$ for training-shift bagging, $3.92$ for oracle-tuned bagging, and $4.13$ for MS-SR.
Thus the benefit of MS-SR is adaptation across regularization scales without a separate validation-based shift search; the results do not establish uniform superiority over well-tuned bagging.

\begin{figure}[t]
    \centering
    \includegraphics[width=\textwidth]{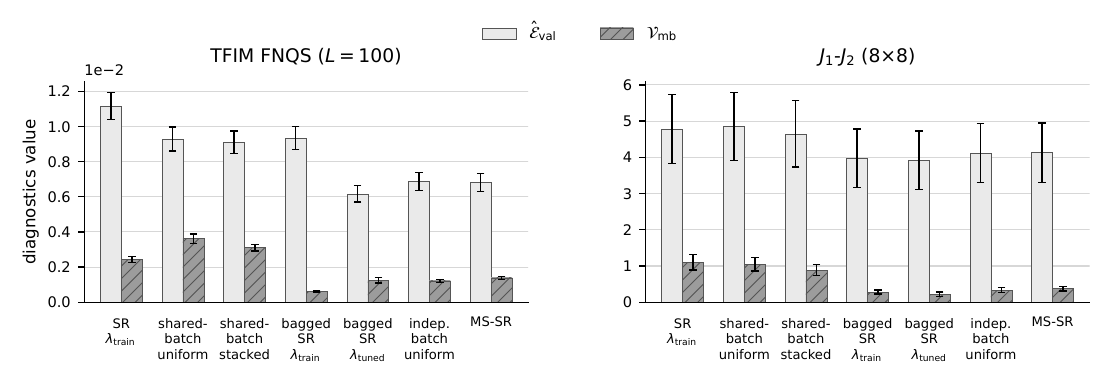}
    \caption{Checkpoint-local MS-SR ablations at $K=4$. Bars show raw validation residuals and multi-batch variances averaged over fixed source checkpoints, with one standard error across checkpoints. Standard SR uses $\lambda_{\mathrm{train}}=10^{-4}$. Bagging is shown at both the training shift and an oracle shift selected on the reporting data. All multi-shift methods use the same NTK-quantile grid. Values retain the doubled-target convention and, for $J_1$--$J_2$, Pauli Hamiltonian units.}
    \label{fig:ch5-ms-sr-ablation}
\end{figure}

Robustness checks with $K\in\{2,3,4,5\}$ reduce the TFIM MS-SR residual from $0.00854$ to $0.00638$ as $K$ grows, with diminishing gains and one extra batch and solve per candidate.
MS-SR beats training-shift bagging at each $K$, while oracle-tuned bagging remains better.
Two alternative quantile grids change the residual by at most $3.4\%$, and checkpoints trained at shifts $10^{-3}$ and $10^{-5}$ give near-parity with the corresponding bagged baseline.

\paragraph{Online training comparisons.}
The main online comparison uses five paired continuations from one shared SR checkpoint at iteration $2500$ of the $8\times8$ $J_1$--$J_2$ model.
This is a separate ViT setup from the symmetry-ramp ablations: it has depth $8$, embedding dimension $72$, and $12$ attention heads, without explicit lattice-translation or rotation symmetrization.
Both methods use Cholesky solves, $N_s=8192$ samples per batch, and constant learning rate $\eta=0.01$.
The Hamiltonian and energy values use the same Pauli convention and zero-magnetization sector as the large-scale $J_1$--$J_2$ ablations.
The comparison concerns later refinement after initial standard-SR training.

\begin{figure}[t]
    \centering
    \captionsetup{font=small}
    \begin{subfigure}[t]{0.50\textwidth}
        \vspace{0pt}
        \centering
        \captionsetup{position=top}
        \caption{Training history}
        \label{fig:ch5-online-history}
        \includegraphics[width=\linewidth]{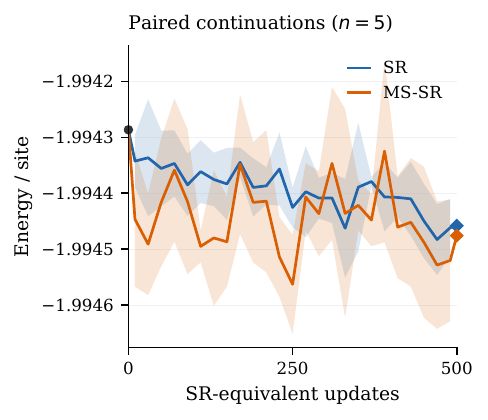}
    \end{subfigure}\hfill
    \begin{subfigure}[t]{0.47\textwidth}
        \vspace{0pt}
        \centering
        \captionsetup{position=top}
        \caption{Endpoint energies per site}
        \label{fig:ch5-online-endpoints}
        \renewcommand{\arraystretch}{2}
        \setlength{\tabcolsep}{3pt}
\footnotesize
\begin{tabular}{@{}rrrr@{}}
\toprule
Pair & SR & MS-SR & $\Delta\;(10^{-5})$ \\
\midrule
1 & $\mathbf{-1.9944675}$ & $-1.9944307$ & $+3.69 \pm 1.09$ \\
2 & $-1.9944450$ & $\mathbf{-1.9945160}$ & $-7.10 \pm 1.07$ \\
3 & $\mathbf{-1.9944733}$ & $-1.9944722$ & $+0.11 \pm 1.11$ \\
4 & $-1.9944593$ & $\mathbf{-1.9944668}$ & $-0.75 \pm 1.04$ \\
5 & $-1.9944451$ & $\mathbf{-1.9944934}$ & $-4.83 \pm 1.02$ \\
\midrule
Mean & $-1.9944580$ & $-1.9944758$ & $-1.78$ \\
\bottomrule
\end{tabular}

    \end{subfigure}
    \caption{Online SR and MS-SR training on the $8\times8$ $J_1$--$J_2$ model. Five paired continuations start from a single shared SR checkpoint; energies per site are in Pauli units. \textbf{(a)} Training curves show means of traces binned in intervals of $20$ nominal SR-equivalent updates, with $\pm1$ across-continuation sample standard deviation. Diamonds mark independent endpoint means. The horizontal axis counts one candidate per SR step and four per MS-SR step; the historical implementation's extra solve and sampled batches are excluded from this nominal count. \textbf{(b)} Independent endpoint energies for the same five pairs. Bold marks the lower energy. Differences are $\Delta=(E_{\mathrm{MS\text{-}SR}}-E_{\mathrm{SR}})/64$. Differences and individual $\pm$ values are in units of $10^{-5}$, with the latter combining the two Monte Carlo standard errors in quadrature. The across-pair confidence interval is given in the text.}
    \label{fig:ch5-online-training}
\end{figure}

Standard SR uses $\lambda=10^{-4}$.
The historical MS-SR implementation instead uses four fixed shifts $\{10^{-2},10^{-3},10^{-4},10^{-5}\}$ on independent candidate batches and fits stacking weights on an additional independent batch.
It therefore differs from the NTK-quantile, leave-one-batch-out algorithm above.
\Cref{fig:ch5-online-history} plots progress in nominal SR-equivalent updates: one SR step counts as one unit and one four-candidate MS-SR step as four, so $500$ SR steps and $125$ MS-SR steps both reach $500$ units.
This axis counts candidates and does not equate actual compute budgets.
The historical callback also performs a base-driver SR solve on another batch, giving five solves and six sampled batches per MS-SR step: $625$ solves and $750$ batches for the MS-SR continuation, versus $500$ of each for SR.

Each final state is evaluated independently using two fresh-chain replicas, with $8{,}388{,}608$ retained samples in total.
\Cref{fig:ch5-online-endpoints} reports the endpoint energies and paired differences $\Delta=(E_{\mathrm{MS\text{-}SR}}-E_{\mathrm{SR}})/N$, with $N=64$.
MS-SR gives lower energy in three of five pairs, with mean $\overline\Delta=-1.78\times10^{-5}$ and paired $95\%$ Student-$t$ interval $[-7.05,\,3.50]\times10^{-5}$.
The interval includes zero and measures continuation variability conditional on this single source state.
These measurements therefore support a qualified online comparison, without establishing a reliable final-energy advantage across independently trained source states or at equal wall time.

\paragraph{Update-construction cost.}
A separate benchmark at the TFIM step-$2000$ checkpoint measures the NTK-quantile, leave-one-batch-out implementation of MS-SR on one H100 80GB GPU, with $N_s=12{,}000$ and $K=4$.
All methods use the same eigendecomposition-based pseudoinverse solver.
Timing includes sampling, local energies, candidate solves, and construction of the final direction, with parameters held fixed.
The first NTK eigendecomposition supplies the quantile shifts and is reused for the first candidate solve; stacking reuses the four candidate batches.
The medians over three paired device/seed settings are shown in \Cref{tab:ch5-update-cost}.
MS-SR takes about four times as long as one SR update and adds approximately $1.03\%$ over bagged SR in this implementation.

\begin{table}[t]
    \centering
    \begin{tabular}{lrrr}
        \toprule
        Method & Batches/solves & Time (s) & Relative to SR \\
        \midrule
        SR & 1/1 & 13.38 & $1.00\times$ \\
        Bagged SR & 4/4 & 53.55 & $4.00\times$ \\
        MS-SR & 4/4 & 54.09 & $4.04\times$ \\
        \bottomrule
    \end{tabular}
    \caption{Steady-state update-construction cost on the TFIM foundation NQS using one H100 80GB GPU. Values are medians over three paired device/seed settings, each with two warm-ups and five synchronized timed repetitions. Bagged SR uses the training shift and excludes the cost of shift tuning.}
    \label{tab:ch5-update-cost}
\end{table}

\section{Summary}
\label{sec:ch5-summary}

This chapter analyzed SR in a regime increasingly common for modern NQS: expressive, overparameterized networks trained from finite Monte Carlo samples.
The main conclusion is that SR should be viewed as statistical spectral filtering.
At a fixed wave function, SR is ridge regression from tangent features to centered local energy.
The tangent-space expressivity gap is orthogonal to the tangent space in population, but finite batches make it behave as residual noise that empirical SR can fit.
The diagonal shift therefore controls a bias-variance tradeoff: it suppresses variance from sampled gap residuals while shrinking useful components of the ideal tangent-space update.

The experiments support this mechanism across scales.
Exact $4\times4$ amplitude-regression experiments separate the two effects of overparameterization: a larger tangent space helps when it reduces the expressivity gap, but it can hurt when it overfits a comparable gap.
Large-scale foundation-NQS diagnostics show the predicted U-shaped validation residual and decreasing multi-batch variance as the solve-time shift increases.
Finally, the filtering view leads to MS-SR, whose independent multi-shift solves reduce validation residuals and update variance relative to fixed-shift SR.


\chapter{Conclusion}
\label{chap:conclusion}

This thesis studied the interface between quantum computing and artificial intelligence from two directions.
Part~I asked how quantum models can be made useful for supervised learning; Part~II asked how machine learning can help with quantum problems such as real-time decoding and quantum many-body physics.
Each chapter addressed a different bottleneck at this interface: representation, generalization, decoding latency, and finite-sample optimization.

In Part~I, \Cref{chap:nqe} showed that the data embedding sets an empirical-risk floor for any downstream quantum classifier, and that Neural Quantum Embedding can lower this floor by reshaping the representation before the data enter the quantum circuit: on binary MNIST the trace distance between class ensembles increased from approximately $0.27$ to $0.84$, translating into classification accuracy improving from $52.7\%$ to $96.1\%$ on IBM hardware. The DQC1/NMR extension reached $98\%$ accuracy on an NMR quantum device, compared to $54\%$ without NQE.
\Cref{chap:generalization} addressed the complementary finite-data question: margin-based metrics predicted generalization in quantum phase recognition more reliably than parameter-based metrics, and the connection between margin mean and trace distance links the two chapters---embeddings with larger state distinguishability can make larger margins attainable.

In Part~II, \Cref{chap:decoder} used machine learning for quantum error correction, where the bottleneck is real-time inference under physical timing constraints.
On Sycamore memory-experiment data, the Mamba decoder matched the accuracy of the reproduced Transformer baseline while its inference cost scales as $\mathcal{O}(d^2)$ rather than $\mathcal{O}(d^4)$ in the code distance.
Under simulated real-time decoding with decoder-induced noise, this latency advantage became an accuracy advantage, improving the finite-size effective threshold from $p_\text{th} \approx 0.0097$ to $p_\text{th} \approx 0.0104$.

\Cref{chap:nqs} studied neural quantum states, where the bottleneck is finite-sample optimization.
At a fixed wave function, stochastic reconfiguration is tangent-space ridge regression from sampled features to centered local energies, and the diagonal shift acts as a statistical spectral filter rather than only a numerical stabilizer.
At fixed large-system checkpoints, multi-shift stochastic reconfiguration (MS-SR) reduced the validation residual and multi-batch variance relative to standard SR at the training shift by approximately $39\%$/$43\%$ on TFIM and $14\%$/$66\%$ on $J_1$--$J_2$.
In five paired online continuations from one shared $8\times8$ $J_1$--$J_2$ checkpoint, independent endpoint evaluations favored MS-SR in three pairs, with mean energy difference per site $-1.78\times10^{-5}$ and paired $95\%$ Student-$t$ confidence interval $[-7.05,3.50]\times10^{-5}$.

The overall conclusion of this thesis is that quantum computing and AI benefit from a two-way exchange, in which each field supplies principled tools for the other's hardest problems.
Quantum computing challenges AI to operate with new mathematical objects and physical constraints; AI challenges quantum computing to adopt data-driven representations, scalable architectures, and statistical regularization for problems such as real-time decoding and many-body physics.
Meeting these challenges together will define the next generation of both quantum technology and machine intelligence.


\setlength\bibitemsep{\baselineskip}  
\clearpage
\addcontentsline{toc}{chapter}{References}  
\printbibliography[title={References}]


\clearpage
\begin{appendices}

\addtocontents{toc}{\protect\renewcommand{\protect\cftchappresnum}{\appendixname\space}}
\addtocontents{toc}{\protect\renewcommand{\protect\cftchapnumwidth}{7em}}

\chapter{Supplementary Technical Details for Main Contributions}
\label{chap:main-supplementary-details}

This appendix collects selected technical details, emphasizing derivations and reproducibility notes that are useful for reading the thesis but too detailed for the main narrative.

\section{Additional Details for Neural Quantum Embedding}
\label{sec:app-nqe-details}

\subsection{Implicit Fidelity Loss and Trace Distance}
\label{subsec:app-nqe-fidelity-trace}

\Cref{subsec:ch2-nqe-training} explains that the NQE loss is a pairwise fidelity objective used as a tractable proxy for increasing the trace distance between class-averaged embedded states. For a class $y\in\{+,-\}$, the empirical ensemble is
\begin{equation}
    \rho^{y}
    =
    \frac{1}{N^y}
    \sum_{i=1}^{N^y}
    \ket{x_i^y}\bra{x_i^y}.
    \label{eq:app-nqe-class-ensemble}
\end{equation}
The purity of this ensemble is determined by within-class fidelities:
\begin{equation}
    \mathrm{Tr}\!\left[(\rho^y)^2\right]
    =
    \frac{1}{(N^y)^2}
    \sum_{i,j=1}^{N^y}
    \left|\braket{x_i^y|x_j^y}\right|^2.
    \label{eq:app-nqe-purity}
\end{equation}
Thus, the same-label part of the NQE loss increases class purity by pushing within-class fidelities toward one. This matters because the trace distance between mixed class ensembles is upper bounded by the trace distance between purifications, and the bound is tightest when the empirical ensembles approach pure states.

For different-label pairs, the loss pushes cross-class fidelities toward zero. In a balanced binary dataset with paired examples, the convexity of trace distance gives
\begin{equation}
    D_{\mathrm{tr}}\!\left(
        \frac{1}{N}\sum_{i=1}^{N}\ket{x_i^-}\bra{x_i^-},
        \frac{1}{N}\sum_{i=1}^{N}\ket{x_i^+}\bra{x_i^+}
    \right)
    \leq
    \frac{1}{N}
    \sum_{i=1}^{N}
    \sqrt{1-\left|\braket{x_i^+|x_i^-}\right|^2}.
    \label{eq:app-nqe-convex-trace}
\end{equation}
Therefore, reducing cross-class fidelities increases the right-hand side of \Cref{eq:app-nqe-convex-trace}, while increasing same-class fidelities makes the mixed ensembles closer to pure representatives. Taken together, these two effects explain why the implicit fidelity objective is aligned with the trace-distance objective used in the empirical-risk lower bound.

We also relates the trace-distance-oriented linear-loss analysis to mean-squared error objectives often used in QML training. Let $Y=(y_1,\ldots,y_N)$ and $f(X)=(f(x_1),\ldots,f(x_N))$. If
\begin{equation}
    L_1=\|Y-f(X)\|_1,
    \qquad
    L_{\mathrm{MSE}}=\|Y-f(X)\|_2^2,
\end{equation}
then the standard vector-norm inequalities imply
\begin{equation}
    \frac{1}{N}L_1^2
    \leq
    L_{\mathrm{MSE}}
    \leq
    L_1^2.
    \label{eq:app-nqe-linear-mse}
\end{equation}
Consequently, improving the embedding-induced lower bound for the linear misclassification loss also improves the range in which the empirical MSE loss can lie.
\Cref{tab:app-nqe-protocol} summarizes the protocol details that are most relevant for reproducing the Chapter~2 NQE experiments.

\begin{table}[!htbp]
    \centering
    \small
    \caption{Selected reproducibility details for the NQE experiments in \textcite{hur2024nqe}.}
    \label{tab:app-nqe-protocol}
    \begin{adjustbox}{width=\textwidth}
    \begin{tabular}{>{\raggedright\arraybackslash}p{0.20\textwidth}
                  | >{\raggedright\arraybackslash}p{0.38\textwidth}
                  | >{\raggedright\arraybackslash}p{0.36\textwidth}}
        \toprule
        \textbf{Component} & \textbf{Configuration} & \textbf{Purpose in the thesis argument} \\
        \midrule
        PCA-NQE network
        & PCA reduces the input to $n$ features. A fully connected ReLU network with two hidden layers maps these to $2n$ feature-map parameters.
        & Shows that the NQE effect is not tied to a large classical frontend. \\
        \midrule
        CNN-NQE network
        & A two-dimensional CNN processes the original image with pooling stages reducing spatial scale from $28\times28$ to $14\times14$ and $7\times7$, followed by an output layer of size $2n$.
        & Provides the image-native variant used when the input is not first compressed by PCA. \\
        \midrule
        Real-device NQE training
        & Four-qubit hardware runs used SGD for 50 iterations, learning rate $0.1$, batch size 10. The loss was evaluated on ibmq\_toronto using qubits selected by high CNOT fidelity. The ZZ feature map used one repetition with nearest-neighbor two-qubit gates.
        & Explains why the hardware experiment is a shallow-circuit test of embedding quality rather than a deep quantum-classifier benchmark. \\
        \midrule
        Noiseless QCNN classification
        & QCNN classifiers used a general SU(4) convolutional ansatz, Nesterov momentum for 1000 iterations, learning rate $0.01$, batch size 128, and five random initializations.
        & Separates embedding quality from hardware noise and optimization variability. \\
        \midrule
        Hardware QCNN classification
        & Hardware classifiers used a basic convolutional ansatz with two $R_y$ rotations and a CNOT, omitted pooling gates to reduce circuit depth, trained for 50 iterations (lr $0.1$, batch size 10), and evaluated 500 test samples with 1024 shots.
        & Supports the claim that increasing trace distance can matter more than moderate device noise in the reported setting. \\
        \bottomrule
    \end{tabular}
    \end{adjustbox}
\end{table}

\paragraph{Feature-map ansatz robustness.}
The standard ZZ feature map used in the NQE-DQC1 experiments is a Hamiltonian-inspired encoding. The single-repetition circuit block
\begin{equation}
    V(\phi)
    =
    \left[
        \exp\!\left(
            i\sum_k \phi_k Z_k
            +
            i\sum_{k<l}\phi_{k,l}Z_kZ_l
        \right)
        H^{\otimes n}
    \right]^{M}
    \label{eq:app-nqe-zz-ansatz}
\end{equation}
is, via the conjugation identity $HZH=X$, equivalent to a first-order Trotter expansion of time evolution under a Hamiltonian containing both $X$-type and $Z$-type interactions:
\begin{equation}
    H_{XZ}
    =
    \sum_k \alpha_k\bigl(X_k + Z_k\bigr)
    +
    \sum_{k,l}\beta_{k,l}\bigl(X_kX_l + Z_kZ_l\bigr).
    \label{eq:app-nqe-hxz}
\end{equation}
To test whether NQE performance is sensitive to this particular Pauli structure, we also evalutated three alternative Hamiltonian-inspired encoding circuits using $XY$, $YZ$, and $XYZ$ interactions. Each circuit preserves the same layer structure---alternating single-qubit rotations and nearest-neighbor two-qubit entangling terms---with the Pauli operators exchanged:
\begin{align}
    V_{XY}(\phi)
    &=
    \left\{
        \exp\!\left[i\sum_k \phi_k Y_k + \phi_{n+k}Y_kY_{k+1}\right]
        \exp\!\left[i\sum_k \phi_k X_k + \phi_{n+k}X_kX_{k+1}\right]
    \right\}^{M/2},
    \label{eq:app-nqe-vxy}
    \\[2pt]
    V_{YZ}(\phi)
    &=
    \left\{
        \exp\!\left[i\sum_k \phi_k Z_k + \phi_{n+k}Z_kZ_{k+1}\right]
        \exp\!\left[i\sum_k \phi_k Y_k + \phi_{n+k}Y_kY_{k+1}\right]
    \right\}^{M/2},
    \label{eq:app-nqe-vyz}
    \\[2pt]
    V_{XYZ}(\phi)
    &=
    \Biggl\{
        \exp\!\left[i\sum_k \phi_k Z_k + \phi_{n+k}Z_kZ_{k+1}\right]
        \exp\!\left[i\sum_k \phi_k Y_k + \phi_{n+k}Y_kY_{k+1}\right]
        \nonumber\\
    &\hspace{2em}\cdot\,
        \exp\!\left[i\sum_k \phi_k X_k + \phi_{n+k}X_kX_{k+1}\right]
    \Biggr\}^{M/2}.
    \label{eq:app-nqe-vxyz}
\end{align}
The effective Hamiltonians underlying these three circuits are, respectively,
\begin{align}
    H_{XY}  &= \sum_k \alpha_k\bigl(X_k + Y_k\bigr) + \sum_k \beta_k\bigl(X_kX_{k+1} + Y_kY_{k+1}\bigr),
    \label{eq:app-nqe-hxy}\\
    H_{YZ}  &= \sum_k \alpha_k\bigl(Y_k + Z_k\bigr) + \sum_k \beta_k\bigl(Y_kY_{k+1} + Z_kZ_{k+1}\bigr),
    \label{eq:app-nqe-hyz}\\
    H_{XYZ} &= \sum_k \alpha_k\bigl(X_k + Y_k + Z_k\bigr) + \sum_k \beta_k\bigl(X_kX_{k+1} + Y_kY_{k+1} + Z_kZ_{k+1}\bigr).
    \label{eq:app-nqe-hxyz}
\end{align}
All three circuits use depth $M=4$ and are tested on MNIST and Fashion-MNIST in place of the standard $H_{XZ}$ map.

\begin{figure}[ht]
    \centering
    \includegraphics[width=\textwidth]{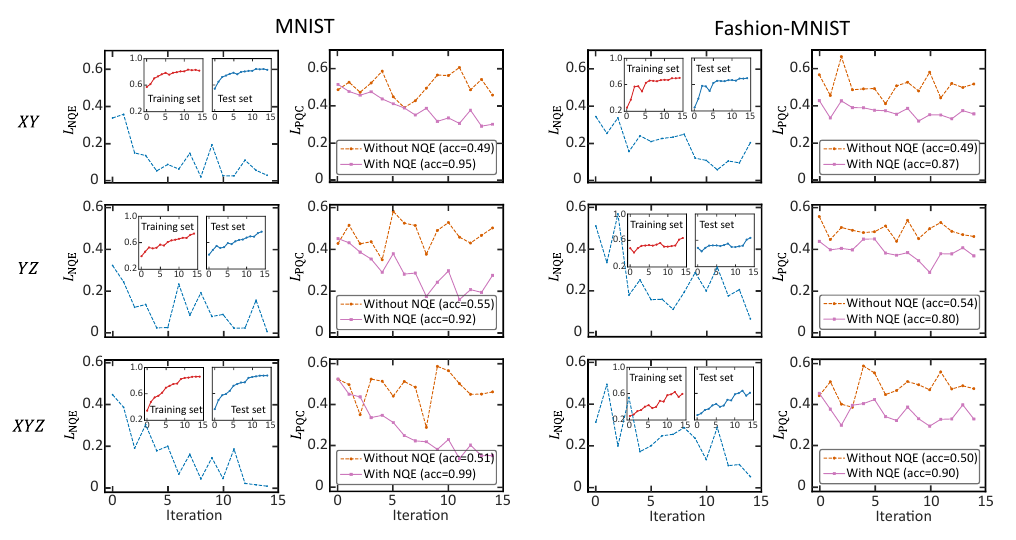}
    \caption{NQE-DQC1 training results for the three alternative Hamiltonian-inspired feature maps ($H_{XY}$, $H_{YZ}$, $H_{XYZ}$) on MNIST (left four columns) and Fashion-MNIST (right four columns). For each ansatz row, the left panel of each dataset shows the NQE loss $L_{\mathrm{NQE}}$ with the class trace distance during training in the inset; the right panel shows the downstream PQC loss $L_{\mathrm{PQC}}$ with (solid) and without (dashed) NQE. All ansatzes use circuit depth $M=4$. The consistent decrease in $L_{\mathrm{NQE}}$ and improvement in $L_{\mathrm{PQC}}$ confirm that the NQE benefit is not specific to the $ZZ$ ($H_{XZ}$) feature map.}
    \label{fig:app-nqe-other-ansatzes}
\end{figure}

\Cref{fig:app-nqe-other-ansatzes} shows the training curves for each ansatz. Across all six dataset--ansatz combinations, the same qualitative behavior holds: $L_{\mathrm{NQE}}$ decreases while the class trace distance (inset) increases, and the downstream PQC loss $L_{\mathrm{PQC}}$ converges to a lower value than the without-NQE baseline. The classification accuracies with NQE are $0.95$, $0.92$, and $0.99$ on MNIST for $H_{XY}$, $H_{YZ}$, and $H_{XYZ}$ respectively, compared with without-NQE baselines of $0.49$, $0.55$, and $0.81$. On Fashion-MNIST, NQE achieves $0.87$, $0.80$, and $0.90$ against baselines of $0.49$, $0.54$, and $0.50$. These results confirm that NQE robustly improves embedding quality regardless of the specific Pauli-interaction structure in the feature-map ansatz.

\paragraph{Multi-class extension.}
For a label set with more than two classes, the NQE-DQC1 objective can be extended by standard one-vs-one or one-vs-rest reductions. A direct multi-class objective is also possible by replacing the binary target with a Kronecker delta:
\begin{equation}
    L_{\mathrm{NQE}}
    =
    \sum_{i,j}
    \left[
        \frac{1}{2^n}
        \mathrm{Tr}\!\left[
            V(g(x_i))V^\dagger(g(x_j))
        \right]
        -
        \delta_{y_i y_j}
    \right]^2.
    \label{eq:app-nqe-multiclass}
\end{equation}
This remains DQC1-compatible because each term only requires estimating a Hilbert--Schmidt inner product between two data-dependent unitaries.

\section{Additional Details for Margin Generalization}
\label{sec:app-margin-details}

\subsection{Proof Structure and Mixed-State Extension}
\label{subsec:app-margin-proof}

\Cref{thm:ch3-margin-bound} is derived by combining a ramp-loss margin argument with covering-number bounds for the QNN-induced function class. First, define the ramp-loss class
\begin{equation}
    \mathcal{F}_\gamma
    =
    \left\{
    (\rho,y)\mapsto
    l_\gamma(\mathcal{M}(h(\rho),y))
    :
    h\in\mathcal{H}
    \right\}.
    \label{eq:app-margin-ramp-class}
\end{equation}
The standard Rademacher-complexity bound gives, with probability at least $1-\delta$, a uniform bound on the expected ramp loss in terms of the empirical ramp loss, the sample Rademacher complexity of $\mathcal{F}_\gamma$, and the concentration term $\sqrt{\ln(2/\delta)/(2m)}$. Since the ramp loss upper bounds the zero-one loss and is upper bounded by the empirical margin error at threshold $\gamma$, the remaining problem is to control $\mathfrak{R}(\mathcal{F}_\gamma|_S)$.

For pure input states, write the classifier output as
\begin{equation}
    g(Ux)
    =
    \left(
        x^\dagger U^\dagger E_1Ux,
        \ldots,
        x^\dagger U^\dagger E_kUx
    \right).
\end{equation}
The composed ramp-margin map is Lipschitz with constant proportional to
$
    E/\gamma 
$
, where
$
    E
    =
    \sqrt{\sum_{i=1}^{k}\|E_i\|_\sigma^2}.
    \label{eq:app-margin-measurement-norm}
$
Thus, the covering number of the restricted ramp-loss class can be controlled by the covering number of the set of transformed training states $\{UX:U\in\mathbb{U}_{\mathrm{QNN}}\}$:
\begin{equation}
    \ln \mathcal{N}\!\left(
        \mathcal{F}_\gamma|_S,
        \epsilon,
        \|\cdot\|_2
    \right)
    \leq
    \left\lceil
        \frac{32mb^2E^2}{\epsilon^2\gamma^2}
    \right\rceil
    \ln(4N^2),
    \label{eq:app-margin-cover}
\end{equation}
where $N=2^n$ is the Hilbert-space dimension and $b$ bounds the distance of the QNN unitary from a reference unitary in the $\|\cdot\|_{2,1}$ sense. Applying Dudley's entropy integral to \Cref{eq:app-margin-cover} yields the Rademacher term
\begin{equation}
    \mathfrak{R}(\mathcal{F}_\gamma|_S)
    =
    \tilde{O}\!\left(
        \frac{bE}{\gamma}\sqrt{\frac{n}{m}}
    \right),
    \label{eq:app-margin-rademacher}
\end{equation}
which gives the theorem in \Cref{eq:ch3-main-bound}.

The same argument extends to mixed input states by vectorization. For
\begin{equation}
    \rho
    =
    \sum_{i,j}\rho_{ij}\ket{i}\bra{j},
    \qquad
    \lvert\rho\rangle\!\rangle
    =
    \sum_{i,j}\rho_{ij}\ket{i}\otimes\ket{j},
\end{equation}
the unitary action satisfies
\begin{equation}
    \lvert U\rho U^\dagger\rangle\!\rangle
    =
    (U\otimes U^*)\lvert\rho\rangle\!\rangle.
    \label{eq:app-margin-vectorization}
\end{equation}
The proof then proceeds with the distance bound applied to $U\otimes U^*$ rather than $U$, and with the measurement-size term evaluated using the Frobenius norms of the POVM elements.

\subsection{Experimental Protocol Notes}
\label{subsec:app-margin-experimental-notes}

\Cref{tab:app-margin-protocol} records the experimental choices from \textcite{hur2024margin} that are useful for interpreting the margin plots in \Cref{chap:generalization}.

\begin{table}[!htbp]
    \centering
    \small
    \caption{Selected reproducibility details for the margin-generalization experiments.}
    \label{tab:app-margin-protocol}
    \begin{adjustbox}{width=\textwidth}
    \begin{tabular}{>{\raggedright\arraybackslash}p{0.22\textwidth}
                  | >{\raggedright\arraybackslash}p{0.72\textwidth}}
        \toprule
        \textbf{Experiment} & \textbf{Protocol details} \\
        \midrule
        Quantum phase recognition
        & The QPR task uses an eight-qubit generalized cluster Hamiltonian with four phases.
          The training set has 20 data points evenly split across the four classes,
          and the test set has 1000 samples.
          Label corruption levels of 0\%, 50\%, and 100\% probe the overfitting regime. \\
        \midrule
        QPR optimization
        & QCNNs are trained with Adam, learning rate $0.001$, and full-batch updates
          for up to 5000 iterations.
          Early stopping compares consecutive 500-iteration loss windows.
          Margin-distribution plots average over 15 repetitions with different training samples. \\
        \midrule
        Ansatz comparisons
        & The main comparison uses QCNNs with distinct two-qubit PQC parameters.
          The supplement repeats the margin-distribution test with shared-parameter QCNNs
          and Strongly Entangling Layers, showing the same qualitative leftward margin shift
          as label corruption increases. \\
        \midrule
        Embedding comparison
        & The classical-data experiment uses the first two classes of MNIST, Fashion-MNIST,
          and Kuzushiji-MNIST. PCA reduces inputs to match the qubit count.
          The compared embeddings are a three-layer ZZ feature map, a trainable quantum
          embedding with $Y$ and $YY$ rotations, and NQE with a fully connected ReLU
          network of layer dimensions $[8,16,32,32,16,8]$. \\
        \bottomrule
    \end{tabular}
    \end{adjustbox}
\end{table}

\section{Additional Details for Stochastic Reconfiguration as Spectral Filtering}
\label{sec:app-sr-details}

\subsection{Fixed-Checkpoint Regression Identities}
\label{subsec:app-sr-identities}

\Cref{chap:nqs} presents the real-valued fixed-checkpoint regression view of SR. The appendix of the SR paper~\cite{hur2026srfilter} provides several identities that clarify the assumptions behind that view.

\paragraph{Complex convention.}
For complex wave functions, define
\begin{equation}
    O_c(x)
    =
    \nabla_\theta\log\psi_\theta(x)
    -
    \mathbb{E}_{\pi_\theta}[\nabla_\theta\log\psi_\theta],
    \qquad
    H_{\mathrm{loc},c}(x)
    =
    H_{\mathrm{loc}}(x)
    -
    \mathbb{E}_{\pi_\theta}[H_{\mathrm{loc}}].
\end{equation}
The Hermitian QGT and force are
\begin{equation}
    Q
    =
    \mathbb{E}_{\pi_\theta}[\overline{O_c(x)}\,O_c(x)^T],
    \qquad
    f
    =
    \mathbb{E}_{\pi_\theta}[\overline{O_c(x)}\,H_{\mathrm{loc},c}(x)].
    \label{eq:app-sr-complex-qgt}
\end{equation}
For complex parameter increments, SR solves $Q\delta=f$ in the QGT seminorm. For real parameter increments in a complex wave function, the equivalent real least-squares system is
\begin{equation}
    \mathrm{Re}[Q]\,\delta
    =
    \mathrm{Re}[f],
    \label{eq:app-sr-real-parameter-system}
\end{equation}
with $S=\mathrm{Re}[Q]$ and $g=\mathrm{Re}[f]$. The regression loss is $\mathbb{E}_{\pi_\theta}|O_c(x)^T\delta-H_{\mathrm{loc},c}(x)|^2$.
The identities below use real-valued features and targets. For the complex extension, scalar squares become modulus squares and the seminorm uses the corresponding real or Hermitian QGT.

\paragraph{Validation residual.}
Let $\delta^*$ be the population least-squares SR direction and define the expressivity gap
\begin{equation}
    \epsilon(x)
    =
    H_{\mathrm{loc},c}(x)-O_c(x)^T\delta^*.
\end{equation}
The normal equations give $\mathbb{E}_{\pi_\theta}[O_c(x)\epsilon(x)]=0$. Therefore, for any fixed update $\delta$,
\begin{equation}
    \mathbb{E}_{\pi_\theta}
    \left[
        \left(O_c(x)^T\delta-H_{\mathrm{loc},c}(x)\right)^2
    \right]
    =
    \|\delta-\delta^*\|_S^2
    +
    \sigma_{\mathrm{gap}}^2,
    \label{eq:app-sr-validation-identity}
\end{equation}
where $\sigma_{\mathrm{gap}}^2=\mathbb{E}_{\pi_\theta}[\epsilon(x)^2]$. This is the identity behind \Cref{eq:ch5-validation-identity}: held-out tangent residuals estimate excess risk plus a constant irreducible gap.

\paragraph{Bias--variance decomposition.}
At a fixed checkpoint, let $Z=\hat{\delta}_\lambda$ be the random shifted-SR update obtained by resampling the Monte Carlo batch, and let $\mu=\mathbb{E}_D[Z]$. Then
\begin{equation}
    \mathbb{E}_D\|Z-\delta^*\|_S^2
    =
    \|\mu-\delta^*\|_S^2
    +
    \mathbb{E}_D\|Z-\mu\|_S^2.
    \label{eq:app-sr-bias-variance}
\end{equation}
The multi-batch diagnostic in \Cref{eq:ch5-multi-batch-variance} estimates the second term by evaluating the prediction-space disagreement among independent updates on a validation batch.

\paragraph{Spectral proxy and local geometry.}
On the non-null QGT subspace, use the eigendecomposition
\begin{equation}
    S=V\mathrm{diag}(s_i)V^T,
    \qquad
    \beta_i^*=(V^T\delta^*)_i.
\end{equation}
The population ridge direction has shrinkage bias
\begin{equation}
    \|\delta_\lambda-\delta^*\|_S^2
    =
    \sum_i
    s_i
    \left(\frac{\lambda}{s_i+\lambda}\right)^2
    (\beta_i^*)^2.
    \label{eq:app-sr-shrinkage}
\end{equation}
Under a scalar residual-noise approximation, the finite-sample variance scales as
\begin{equation}
    \frac{\sigma_{\mathrm{gap}}^2}{N_s}
    \sum_i
    \left(\frac{s_i}{s_i+\lambda}\right)^2,
    \label{eq:app-sr-variance-proxy}
\end{equation}
giving the spectral-risk proxy in \Cref{eq:ch5-spectral-risk}. The same $S$-seminorm also controls local Fubini--Study infidelity. For nearby displacements $a$ and $b$,
\begin{equation}
    1-\left|\braket{\Psi(\theta+a)|\Psi(\theta+b)}\right|^2
    =
    (a-b)^TS(a-b)+O((\|a\|+\|b\|)^3).
    \label{eq:app-sr-fs-local}
\end{equation}
With the full QGT, regression excess risk therefore measures the discrepancy from the population tangent-space imaginary-time step in the local state geometry.
The small-system diagnostics in \Cref{chap:nqs} instead retain only the amplitude features $\nabla_\theta\mathrm{Re}\log\psi_\theta$ and the real local-energy target, while evaluating local energies and Born probabilities from the complex state. Their covariance omits the phase channel, so those amplitude regression risks do not measure full complex-SR error or complete local infidelity.

\paragraph{Sample centering.}
Practical SR uses sample-centered log-derivatives and sample-centered local energies. This is equivalent to ridge regression with an unpenalized intercept:
\begin{equation}
    \min_{a,\delta}
    \frac{1}{n}
    \sum_{j=1}^{n}
    \left(
        a+O(x_j)^T\delta-H_{\mathrm{loc}}(x_j)
    \right)^2
    +
    \lambda\|\delta\|_2^2.
    \label{eq:app-sr-intercept}
\end{equation}
Optimizing over $a$ subtracts the sample means of $O$ and $H_{\mathrm{loc}}$. This removes the constant energy/normalization component but does not remove the $x$-dependent expressivity gap that drives the finite-sample overfitting effect.

\paragraph{Large-system normalization.}
The large-scale implementation uses target $2H_{\mathrm{loc},c}$ and correspondingly doubled SR directions. Its raw validation residuals and multi-batch variances are therefore four times those under the $H_{\mathrm{loc},c}$ convention used in the regression derivation; relative comparisons and method ordering are unchanged.
The large $J_1$--$J_2$ experiments also use Pauli operators. Converting that Hamiltonian and its SR directions to the spin convention $\mathbf S=\boldsymbol\sigma/2$ would divide the squared diagnostics by $16$, independently of the doubled-force convention. The reported figures retain the original normalization.

\Cref{tab:app-sr-protocol} summarizes the reproducibility details from the SR paper~\cite{hur2026srfilter} that are most relevant to the claims in \Cref{chap:nqs}.

\begin{table}[!htbp]
    \centering
    \small
    \caption{Selected protocol details for the SR spectral-filtering experiments.}
    \label{tab:app-sr-protocol}
    \begin{adjustbox}{width=\textwidth,max totalheight=0.9\textheight}
    \begin{tabular}{>{\raggedright\arraybackslash}p{0.22\textwidth}
                  | >{\raggedright\arraybackslash}p{0.72\textwidth}}
        \toprule
        \textbf{Setting} & \textbf{Details} \\
        \midrule
        Exact $4\times4$ $J_1$--$J_2$
        & Archived graph with 32 periodic nearest-neighbor bonds and 24 diagonal
          bonds, omitting eight diagonal bonds across one seam. Couplings are
          $J_1=1$, $J_2=0.5$ in the spin convention $\mathbf S=\boldsymbol\sigma/2$.
          All $2^{16}=65{,}536$ states enter the exact Born sums for the amplitude
          covariance, amplitude-optimal update, and amplitude gap variance.
          The RBM has 1120 real parameters at $P/N_s=0.273$ with $N_s=4096$.
          The ViT has 13,448 real parameters at $P/N_s=3.283$. \\
        \midrule
        Matched-noise and matched-state diagnostics
        & Matched-noise pairs independently trained RBM and ViT checkpoints with
          comparable amplitude gap variance and draws 100 independent batches at
          $\lambda=10^{-9}$. Matched-state first fits a ViT to a partially converged
          RBM wave function to fidelity $F=0.983626$, then compares amplitude
          regression diagnostics from 10 resampled batches. \\
        \midrule
        1D TFIM foundational NQS
        & The large-scale TFIM diagnostic uses a periodic $L=100$ transverse-field
          Ising family with $h\in[0.8,1.2]$. The conditional ViT foundation NQS has
          six layers, model dimension 72, 12 attention heads, patch size 4,
          translational invariance, and 198,144 parameters. Each SR step uses
          12,000 samples from 6000 field replicas with two spin configurations per field. \\
        \midrule
        Large-scale diagnostics
        & The shared-checkpoint run uses training shift $\lambda_{\mathrm{train}}=10^{-4}$.
          Diagnostic checkpoints are steps 1100 through 2000 for TFIM. Solve-time shifts
          are swept over $\{10^{-8},10^{-7},\ldots,10^{-1}\}$. The TFIM U-curve uses
          100 independently resampled updates per shift; ablations use 20 repeats
          per checkpoint. Validation uses 120,000 held-out TFIM samples or 100,000
          held-out $J_1$--$J_2$ samples. \\
        \midrule
        MS-SR ablations
        & Offline ablations use $K=4$ NTK-spectrum quantiles $\{0.9,0.7,0.4,0.1\}$,
          cached per checkpoint from an independent reference batch. Comparisons
          include standard SR, shared-batch uniform and stacked multi-shift,
          independent-batch uniform multi-shift, and full MS-SR with leave-one-batch-out
          stacking. Bagged SR uses either the training shift or a checkpoint-tuned
          shift selected on the reporting data by the eight-point sweep above;
          the tuned variant is an oracle reference. \\
        \bottomrule
    \end{tabular}
    \end{adjustbox}
\end{table}

\chapter{Related Non-First-Author Contributions}
\label{chap:related-nonfirst-author}
This appendix records two closely related collaborative papers in which the NQE methods introduced in \Cref{chap:nqe} are extended and validated in application domains outside the primary thesis narrative. The goal is not to reproduce those papers in detail, but to summarize the connection to the thesis and the main empirical findings.

\Cref{tab:appendix-related-papers} gives a compact overview. Both papers extend the Neural Quantum Embedding perspective developed in \Cref{chap:nqe}: one applies trainable quantum embeddings to ligand-based virtual screening, and the other generalizes NQE to multi-channel image data.

\begin{table}[!htbp]
    \centering
    \small
    \caption{Summary of non-first-author contributions related to the main thesis.}
    \label{tab:appendix-related-papers}
    \begin{adjustbox}{width=\textwidth}
    \begin{tabular}{>{\raggedright\arraybackslash}p{0.22\textwidth}
                  | >{\raggedright\arraybackslash}p{0.16\textwidth}
                  | >{\raggedright\arraybackslash}p{0.18\textwidth}
                  | >{\raggedright\arraybackslash}p{0.20\textwidth}
                  | >{\raggedright\arraybackslash}p{0.24\textwidth}}
        \toprule
        \textbf{Paper} & \textbf{Thesis connection} & \textbf{Data domain} & \textbf{Method} & \textbf{Headline result} \\
        \midrule
        Optimizing Quantum Data Embeddings for Ligand-Based Virtual Screening~\cite{choi2025optimizing_quantum_embeddings}
        & NQE application (\Cref{chap:nqe})
        & Molecular descriptors from LIT-PCBA and COVID-19 ligand datasets
        & NQE with ZZ/XYZ feature maps, QCNN classifiers,
          quantum kernels, and quantum-pretrained classical embeddings
        & NQE increased class trace distances; quantum or hybrid variants
          outperformed classical RBF/linear baselines, especially in
          limited-data or imbalanced settings. \\
        \midrule
        Multi-Channel Convolutional Neural Quantum Embedding~\cite{kim2026multi_channel_cnqe}
        & NQE extension (\Cref{chap:nqe})
        & CIFAR-10 and Tiny ImageNet binary image tasks
        & Convolutional NQE interfaces for multi-channel data,
          followed by QCNN classification
        & Trace distance and classification accuracy were strongly correlated.
          Parameter-efficient multi-channel interfaces achieved competitive
          accuracy with fewer trainable parameters. \\
        \bottomrule
    \end{tabular}
    \end{adjustbox}
\end{table}

\section{Quantum Embeddings for Ligand-Based Virtual Screening}
\label{sec:appendix-ligand-nqe}

\textcite{choi2025optimizing_quantum_embeddings} studied whether optimized quantum data embeddings can improve ligand-based virtual screening, a setting where labeled biological data are often limited and class-imbalanced. The study used molecular descriptors computed from SMILES strings and evaluated two benchmark collections: LIT-PCBA, containing multiple biological targets with severe activator/inactivator imbalance, and a smaller COVID-19 ligand dataset. This application is naturally connected to \Cref{chap:nqe} because the central question is whether trainable quantum embeddings can produce more useful molecular representations than fixed or purely classical embeddings.

For the LIT-PCBA experiments, the paper compared NQE models using ZZ and XYZ quantum feature maps against a classical neural-network parameterized RBF embedding. The NQE-pretrained embeddings were then evaluated with QCNN classifiers, while the classical embeddings were evaluated with comparable shallow classical classifiers. In the ZZ-feature-map setting, NQE increased the trace distance between class ensembles after training across the reported conditions. For example, the GBA target in the balanced setting increased from approximately $5 \times 10^{-4}$ to about $0.55$ on the test set. This mirrors the main mechanism of \Cref{chap:nqe}: improving the embedding geometry can lower the effective discrimination barrier faced by the downstream classifier.

\begin{figure}[t]
    \centering
    \includegraphics[width=\textwidth]{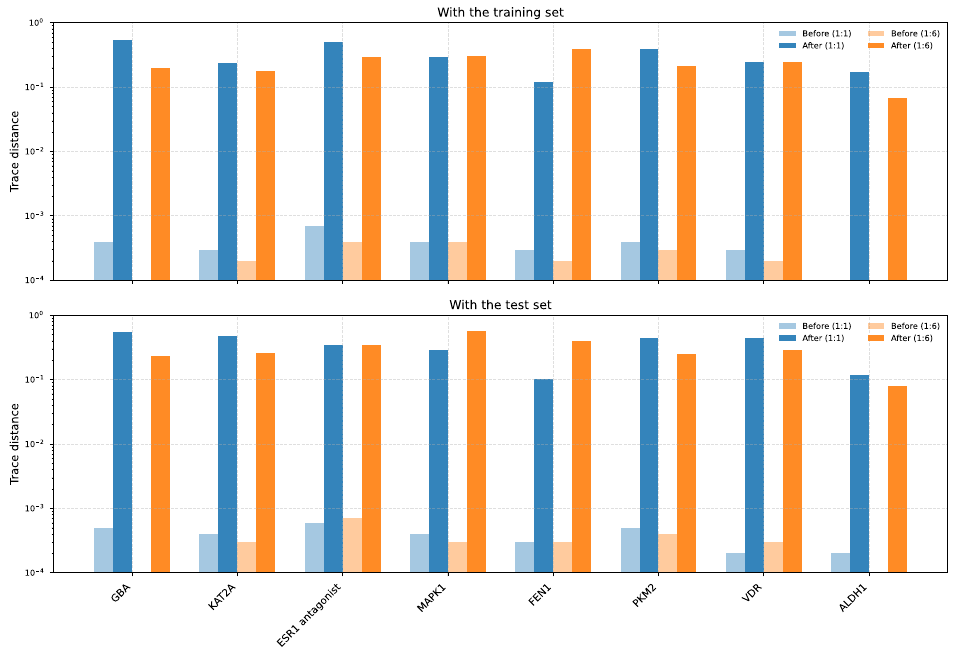}
    \caption{Trace-distance increase obtained by NQE training with the ZZ feature map on LIT-PCBA targets. The plot compares trace distances before and after NQE training for balanced (1:1) and imbalanced (1:6) activator/inactivator settings, separately for training and test sets. The log scale highlights that the untrained embeddings produce nearly indistinguishable class ensembles, while NQE raises the class separation by several orders of magnitude for many targets.}
    \label{fig:appendix-ligand-nqe-trace}
\end{figure}

\Cref{fig:appendix-ligand-nqe-trace} illustrates the most direct connection to the trace-distance argument in \Cref{chap:nqe}. Before training, the ZZ feature map typically produces class-ensemble trace distances near $10^{-4}$--$10^{-3}$ for these molecular descriptors. After NQE training, the distances move into the $10^{-1}$ range or higher for several targets, including GBA, ESR1 antagonist, MAPK1, FEN1, PKM2, and VDR. The effect is visible in both the training and test panels, indicating that the learned embedding is not merely separating the sampled training pairs but also improving the geometry of held-out ligand examples.

The paper also explored quantum-pretrained classical embeddings, where neural networks trained through NQE were reused in classical classifiers. These transfer-style variants often improved performance in limited-data or class-imbalanced cases, suggesting that the representation learned during quantum embedding optimization can be useful even outside a fully quantum classifier. On the COVID-19 dataset, projected quantum kernels with the ZZ feature map achieved balanced accuracies of $0.83$ and $0.80$ for 4- and 8-qubit configurations, respectively, compared with classical SVM baselines in the range $0.59$--$0.65$. The main takeaway is that NQE-style embedding optimization can be adapted to molecular screening as a data-efficient representation-learning tool, although hardware execution and biological interpretability remain directions for future work.

\section{Multi-Channel Convolutional Neural Quantum Embedding}
\label{sec:appendix-multichannel-cnqe}

\textcite{kim2026multi_channel_cnqe} extended the NQE framework to multi-channel image data through convolutional neural quantum embedding (CNQE). Whereas the original NQE experiments in \Cref{chap:nqe} focused mainly on low-dimensional or single-channel inputs after preprocessing, CNQE studies how a classical convolutional interface should map multi-channel data, such as RGB images, into quantum feature-map parameters. The work introduced three interface structures, denoted $g_a$, $g_b$, and $g_c$, that differ in whether channels are processed jointly or separately before quantum embedding.

The main theoretical and empirical point is that the classical-to-quantum interface controls both parameter efficiency and the separability of the resulting quantum states. The paper compared fidelity-based and Hilbert--Schmidt-based NQE losses, multiple embedding circuits, and QCNN classifiers on binary tasks from CIFAR-10 and Tiny ImageNet. Across the tested configurations, trace distance and downstream classification accuracy were strongly correlated, with Pearson correlation $r = 0.7926$ and Spearman correlation $\rho = 0.8184$ reported across model configurations. This supports the trace-distance view developed in \Cref{chap:nqe}: embeddings that better separate class ensembles tend to give the classifier a better achievable operating point.

\begin{figure}[t]
    \centering
    \includegraphics[width=0.8\textwidth]{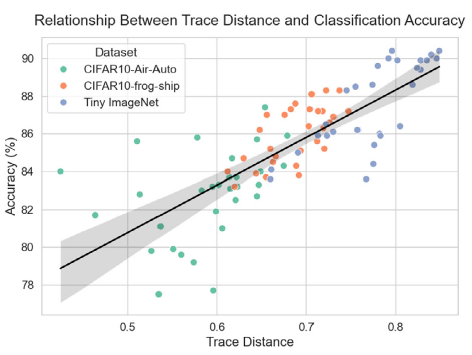}
    \caption{Relationship between class trace distance and QCNN classification accuracy across CNQE configurations on CIFAR-10 and Tiny ImageNet binary tasks. Each point corresponds to a choice of dataset, classical-to-quantum interface, loss function, and embedding circuit. The positive trend supports the use of trace distance as an embedding-quality diagnostic for multi-channel NQE.}
    \label{fig:appendix-cnqe-trace-accuracy}
\end{figure}

\Cref{fig:appendix-cnqe-trace-accuracy} is the key empirical diagnostic from the CNQE study. The figure does not show a single model comparison; instead, it aggregates many design choices and asks whether the NQE quantity optimized before classifier training remains predictive once a QCNN classifier is trained afterward. The observed monotone trend is important because it links architectural engineering choices, such as channel-wise versus joint-channel interfaces, back to the same state-discrimination quantity used throughout \Cref{chap:nqe}.

Empirically, CNQE-QCNN models achieved high accuracy on the selected multi-channel tasks, including results near $95\%$ on CIFAR-10 frog--ship and Tiny ImageNet school-bus--maypole binary classification tasks in favorable configurations. The channel-separated interface $g_c$ was particularly interesting: although it produced smaller trace distances than $g_a$ and $g_b$ in some statistical comparisons, it used the fewest trainable parameters and still achieved competitive classification accuracy. The paper therefore broadens NQE from a basic embedding optimizer into a practical design space for image-like, multi-channel inputs.

\end{appendices}

\end{document}